\documentclass[preprint,3p,12pt]{elsarticle}

\usepackage[normalem]{ulem}  

\renewcommand\sout{\bgroup \color{red} \ULdepth=-.5ex \ULset}

\usepackage{amssymb}
\usepackage{amsthm}
\usepackage{color}

\journal{Physica D}

\def\Vec#1{\mbox{\boldmath$#1$}}

\begin{document}

\begin{frontmatter}

\title{
  Causal hydrodynamics
  from kinetic theory
  by doublet scheme in renormalization-group method
}

\author{Kyosuke Tsumura}
\ead{kyosuke.tsumura@fujifilm.com}
\address{
  Analysis Technology Center,
  Research \& Development Management Headquarters,
  Fujifilm Corporation,
  Kanagawa 250-0193, Japan
}
\author{Yuta Kikuchi}
\author{Teiji Kunihiro}
\address{
    Department of Physics,
    Kyoto University,
    Kyoto 606-8502, Japan
}

\begin{abstract}
We develop a general framework in the renormalization-group (RG) method
for extracting a mesoscopic dynamics from an evolution equation
by incorporating some excited (fast) modes as additional components to the invariant manifold spanned by zero modes.
We call this framework the doublet scheme.
The validity of the doublet scheme is first tested and demonstrated by 
taking the Lorenz model as a simple three-dimensional dynamical system; 
it is shown that the two-dimensional reduced dynamics on 
the attractive manifold composed of the would-be zero and a fast modes
are successfully obtained in a natural way. 
We then apply the doublet scheme to
construct causal hydrodynamics
as a mesoscopic dynamics of kinetic theory, i.e., the Boltzmann equation,
in a systematic manner with no ad-hoc assumption.
It is found that
our equation has the same form as  thirteen-moment causal hydrodynamic equation,
but the microscopic formulae of the transport coefficients and relaxation times are different.
In fact,
in contrast to the Grad equation,
our equation leads to the same expressions for the transport coefficients
as given by the Chapman-Enskog expansion method
and suggests novel formulae of the relaxation times
expressed in terms of relaxation functions
which allow a natural physical interpretation of the relaxation times.
Furthermore,
our theory nicely gives the explicit forms of the distribution function and the 
thirteen hydrodynamic variables in terms of the linearized collision operator, 
which in turn clearly suggest the proper ansatz forms of them 
to be adopted in the method of moments.
\end{abstract}


\end{frontmatter}
\section{
  Introduction
}
\label{sec:ChapB-1}
Dissipative hydrodynamic equation is a powerful means for
describing the low-frequency and long-wavelength dynamics of many-body systems,
which are close to equilibrium state.
A typical equation is the Navier-Stokes equation,
whose dynamical variables are five fields consisting of temperature, density, and fluid velocity.

One of the problems inherent in the Navier-Stokes equation
is instantaneous propagation of information,
i.e., the lack of causality,
which is attributed to parabolicity of the equation \cite{intro003,intro004,intro005-A,intro005-B}.
Here,
the parabolicity is a character of diffusion equations
containing
first-order (second-order) temporal-derivative (spatial-derivative) terms
of dynamical variables.
This character plagues
a relativistically covariant extension of the Navier-Stokes equation.

In 1949, Grad \cite{intro001} showed that
the lack of causality could be circumvented
within the framework of kinetic theory, i.e., the Boltzmann equation
by employing a method of moments,
where an ad-hoc but seemingly plausible ansatz
for the functional forms of the distribution function
and the moments
leads to a closed system of differential equations 
as the hydrodynamic equations.
In particular, for the thirteen-moment approximation
to the functional forms,
the resultant equation is similar to the Navier-Stokes equation
but respects the causality,
because the character of the equation
is hyperbolic instead of parabolic,
with finite propagation speeds.
This thirteen-moment causal equation is called the Grad equation,
whose dynamical variables are
thirteen fields,
i.e.,
temperature, density, fluid velocity,
viscous pressure, and heat flux.

In 1996,
Jou and his collaborators \cite{mesoscopic,intro018} called
the description by the Grad equation
\textit{mesoscopic}
since it occupies an intermediate level
between the descriptions by the Navier-Stokes equation and the Boltzmann equation.
In fact,
the Grad equation has been applied to
various kinetic problems, e.g., in plasma and in photon transport \cite{balescu},
whose dynamics often cannot be described by the Navier-Stokes equation
since the systems are not close enough to the equilibrium state.

However, it has recently turned out that
the dynamics described by the Grad equation
is not consistent with
the Boltzmann equation
in the mesoscopic scales of space and time.
In fact,
numerical simulations \cite{torrilhon} have shown that
the solutions to the Grad equation
are in disagreement with 
those to the Boltzmann equation in the mesoscopic regime.
Although
a lot of efforts
have been made to 
reformulate the method of moments 
so as to get solutions more faithful to the Boltzmann equation \cite{torrilhon},
the resultant equations 
are found to still violate the causality, 
unfortunately \cite{intro013-A,intro013-B,intro014}.
Although various extensions 
of Grad's theory 
has been proposed \cite{intro015,intro016,intro017}
so as to circumvent the causality problem,
the whole consistency between the resultant equations
and the solution of the Boltzmann equation in the mesoscopic regime
seems not yet achieved. 
Thus,
it is still a challenge to construct
the causal hydrodynamic equation consistent with the Boltzmann equation.

The purpose of this paper is to construct
the hydrodynamic equation that respects both the causality and 
the consistency with the Boltzmann equation in the mesoscopic regime.
A natural approach to this end 
is to solve the Boltzmann equation faithfully  
in a way valid up to the mesoscopic regime
and extract from the solution
a simpler equation describing the mesoscopic dynamics of the Boltzmann equation.
Here,
we note that
the faithful solution of the Boltzmann equation should lead to an equation that
is free from
any ansatz for
the functional forms of the distribution function
and
respects the causality
that the underlying microscopic theory possesses.

The problem we are facing is to 
solve a non-linear differential equation such as the Boltzmann equation in an asymptotic regime
and extract an effective action or equation with a simpler form than the original one,
which reproduces  solutions in the asymptotic regime.
This is a reduction of the dynamics, and there are some powerful reduction 
theories of the dynamics \cite{kuramoto}.
As a reduction theory,
we take the ``renormalization-group (RG) method''
\cite{rgm001,rgm002,rgm003,rgm004,rgm005,rgm006,rgm007,rgm008,env001,env002,env005,qm,env006,env007,env008}.
The RG method as formulated in Refs. \cite{env001,env002,env005,qm,env006,env007,env008}
is a powerful tool for reducing evolution equations
based on the notion of
attractive manifold or invariant manifold \cite{geometrical},
which
the dynamical variables approach to and after some time are eventually confined in.
In fact,
the RG method \cite{env001,env002,env005,qm,env006,env007,env008}
has been applied to reduce kinetic equations
to a slower dynamics
with fewer degrees of freedom,
which is realized on the invariant manifold asymptotically.

Hatta and the authors \cite{env007,env008} used the RG method
to derive the Navier-Stokes equation from the Boltzmann equation.
An essential point in the derivation of the Navier-Stokes equation
was to utilize five zero modes of the linearized collision operator,
which form the invariant manifold on which hydrodynamics is defined;
the would-be constant five zero modes,
corresponding to temperature, density, and fluid velocity,
acquire the time dependence on the manifold by the RG equation.

\begin{figure}
 \begin{minipage}{\hsize}
 \begin{center}
  \includegraphics[width=8cm]{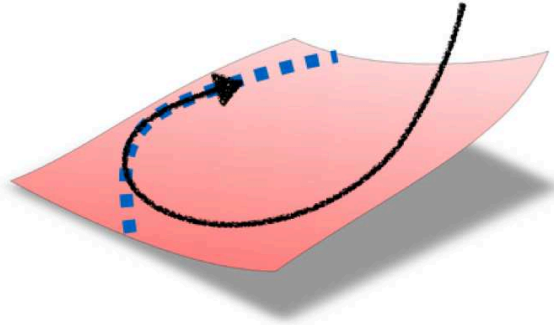}
 \end{center}
 \end{minipage}
 \caption{\label{fig:A}
A schematic 
picture of
the invariant/attractive manifold spanned by the zero 
and excited modes.
The solid and dashed lines denote 
the orbit of an exact solution to the microscopic (Boltzmann) equation,
and the asymptotic one  on the 
invariant/attractive manifold 
spanned only by the zero modes, respectively,
while
the surface shows the invariant/attractive manifold 
extended by incorporating
excited modes.
Under the time evolution of the system,
the exact solution starting from a point away from the surface
is rapidly attracted along the solid line
to the surface, then 
after performing a fast motion on it,
the solution approaches  
the dashed line,
and eventually shows a slow motion confined on it. 
}
\end{figure}

Thus,
a basic observation presented in the extraction of the mesoscopic dynamics from the Boltzmann equation
is to include some excited (fast) modes
as additional components of the invariant/attractive manifold,
because the mesoscopic dynamics is faster than hydrodynamics
that is defined on the invariant manifold spanned by the five zero modes:
Figure \ref{fig:A} gives a
schematic picture of the invariant/attractive manifold 
composed of the (would-be) zero and excited modes.

Which excited modes should we adopt?
In this paper,
we try to determine the excited modes
based on the following
basic principle in the reduction theory of the dynamics \cite{kuramoto}:
The resultant dynamics should be as simple as possible because we are interested to 
reduce the dynamics to a simpler one.
Here, we note that
this principle
is the very spirit of the reduction theory of the dynamics \cite{kuramoto},
and here the term ``simple" is used to express that
the resultant dynamics is described with a fewer number of dynamical variables
and is given by an equation composed of a fewer number of terms.

We demonstrate that
this
principle leads
to the \textit{doublet scheme} in the RG method,
which uniquely determines the number and form of the excited modes
that should be included in the invariant/attractive manifold
on which the mesoscopic dynamics of the Boltzmann equation is defined:
The doublet scheme can be applied to a wide class of evolution equations.
We also demonstrate that the
mesoscopic dynamics obtained by the RG equation
contains thirteen dynamical variables
and respects the causality.
We show that
the form of the resultant equation
is the same as that of the Grad equation,
but the microscopic formulae of the coefficients,
e.g., the transport coefficients and relaxation times, are different,
and our theory leads to the same expressions for the transport coefficients
as given by the Chapman-Enskog method \cite{chapman-enskog}
and also novel formulae of the relaxation times
in terms of relaxation functions,
which allow a natural physical interpretation of the relaxation times.
Moreover,
the distribution function and the moments which are explicitly
constructed in our theory
provide a proper new ansatz
for the functional forms of the distribution function and the moments 
in the method of moments proposed by Grad.

We here remark that
some results in the present paper
have been announced in the proceedings \cite{TK2012PTP}
by the present authors.
In the present paper,
we shall 
make a detailed and complete account on
the derivations of the causal hydrodynamic equations
together with those of 
the microscopic expressions of the 
transport coefficients and relaxation times.

This paper is organized as follows:
In Sec. \ref{sec:ChapB-3},
we describe the doublet scheme in the RG method.
In Sec. \ref{sec:C},
we analyze the Lorenz model \cite{lorentz} in the doublet scheme
and demonstrate
the validity of the doublet scheme
as a method for constructing the invariant/attractive manifold that incorporates
the excited modes as well as the would-be zero modes.

In Sec. \ref{sec:ChapB-2},
we give
a brief but self-contained account of the Boltzmann equation
and Grad's thirteen-moment approximation in the method of moments.
In Sec. \ref{sec:ChapB-4},
we present the causal hydrodynamic equation
and the microscopic representations of the transport coefficients and relaxation times
that are obtained with the doublet scheme in the RG method.
The last section is devoted to a summary and concluding remarks.
In \ref{sec:ChapB-6},
we give some formulae used for the construction of the doublet scheme.
In \ref{sec:appA},
we prove that
the mesoscopic dynamics obtained by  the doublet scheme
is reduced to the slow dynamics
described solely by
the zero modes in the asymptotic regime after a time.
In \ref{sec:ChapB-9},
we present the detailed derivation
of the causal hydrodynamics as the mesoscopic dynamics of the Boltzmann equation.

\section{
  Derivation of mesoscopic dynamics
  from generic evolution equation with doublet scheme in RG method
}
\label{sec:ChapB-3}
In this section,
we develop a method on the basis of the RG method
to extract the mesoscopic dynamics from a generic evolution equation
by constructing the invariant/attractive manifold incorporating some appropriate excited modes
as well as the zero modes of its linearized evolution operator
based on the following
general principle of the reduction theory of the dynamics \cite{kuramoto}:
The resultant dynamics should be as simple as possible
because we are interested 
in reducing the dynamics to a simpler one.
As mentioned in Sec. \ref{sec:ChapB-1},
we use
this principle
to derive an equation describing the mesoscopic dynamics,
where the number of dynamical variables and terms are reduced
as few as possible.
We will see that
this principle uniquely determines
the number and form of the excited modes
that should be included in the invariant/attractive manifold
on which the mesoscopic dynamics of the evolution equation is defined.

\subsection{
  Evolution equation
}
\label{sec:ChapB-3-1}
As a generic evolution equation,
we treat a system of differential equations
with two non-linear terms,
which represent the relaxation to a static solution
and weak perturbation, respectively.
The equation reads
\begin{eqnarray}
  \label{eq:ChapB-3-1-001}
  \frac{\partial}{\partial t}X_i = G_i(X_1,\,\cdots,\,X_N)
  + \epsilon\,F_i(X_1,\,\cdots,\,X_N),\,\,\,\,\,\,i=1,\,\cdots,\,N,
\end{eqnarray}
which is also rewritten in a more convenient vector form
\begin{eqnarray}
  \label{eq:ChapB-3-1-002}
  \frac{\partial}{\partial t}\Vec{X} = \Vec{G}(\Vec{X}) + \epsilon\,\Vec{F}(\Vec{X}).
\end{eqnarray}
In Eq. (\ref{eq:ChapB-3-1-002}),
the dynamical variables are represented by $N$-component vector $\Vec{X}$ 
($1 < N \le \infty$),
whereas $\Vec{G}(\Vec{X})$ and $\Vec{F}(\Vec{X})$ are
non-linear functions 
of $\Vec{X}$,
and $\epsilon$ is introduced as an indicator of the smallness of $\Vec{F}(\Vec{X})$
that is finally set equal to $1$; the vector $\Vec{X}(t)$ governed 
by Eq. (\ref{eq:ChapB-3-1-002}) without $\Vec{F}(\Vec{X})$
relaxes to the static solution $\Vec{X}^\mathrm{eq}$ under time evolution as
\begin{eqnarray}
  \label{eq:ChapB-3-1-003}
  \Vec{X}(t\rightarrow\infty) \rightarrow \Vec{X}^\mathrm{eq},
\end{eqnarray}
which is given as a solution to
\begin{eqnarray}
  \label{eq:ChapB-3-1-004}
  \Vec{G}(\Vec{X}^\mathrm{eq}) = 0.
\end{eqnarray}
Here,
we suppose that
the static solution $\Vec{X}^\mathrm{eq}$ is unique and forms
a well-defined $M_0$-dimensional invariant manifold with $M_0$ being smaller than or equal to $N$.
This means that
$\Vec{X}^\mathrm{eq}$ is parametrized
by $M_0$ integral constants $C_{\alpha}$ with $\alpha = 1,\,\cdots,\,M_0$;
\begin{eqnarray}
  \label{eq:ChapB-3-1-005}
  \Vec{X}^\mathrm{eq} = \Vec{X}^\mathrm{eq}(C_{1},\,\cdots,\,C_{M_0}).
\end{eqnarray}

We first define the linearized evolution operator $A$ by
\begin{eqnarray}
  \label{eq:ChapB-3-1-007}
  A_{ij} \equiv \frac{\partial}{\partial X_j} G_i(X_1,\,\cdots,\,X_N) \Bigg|_{\Vec{X}=\Vec{X}^{\mathrm{eq}}}.
\end{eqnarray}
We note that
in accordance with Eq. (\ref{eq:ChapB-3-1-005}),
$A$ has $M_0$ eigenvectors belonging to the zero eigenvalue, i.e., zero modes,
and the dimension of the kernel of $A$ is $M_0$;
i.e., \textrm{dim}\,$[\textrm{ker}\,A]=M_0$.
In fact,
by differentiating Eq. (\ref{eq:ChapB-3-1-004}) with respect to the $M_0$ integral constants $C_{\alpha}$,
we have
\begin{eqnarray}
  \label{eq:ChapB-3-1-006}
  A \, \partial\Vec{X}^\mathrm{eq}/\partial C_{\alpha} = 0,
\end{eqnarray}
which means that $\Vec{\varphi}^{\alpha}_0$ defined by
\begin{eqnarray}
  \label{eq:ChapB-3-1-008}
  \Vec{\varphi}^{\alpha}_0 = \partial\Vec{X}^\mathrm{eq}/\partial C_{\alpha},
\end{eqnarray}
are the $M_0$ zero modes.
The invariant manifold is spanned by $\Vec{\varphi}^{\alpha}_0$ with $\alpha = 1,\,\cdots,\,M_0$.

We define the projection operator $P_0$ onto the kernel of $A$,
which is called the P${}_0$ space,
and the projection operator $Q_0$ onto the Q${}_0$ space
as the complement to the P${}_0$ space:
With the use of an inner product which satisfies the positive definiteness of the norm as
$\langle\, \Vec{\psi}\,,\,\Vec{\psi} \,\rangle > 0$ with
$\Vec{\psi} \ne 0$,
we define
\begin{eqnarray}
  \label{eq:ChapB-3-1-010}
  P_0 \, \Vec{\psi} \equiv
  \sum_{\alpha,\beta=1}^{M_0} \, \Vec{\varphi}^{\alpha}_0 \, \eta^{-1}_{0\alpha\beta} \,
  \langle\,\Vec{\varphi}^{\beta}_0\,,\,\Vec{\psi}\,\rangle,\,\,\,
  Q_0 \equiv 1 - P_0,
\end{eqnarray}
where 
$\eta^{-1}_{0\alpha\beta}$ is the inverse matrix of
the P${}_0$-space metric matrix
$\eta^{\alpha\beta}_0$ defined by
\begin{eqnarray}
  \label{eq:ChapB-3-1-012}
  \eta^{\alpha\beta}_0
  \equiv \langle\,\Vec{\varphi}^{\alpha}_0\,,\,\Vec{\varphi}^{\beta}_0\,\rangle.
\end{eqnarray}
It is easily verified that
the following identity is satisfied:
\begin{eqnarray}
  \label{eq:innerproduct_with_zeromodes}
  \langle\,\Vec{\varphi}^{\alpha}_0\,,\,P_0\,\Vec{\psi}\,\rangle =
  \langle\,\Vec{\varphi}^{\alpha}_0\,,\,\Vec{\psi}\,\rangle,
\end{eqnarray}
with $\Vec{\psi}$ being an arbitrary vector.

We assume that
the other eigenvalues of $A$ are real negative and
 negative eigenvalues closest to zero are discrete.
Accordingly
we 
suppose that
with the inner product
$A$ is self-adjoint,
\begin{eqnarray}
  \label{eq:ChapB-3-1-013}
  \langle\, A\,\Vec{\psi}\,,\,\Vec{\chi} \,\rangle
  = \langle\, \Vec{\psi}\,,\,A\,\Vec{\chi} \,\rangle,
\end{eqnarray}
where $\Vec{\psi}$ and $\Vec{\chi}$ are arbitrary vectors.
We will see that
this self-adjoint nature of $A$ plays an essential role
in making the form of the resultant equation simpler. 

\subsection{
  Approximate solution around arbitrary initial time
}
\label{sec:ChapB-3-2}
To obtain the mesoscopic dynamics of Eq. (\ref{eq:ChapB-3-1-002}),
first we apply the perturbation theory
to construct a solution to Eq. (\ref{eq:ChapB-3-1-002}),
which represents the motion caused by the zero modes and the excited modes.
Let $\Vec{X}(t)$ be an yet unknown exact solution to Eq. (\ref{eq:ChapB-3-1-002})
with an initial condition given at, say, $t=-\infty$:
The solution forms 
an orbit $\Vec{X}(t)$ parametrized by $t$.
Then let us pick up an arbitrary point $\Vec{X}(t_0)$ on the orbit.
In accordance with the general formulation of the RG method \cite{env001,env002,qm,env006},
we try to construct a perturbative solution $\Vec{\tilde{X}}(t\,;\,t_0)$ 
with the initial value  set to $\Vec{X}(t_0)$ at $t=t_0$:
\begin{eqnarray}
  \label{eq:ChapB-3-2-001}
  \Vec{\tilde{X}}(t = t_0\,;\,t_0) = \Vec{X}(t_0).
\end{eqnarray}
Here, we have made explicit the $t_0$ dependence
of $\Vec{\tilde{X}}(t = t_0\,;\,t_0)$.
It is noted that
in the RG method
the initial value $\Vec{X}(t_0)$
and
the RG equation applied to the perturbative solution $\Vec{\tilde{X}}(t\,;\,t_0)$
provide the invariant/attractive manifold
and
the reduced dynamics defined on it, respectively.

The initial value as well as the perturbative solution
are expanded
with respect to $\epsilon$ as follows:
\begin{eqnarray}
  \label{eq:ChapB-3-2-002}
  \Vec{\tilde{X}}(t\,;\,t_0) &=& \Vec{\tilde{X}}_0(t\,;\,t_0)
  + \epsilon\,\Vec{\tilde{X}}_1(t\,;\,t_0)
  + \epsilon^2\,\Vec{\tilde{X}}_2(t\,;\,t_0)
  +\cdots,\\
  \label{eq:ChapB-3-2-003}
  \Vec{X}(t_0) &=& \Vec{X}_0(t_0)
  + \epsilon\,\Vec{X}_1(t_0)
  + \epsilon^2\,\Vec{X}_2(t_0)
  +\cdots.
\end{eqnarray}
The respective initial conditions at $t=t_0$ are set up as
\begin{eqnarray}
  \label{eq:ChapB-3-2-004}
  \Vec{\tilde{X}}_l(t = t_0\,;\,t_0) = \Vec{X}_l(t_0),
  \,\,\,\,\,\,l =
  0,\,1,\,2,\,\cdots.
\end{eqnarray}
In the expansion,
the zeroth-order initial value $\Vec{\tilde{X}}_0(t_0\,;\,t_0) = \Vec{X}_0(t_0)$ is
supposed to be as close as possible to
the exact solution.

\begin{figure}
 \begin{minipage}{\hsize}
 \begin{center}
  \includegraphics[width=16cm]{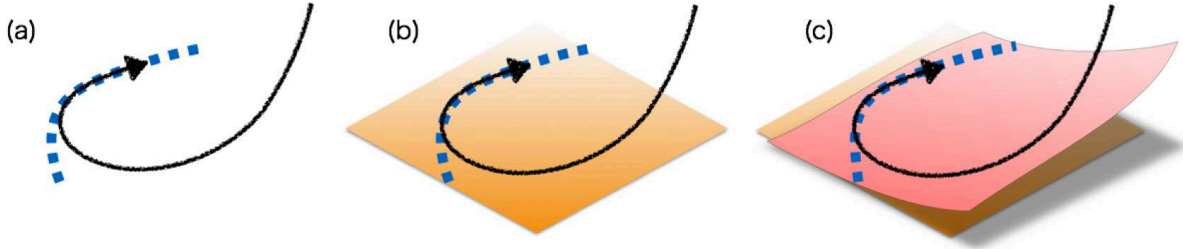}
 \end{center}
 \end{minipage}
 \caption{
 \label{fig:B}
 A geometrical illustration 
 of the perturbative construction of the invariant/attractive manifold 
 for the mesoscopic dynamics.
 In all the panels,
 the solid lines denote
 the orbit of an exact solution to the generic evolution equation (\ref{eq:ChapB-3-1-002}),
 while the dashed ones represent the asymptotic one 
 confined on the manifold spanned by the zero modes,
 which can be identified by the zeroth-order analysis;
 this manifold is represented as a surface presented in panel (b).
 The new surface shown in panel (c) represents the manifold modified (extended) 
 by incorporating some excited modes properly so as to a closed dynamics 
 is obtained up to the second-order of the perturbation,
 which is identified with the mesoscopic dynamics. 
 }
\end{figure}

Substituting the above expansions into Eq. (\ref{eq:ChapB-3-1-002}),
we obtain the series of the perturbative equations with respect to $\epsilon$.
Now we are interested in the slow  behavior of the solution in the asymptotic regime, and
we suppose that
the zeroth-order solution is
given by a static solution spanned by the zero modes;
excited modes and the slippery behavior of the zero modes (as given by secular terms)
are to be incorporated
in the first- and higher-order solutions.
Here,
we carry out the perturbative analysis up to the second order,
which is necessary to obtain the mesoscopic dynamics.

Before entering  the perturbative analysis, we illustrate  in a geometrical way 
in Fig. \ref{fig:B}
the way how the invariant/attractive manifold spanned solely by the zero modes
is extended or improved  so as to accommodate whole the mesoscopic dynamics.

The zeroth-order equation reads
\begin{eqnarray}
  \label{eq:ChapB-3-2-005}
  \frac{\partial}{\partial t}\Vec{\tilde{X}}_0(t\,;\,t_0) = \Vec{G}(\Vec{\tilde{X}}_0(t\,;\,t_0)).
\end{eqnarray}
As mentioned above,
we are interested in the slow motion
that would be realized asymptotically as $t\rightarrow \infty$.
Thus
we try to find a stationary solution 
satisfying
\begin{eqnarray}
  \frac{\partial}{\partial t}\Vec{\tilde{X}}_0(t\,;\,t_0) = 0,
\end{eqnarray}
which is realized when 
$\Vec{\tilde{X}}_0(t\,;\,t_0)$ is the fixed point,
\begin{eqnarray}
  \Vec{G}(\Vec{\tilde{X}}_0(t\,;\,t_0)) = 0.
\end{eqnarray}
This is nothing but Eq. (\ref{eq:ChapB-3-1-004}),
and thus
$\Vec{\tilde{X}}_0(t\,;\,t_0)$ is identified with $\Vec{X}^{\mathrm{eq}}$
because of the uniqueness of the static solution or
fixed point which we have assumed:
\begin{eqnarray}
  \label{eq:ChapB-3-2-010}
  \Vec{\tilde{X}}_0(t\,;\,t_0) = \Vec{X}^\mathrm{eq}(t_0),
\end{eqnarray}
accordingly
\begin{eqnarray}
  \label{eq:ChapB-3-2-011}
  \Vec{X}_0(t_0) = \Vec{\tilde{X}}_0(t=t_0\,;\,t_0) = \Vec{X}^\mathrm{eq}(t_0).
\end{eqnarray}
We note that $\Vec{X}^\mathrm{eq}(t_0)$ depends on $t_0$
through the would-be $M_0$ integral constants $C_{\alpha}(t_0)$ with $\alpha=1,\,\cdots,\,M_0$
defined in Eq. (\ref{eq:ChapB-3-1-005});
\begin{eqnarray}
  \label{eq:ChapB-3-2-008}
  \Vec{X}^\mathrm{eq}(t_0) = 
  \Vec{X}^\mathrm{eq}(C_{1}(t_0),\,\cdots,\,C_{M_0}(t_0)).
\end{eqnarray}
In the following,
we suppress the initial time $t_0$ when no misunderstanding is expected.

\subsection{
  First-order solution and
  doublet scheme
}
\label{sec:ChapB-3-3}
The first-order equation reads
\begin{eqnarray}
  \label{eq:ChapB-3-4-001}
  \frac{\partial}{\partial t}\Vec{\tilde{X}}_1(t) = A \, \Vec{\tilde{X}}_1(t) + \Vec{F}_0,
\end{eqnarray}
where the inhomogeneous term $\Vec{F}_0$ is defined as
\begin{eqnarray}
  \label{eq:inhomo}
  \Vec{F}_0 \equiv
  \Vec{F}(\Vec{X}^{\mathrm{eq}}).
\end{eqnarray}

With the use of the projection operators $P_0$ and $Q_0$,
we can obtain a
general solution to Eq. (\ref{eq:ChapB-3-4-001}) as
\begin{eqnarray}
  \label{eq:first-eq1}
  \Vec{\tilde{X}}_1(t\,;\,t_0)
  = \mathrm{e}^{A(t-t_0)}\,\Vec{\phi} + (t-t_0)\,P_0\,\Vec{F}_0 + (\mathrm{e}^{A(t-t_0)} - 1) \, A^{-1} \, Q_0 \, \Vec{F}_0,
\end{eqnarray}
with
\begin{eqnarray}
  \Vec{X}_1(t_0) = \Vec{\tilde{X}}_1(t=t_0\,;\,t_0) = \Vec{\phi},
\end{eqnarray}
where $\Vec{\phi}$ is the integral constant.
Without loss of generality,
we can suppose that
$\Vec{\phi}$ contains no zero modes
because such zero modes can be eliminated
by the redefinition of the zeroth-order initial value $\Vec{X}^{\mathrm{eq}}$.
We note the appearance of the secular term
proportional to $t-t_0$ in Eq. (\ref{eq:first-eq1}),
which apparently invalidate the perturbative solution when $|t-t_0|$ becomes large.

For later convenience,
let us expand $\mathrm{e}^{A(t-t_0)}$ with respect to $t-t_0$
and retain the terms up to the first order as
\begin{eqnarray}
  \label{eq:first-eq1_expanded}
  \Vec{\tilde{X}}_1(t\,;\,t_0)
  = \Vec{\phi}
  + (t-t_0)\,(
  A\,\Vec{\phi}
  + P_0\,\Vec{F}_0 + Q_0 \, \Vec{F}_0).
\end{eqnarray}
Here,
the neglected terms of $O((t-t_0)^2)$ are irrelevant
when we impose the RG equation,
which can be identified with the envelope equation \cite{env001}
and thus the
globally improved solution
is constructed
by patching the tangent line of the perturbative solution at the arbitrary initial time $t-t_0$.

Now
the problem is how to extend the vector space
beyond that spanned by the zero modes
to accommodate the excited modes that are responsible
for the
closed
mesoscopic dynamics;
let us call such the closed vector space the P$_1$ space, which is a subspace
of the Q$_0$ space.
The vector fields belonging to the 
P$_1$ space constitute the basic variables
for the description
of the mesoscopic dynamics
together with the zero modes.
Although it is not apparent a priori whether such a vector space exists,
the form of the perturbative solution (\ref{eq:first-eq1_expanded}) itself
nicely suggests the way how to construct the P$_1$ space,
if we recall that
the basic principle of the reduction theory of dynamical systems
is to reduce the original equation to 
a closed system composed of a as small as possible number  of variables and equations.
In our case, this principle is implemented by requiring 
that the tangent space of the perturbative solution at $t=t_0$
should be spanned by a as small as possible number of independent vectors. 
In fact the solution (\ref{eq:first-eq1_expanded}) is nicely written as a linear combination of
the vector $P_0\,\Vec{F}_0$ belonging to the P$_0$ space and the three new vectors, i.e.,
$\Vec{\phi}$, $A\,\Vec{\phi}$, and $Q_0 \, \Vec{F}_0$.
Thus,
we find that the minimal P$_1$ space that is closed can be constructed if
the following conditions are satisfied:
\begin{itemize}
\item
$A\,\Vec{\phi}$ and $Q_0 \, \Vec{F}_0$ belong to a common vector space.
\item
The P${}_1$ space is spanned by
the bases
of
the union of
$\Vec{\phi}$ and $A\,\Vec{\phi}$.
\end{itemize}
The first condition is  restated as that
$\Vec{\phi}$ and $A^{-1}\,Q_0 \, \Vec{F}_0$
should belong to a common vector space;
note that $\Vec{\phi}$ was defined 
to be orthogonal to the zero modes of $A$ and hence belong to the Q$_0$ space.

Thus one sees that the problem is reduced
to calculate $A^{-1}\,Q_0 \, \Vec{F}_0$
and examine
whether it is spanned by a finite number of independent vector fields.
It is, however, not possible to perform this task for a generic $\Vec{F}_0$.
Since our primary purpose is to develop a general theory that is applicable 
to the reduction of the Boltzmann equation,
we here just take the case where
$A^{-1}\,Q_0 \, \Vec{F}_0$
can be expressed in
the following form,
as in the case for the Boltzmann equation
and the Lorenz model,
\begin{eqnarray}
\label{eq:AQF}
A^{-1}\,Q_0 \, \Vec{F}_0 =
\sum_{\mu=1}^{M_1} \, (A^{-1} \, \Vec{\varphi}^{\mu}_1) \, f_{\mu},
\end{eqnarray}
where
$f_{\mu}$
with $\mu = 1,\,\cdots,\,M_1$ 
are linear independent functions of $\Vec{C}=(C_{1},\,\cdots,\,C_{M_0})$;
$f_{\mu}=f_{\mu}(\Vec{C})$.
Here the linear independence of $f_{\mu}$ means that the following statement is satisfied:
\begin{eqnarray}
  \label{eq:linear_independence}
  \sum_{\mu=1}^{M_1} \, \alpha^{\mu} \, f_{\mu}(\Vec{C}) = 0,\,\,\,\forall \Vec{C}
  \Longrightarrow \alpha^{1} = \cdots = \alpha^{M_1} = 0.
\end{eqnarray}
It is noted that
$\Vec{\varphi}^{\mu}_1$ can be generic vectors that are not necessarily an
eigenvector of $A$.
From now on,
we shall only consider a generic case where any $\Vec{\varphi}^{\mu}_1$ is not
an eigenvector of $A$.

We can take the $M_1$ vectors $A^{-1}\,\Vec{\varphi}^{\mu}_1$
as
the bases of the vector space
spanned
by $A^{-1}\,Q_0 \, \Vec{F}_0$ and $\Vec{\phi}$.
Accordingly,
$\Vec{\phi}$ can be written as a linear combination
of these bases as
\begin{eqnarray}
  \label{eq:ChapB-3-3-014}
  \Vec{\phi} = \sum_{\mu=1}^{M_1} \, (A^{-1} \, \Vec{\varphi}^{\mu}_1) \, C^\prime_{\mu}.
\end{eqnarray}
Here we have introduced the $M_1$ integral constants
$C^\prime_{\mu}$
as mere coefficients of the basis vectors,
which have the $t_0$ dependence $C^\prime_\mu(t_0)$ as $C_\alpha(t_0)$ does.
We stress that
the form of $\Vec{\phi}$ given in Eq. (\ref{eq:ChapB-3-3-014})
is the most
general expression
that makes $A\,\Vec{\phi}$ and $Q_0\,\Vec{F}_0$ belong to
a
common space
provided that $A^{-1}\,Q_0 \, \Vec{F}_0$ takes the form given in Eq. (\ref{eq:AQF}).

As is clear now, we see that the P${}_1$ space is identified
with the vector space spanned by
$A^{-1} \, \Vec{\varphi}^{\mu}_1$
and $\Vec{\varphi}^{\mu}_1$ with $\mu=1,\,\cdots,\,M_1$.
The pair of
$A^{-1} \, \Vec{\varphi}^{\mu}_1$ and $\Vec{\varphi}^{\mu}_1$
are called the \textit{doublet modes}.
The Q${}_0$ space is now decomposed into the P${}_1$ space
spanned by the doublet modes and the Q${}_1$ space
which is the complement to the P${}_0$ and P${}_1$ spaces.
The corresponding projection operators are denoted
as $P_1$ and $Q_1$, respectively,
whose definitions are given by
\begin{eqnarray}
  \label{eq:ChapB-3-3-005}
  P_1 \, \Vec{\psi} &\equiv&
  \sum_{m,n=0,1} \,\sum_{\mu,\nu=1}^{M_1} \,
  A^{-m}\,\Vec{\varphi}_1^{\mu}
  \,\eta^{-1}_{1m\mu,n\nu} \,
  \langle\,A^{-n}\,\Vec{\varphi}^{\nu}_1\,,\,\Vec{\psi}\,\rangle,\\
  Q_1 &\equiv& Q_0 - P_1.
\end{eqnarray}
Here,
$\eta^{-1}_{1m\mu,n\nu}$ has been introduced as
the inverse matrix of the P${}_1$-space metric matrix given by
\begin{eqnarray}
  \label{eq:ChapB-3-3-007}
  \eta^{m\mu,n\nu}_1 \equiv \langle\, A^{-m}\,\Vec{\varphi}^{\mu}_1 \,,\, A^{-n}\,\Vec{\varphi}^{\nu}_1 \,\rangle.
\end{eqnarray}
We note that
the direct use of the definitions
(\ref{eq:ChapB-3-3-005}) and (\ref{eq:ChapB-3-3-007}) leads
to the following equalities:
\begin{eqnarray}
  \label{eq:ChapB-3-7-003}
  \langle\,\Vec{\varphi}^{\mu}_1\,,\,P_1\,\Vec{\psi}\,\rangle
  = \langle\,\Vec{\varphi}^{\mu}_1\,,\,\Vec{\psi}\,\rangle,\,\,\,
  \langle\,A^{-1}\,\Vec{\varphi}^{\mu}_1\,,\,P_1\,\Vec{\psi}\,\rangle
  = \langle\,A^{-1}\,\Vec{\varphi}^{\mu}_1\,,\,\Vec{\psi}\,\rangle,
\end{eqnarray}
with $\Vec{\psi}$ being an arbitrary vector.

\subsection{
   Second-order solution
}
\label{sec:ChapB-3-5}
The second-order equation is written as
\begin{eqnarray}
  \label{eq:ChapB-3-5-001}
  \frac{\partial}{\partial t}\Vec{\tilde{X}}_2(t) = A\,\Vec{\tilde{X}}_2(t) + \Vec{K}(t-t_0),
\end{eqnarray}
with the time-dependent inhomogeneous term given by
\begin{eqnarray}
  \label{eq:ChapB-3-5-011}
  \Vec{K}(t-t_0) \equiv
  \frac{1}{2}\,B \, \big[ \Vec{\tilde{X}}_1(t\,;\,t_0)\,,\,\Vec{\tilde{X}}_1(t\,;\,t_0) \big]
  + F_1 \, \Vec{\tilde{X}}_1(t\,;\,t_0).
\end{eqnarray}
Here, we have introduced $B$ and $F_1$ whose components are given by
\begin{eqnarray}
  \label{eq:ChapB-3-5-002}
  B_{ijk} \equiv \frac{\partial^2}{\partial X_j\,\partial X_k}G_i(X)\Bigg|_{\Vec{X}=\Vec{X}^\mathrm{eq}},\,\,\,
  F_{1ij} \equiv \frac{\partial}{\partial X_j}F_i(X)\Bigg|_{\Vec{X}=\Vec{X}^\mathrm{eq}}.
\end{eqnarray}
In Eq. (\ref{eq:ChapB-3-5-011}),
we have used the notation 
\begin{eqnarray}
  (B \, \big[ \Vec{\psi}\,,\,\Vec{\chi} \big])_i
  = \sum_{j=1}^N\,\sum_{k=1}^N\,B_{ijk}\,\psi_j\,\chi_k,
\end{eqnarray}
with $\psi_i = (\Vec{\psi})_i$ and $\chi_i = (\Vec{\chi})_i$.

To obtain appropriate initial values and solutions
with the motion coming from the P${}_0$ and P${}_1$ spaces to Eq. (\ref{eq:ChapB-3-5-001}),
we utilize the formulae (\ref{eq:ChapB-6-005}) and (\ref{eq:ChapB-6-006}) given
in
\ref{sec:ChapB-6}:
By setting $\Vec{R}(t-t_0) = \Vec{K}(t-t_0)$ in Eqs. (\ref{eq:ChapB-6-005}) and (\ref{eq:ChapB-6-006}),
we find that the initial value and solution to Eq. (\ref{eq:ChapB-3-5-001}) read
\begin{eqnarray}
  \label{eq:ChapB-3-5-004}
  \Vec{\tilde{X}}_2(t_0) =
  - Q_1 \, \mathcal{G}(s) \,
  Q_0 \, \Vec{K}(s)\Big|_{s=0},
\end{eqnarray}
and
\begin{eqnarray}
  \label{eq:ChapB-3-5-005}
  \Vec{\tilde{X}}_2(t\,;\,t_0)
  &=&
  (t-t_0) \, P_0 \, \Vec{K}(0)
  +
  (t-t_0)\,(A - \partial/\partial s)
  \,P_1 \,
  \mathcal{G}(s) \,
  Q_0 \, \Vec{K}(s)\Big|_{s=0}\nonumber\\
  &&{}-
  (1 + (t-t_0)\,\partial/\partial s) \, Q_1 \,
  \mathcal{G}(s) \,
  Q_0 \, \Vec{K}(s)\Big|_{s=0},
\end{eqnarray}
respectively.
The derivation of this solution is presented in \ref{sec:ChapB-6},
where the complete expression of the solution not restricted
to $t\sim t_0$ is given.
In Eq. (\ref{eq:ChapB-3-5-005}),
we have retained only terms up
to the first order of $(t-t_0)$,
and introduced a ``propagator''
\begin{eqnarray}
\label{eq:propagator}
\mathcal{G}(s) \equiv (A - \partial/\partial s)^{-1}.
\end{eqnarray}
We notice again the appearance of secular terms in Eq. (\ref{eq:ChapB-3-5-005}).

Summing up the perturbative solutions up to the second order with respect to $\epsilon$,
we have the full expression of
the initial value
and the approximate solution around $t \sim t_0$ to the second order:
\begin{eqnarray}
  \label{eq:ChapB-3-5-016}
  \Vec{X}(t_0) &=&
  \Vec{X}^\mathrm{eq} + \epsilon \, \Vec{\phi}
  - \epsilon^2 \, Q_1 \, \mathcal{G}(s) \,
  Q_0 \, \Vec{K}(s)\Big|_{s=0} + O(\epsilon^3),
\end{eqnarray}
and
\begin{eqnarray}
  \label{eq:ChapB-3-5-017}
  \Vec{\tilde{X}}(t\,;\,t_0)
  &=&
  \Vec{X}^\mathrm{eq} + \epsilon \, \Bigg[ \Vec{\phi} + (t-t_0)\,A\,\Vec{\phi}
  + 
  (t-t_0) \, P_0 \, \Vec{F}_0
  +
  (t-t_0) \, Q_0 \, \Vec{F}_0 \Bigg]\nonumber\\
  &&{}+
  \epsilon^2 \, \Bigg[
  (t-t_0)
  \, P_0 \, \Vec{K}(0)
  +
  (t-t_0)\,(A - \partial/\partial s)
  \,P_1 \,
  \mathcal{G}(s) \,
  Q_0 \, \Vec{K}(s)\Big|_{s=0}\nonumber\\
  &&{}-
  (1 + (t-t_0) \, \partial/\partial s) \,Q_1 \,
  \mathcal{G}(s) \,
  Q_0 \, \Vec{K}(s)\Big|_{s=0} \Bigg]
  + O(\epsilon^3).
\end{eqnarray}
We note that
in Eq. (\ref{eq:ChapB-3-5-017})
the fast motion caused by the Q${}_1$ space has been eliminated
by adopting the appropriate initial value (\ref{eq:ChapB-3-5-016}).
  
\subsection{
   RG improvement of perturbative expansion
}        
\label{sec:ChapB-3-6}
We emphasize that
the solution (\ref{eq:ChapB-3-5-017}) contains the secular terms
that apparently invalidates the perturbative expansion
for $t$ away from the initial time $t_0$.
The point of the RG method lies in the fact that
we can utilize the secular terms to obtain a solution valid in a global domain
as discussed in Refs. \cite{env001,env002,env005,qm,env006,env007,env008}.
By applying the RG equation to the local solution (\ref{eq:ChapB-3-5-017}),
we can convert 
the set of the locally valid approximate solutions to the solution valid in a global domain:
\begin{eqnarray}
  \label{eq:ChapB-3-6-001}
  \frac{\partial}{\partial t_0}
  \Vec{\tilde{X}}(t\,;\,t_0)\Bigg|_{t_0=t}
  = 0,
\end{eqnarray}
which is reduced to
\begin{eqnarray}
  \label{eq:ChapB-3-6-002}
  &&\frac{\partial}{\partial t}\Vec{X}^\mathrm{eq} + \epsilon \, \Bigg[  - A \, \Vec{\phi}
  + \frac{\partial}{\partial t}\Vec{\phi}
  - P_0 \, \Vec{F}_0 - Q_0 \, \Vec{F}_0 \Bigg]\nonumber\\
  &&{} + \epsilon^2 \, \Bigg[
  - P_0 \, \Vec{K}(0)
  - (A - \partial/\partial s) \, P_1 \,
    \mathcal{G}(s) \, Q_0 \, \Vec{K}(s)\Big|_{s=0}\nonumber\\
  &&{}+ (\partial/\partial s) \, Q_1 \, \mathcal{G}(s)
  \, Q_0 \, \Vec{K}(s)\Big|_{s=0}
    \Bigg]
  + O(\epsilon^3) = 0.
\end{eqnarray}
It is noted that
the RG equation (\ref{eq:ChapB-3-6-002}) gives the equation of motion
governing the dynamics of the would-be integral 
constant $C_{\alpha}$ in $\Vec{X}^\mathrm{eq}$ and $C^\prime_{\mu}$ in $\Vec{\phi}$.
The
globally improved solution
can be obtained as
the initial value (\ref{eq:ChapB-3-5-016}) 
\begin{eqnarray}
  \label{eq:ChapB-3-6-004}
  \Vec{X}^{\mathrm{global}}(t) &\equiv& \Vec{X}(t_0=t)\nonumber\\
  &=&
  \Vec{X}^\mathrm{eq} + \epsilon \, \Vec{\phi}
  - \epsilon^2 \, Q_1 \, \mathcal{G}(s) \, 
  Q_0 \, \Vec{K}(s)\Big|_{s=0} \Bigg|_{t_0 = t}
  + O(\epsilon^3),
\end{eqnarray}
where the exact solution to Eq. (\ref{eq:ChapB-3-6-002}) is inserted.
It is noteworthy that
we have derived the mesoscopic dynamics of Eq. (\ref{eq:ChapB-3-1-002})
in the form of the pair of Eqs. (\ref{eq:ChapB-3-6-002}) and (\ref{eq:ChapB-3-6-004}):
Equation (\ref{eq:ChapB-3-6-004}) is nothing but
the invariant/attractive manifold of Eq. (\ref{eq:ChapB-3-1-002}),
and Eq. (\ref{eq:ChapB-3-6-002}) describes
the mesoscopic dynamics defined on it.

\subsection{
   Reduction of RG equation to simpler form
}            
\label{sec:ChapB-3-7}
We find  that
the RG equation (\ref{eq:ChapB-3-6-002})
includes terms belonging to the Q${}_1$ space
that do not constitute the modes responsible for the mesoscopic dynamics.
Although these modes
could be incorporated as noise terms to make a stochastic
mesoscopic dynamics, such an attempt is beyond the scope of the present work. 
Here we simply average out them
to have the mesoscopic dynamics as a regular differential equation.
This averaging can be made by taking the inner product of
Eq. (\ref{eq:ChapB-3-6-002}) with the zero modes
$\Vec{\varphi}_0^\alpha$
and the excited modes $A^{-1}\,\Vec{\varphi}_1^\mu$
used in the definition of $\Vec{\phi}$.

To this end,
we first multiply the projection operators $P_0$ and $P_1$ from the left-hand side 
of Eq. (\ref{eq:ChapB-3-6-002})
and we have
\begin{eqnarray}
  \label{eq:ChapB-3-7-001}
  &&P_0 \, \frac{\partial}{\partial t}\Vec{X}^\mathrm{eq}
  + \epsilon \, \Bigg[ P_0 \, \frac{\partial}{\partial t}\Vec{\phi}
  - P_0 \, \Vec{F}_0\Bigg] - \epsilon^2 \, P_0 \, \Vec{K}(0) + O(\epsilon^3)
  = 0,\\
  \label{eq:ChapB-3-7-002}
  &&
  \epsilon \, \Bigg[- P_1 \, A \, \Vec{\phi}
  + P_1 \, \frac{\partial}{\partial t}\Vec{\phi}
  - P_1 \, Q_0 \, \Vec{F}_0 \Bigg]\nonumber\\
  &&{}
  - \epsilon^2 \, P_1 \, (A - \partial/\partial s) \, P_1\,
  \mathcal{G}(s)
  \, Q_0 \, \Vec{K}(s)\Big|_{s=0}
   + O(\epsilon^3)
  =  0,
\end{eqnarray}
respectively.
Here,
we have used the fact that $\partial\Vec{X}^\mathrm{eq}/\partial t$ belongs to the P${}_0$ space:
\begin{eqnarray}
  \label{eq:dxstdt_is_zeromode}
  \frac{\partial}{\partial t}\Vec{X}^\mathrm{eq}
  = \sum_{\alpha=1}^{M_0} \, \Vec{\varphi}^{\alpha}_0 \, \frac{\partial}{\partial t}C_{\alpha},
\end{eqnarray}
which is derived from Eqs. (\ref{eq:ChapB-3-1-008}) and (\ref{eq:ChapB-3-2-008}).

Then,
by taking the inner product of
Eqs. (\ref{eq:ChapB-3-7-001}) and (\ref{eq:ChapB-3-7-002})
with
$\Vec{\varphi}^{\alpha}_0$ and $A^{-1}\,\Vec{\varphi}^{\mu}_1$,
respectively,
we arrive at
\begin{eqnarray}
  \label{eq:ChapB-3-7-005}
  \langle\,\Vec{\varphi}^{\alpha}_0\,,\,\frac{\partial}{\partial t}(\Vec{X}^\mathrm{eq}
  + \epsilon \, \Vec{\phi})\,\rangle
  - \epsilon \,
  \langle\,\Vec{\varphi}^{\alpha}_0\,,\,
  \Vec{F}_0 + \epsilon \,F_1\,\Vec{\phi}\,\rangle
  &=& \epsilon^2 \,\frac{1}{2} \, \langle\,\Vec{\varphi}^{\alpha}_0\,,\,
  B\,\big[ \Vec{\phi} \,,\, \Vec{\phi} \big] \,\rangle,\\
  \label{eq:ChapB-3-7-006}
  \epsilon \, \langle\,A^{-1}\,\Vec{\varphi}^{\mu}_1\,,\,\frac{\partial}{\partial t}\Vec{\phi}\,\rangle
  - \epsilon \,
  \langle\,A^{-1}\,\Vec{\varphi}^{\mu}_1\,,\,
  \Vec{F}_0 + \epsilon \, F_1\,\Vec{\phi}
  \,\rangle
  &=& \epsilon \, \langle\,A^{-1}\,\Vec{\varphi}^{\mu}_1\,,\,A\,\Vec{\phi}\,\rangle\nonumber\\
  &&{}+ \epsilon^2 \, \frac{1}{2} \, \langle\,A^{-1}\,\Vec{\varphi}^{\mu}_1\,,\,
  B\,\big[ \Vec{\phi}\,,\,\Vec{\phi} \big] \,\rangle,
\end{eqnarray}
where we have omitted $O(\epsilon^3)$.
In the derivation of
Eq. (\ref{eq:ChapB-3-7-006}),
we have used the following identity:
\begin{eqnarray}
  \label{eq:ChapB-3-7-008}
  &&\langle\,A^{-1}\,\Vec{\varphi}^{\mu}_1\,,\,
  (A - \partial/\partial s) \, P_1 \, \mathcal{G}(s) \, Q_0 \, \Vec{K}(s)\Big|_{s=0}
  \,\rangle\nonumber\\
  &=&
  \langle\,(A - \partial/\partial s)\,A^{-1}\,\Vec{\varphi}^{\mu}_1\,,\,
  P_1 \, \mathcal{G}(s) \, Q_0 \, \Vec{K}(s)\Big|_{s=0}
  \,\rangle\nonumber\\
  &=&
  \langle\,(\Vec{\varphi}^{\mu}_1 - A^{-1}\,\Vec{\varphi}^{\mu}_1\,\partial/\partial s)\,,\,
  P_1 \, \mathcal{G}(s) \, Q_0 \, \Vec{K}(s)\Big|_{s=0}
  \,\rangle\nonumber\\
  &=&
  \langle\,(\Vec{\varphi}^{\mu}_1 - A^{-1}\,\Vec{\varphi}^{\mu}_1\,\partial/\partial s)\,,\,
  \mathcal{G}(s) \, Q_0 \, \Vec{K}(s)\Big|_{s=0}
  \,\rangle\nonumber\\
  &=&
  \langle\,(A - \partial/\partial s)\,A^{-1}\,\Vec{\varphi}^{\mu}_1\,,\,
  \mathcal{G}(s) \, Q_0 \, \Vec{K}(s)\Big|_{s=0}
  \,\rangle\nonumber\\
  &=&
  \langle\,A^{-1}\,\Vec{\varphi}^{\mu}_1\,,\,
  (A - \partial/\partial s)\,\mathcal{G}(s) \, Q_0 \, \Vec{K}(s)\Big|_{s=0}
  \,\rangle\nonumber\\
  &=&
  \langle\,A^{-1}\,\Vec{\varphi}^{\mu}_1\,,\,
  Q_0 \, \Vec{K}(s)\Big|_{s=0}
  \,\rangle\nonumber\\
  &=&
  \langle\,A^{-1}\,\Vec{\varphi}^{\mu}_1\,,\,
  \Vec{K}(0)
  \,\rangle\nonumber\\
  &=&
  \frac{1}{2} \, \langle\,A^{-1}\,\Vec{\varphi}^{\mu}_1\,,\,
  B\,\big[ \Vec{\phi} \,,\, \Vec{\phi} \big]
  \,\rangle
  +
  \langle\,A^{-1}\,\Vec{\varphi}^{\mu}_1\,,\,
  F_1\,\Vec{\phi}
  \,\rangle.
\end{eqnarray}
In Eq. (\ref{eq:ChapB-3-7-008}),
we have used the self-adjoint nature of $A$ shown in Eq. (\ref{eq:ChapB-3-1-013}),
the identities given by (\ref{eq:ChapB-3-7-003}),
the definition of $\mathcal{G}(s)$ given by Eq. (\ref{eq:propagator}),
and the relation derived from Eq. (\ref{eq:ChapB-3-5-011});
$\Vec{K}(0) =
\frac{1}{2}\,B\,\big[ \Vec{\phi}\,,\, \Vec{\phi}\big] + F_1 \, \Vec{\phi}$.
We note that
the pair of
Eqs. (\ref{eq:ChapB-3-7-005}) and (\ref{eq:ChapB-3-7-006}) is also
the equation of motion governing $C_{\alpha}$ in $\Vec{X}^\mathrm{eq}$ and $C^\prime_{\mu}$ in $\Vec{\phi}$,
which is much simpler than Eq. (\ref{eq:ChapB-3-6-002}).

Two remarks are in order here:
(i) The mesoscopic dynamics given by Eqs. (\ref{eq:ChapB-3-7-005}) and (\ref{eq:ChapB-3-7-006})
is consistent with the slow dynamics described
solely by the zero modes
in the asymptotic regime.
A proof for this natural property of the mesoscopic dynamics is 
presented in \ref{sec:appA}.
(ii)
The doublet scheme in the RG method
itself
has a universal nature
and can be applied to derive a mesoscopic dynamics from a wide class of evolution equations,
as far as
the equation can be written as Eq. (\ref{eq:ChapB-3-1-002})
and the linearized evolution operator $A$ is self-adjoint
as shown in Eq. (\ref{eq:ChapB-3-1-013}).
Since the Boltzmann equation and the Lorenz model satisfy these conditions
as will be seen in \ref{sec:ChapB-4-1} and Sec. \ref{sec:C}, respectively,
the mesoscopic dynamics of them can be extracted by the doublet scheme.
We note that
when all the eigenvalues of the linear operator $A$ are
real numbers, there exists an inner product with which $A$ can be made
self-adjoint.
An extension of the doublet scheme
to the case where $A$ has complex eigenvalues
is left as a future problem.

\section{
Example: mesoscopic dynamics of the  Lorenz model
}
\label{sec:C}
In this section,
we demonstrate how successfully
the doublet scheme in the RG method developed in Sec. \ref{sec:ChapB-3}
works to construct the invariant/attractive manifold that incorporates the excited modes as well as
the would-be zero modes.
To this end,
we adopt 
a simple finite-dimensional dynamical system, i.e.,
the Lorenz model for thermal convection \cite{lorentz}. 

The Lorenz model
is given by
\begin{eqnarray}
\label{eq:lorentzeq_for_x}
\dot{x} &=& \sigma (-x+y),\\
\label{eq:lorentzeq_for_y}
\dot{y} &=& rx - y - xz,\\
\label{eq:lorentzeq_for_z}
\dot{z} &=& xy - b z,
\end{eqnarray}  
where $x$, $y$, and $z$ denote
the
dynamical variables
and
$\sigma>0$, $r>0$, and $b>0$
are
model parameters.
For $0<r<1$
there exists one steady state given by
\begin{eqnarray}
\label{eq:steady_state_A}
(A) \,\,\,\, (x,y,z) &=& (0,0,0),
\end{eqnarray}  
while
for $1<r$
the steady states are (A) and 
\begin{eqnarray}
\label{eq:steady_state_B}
(B) \,\,\,\, (x,y,z) &=& (+\sqrt{b(r-1)},+\sqrt{b(r-1)},r-1),\\
\label{eq:steady_state_C}
(C) \,\,\,\, (x,y,z) &=& (-\sqrt{b(r-1)},-\sqrt{b(r-1)},r-1).
\end{eqnarray}  
The
linear stability analysis \cite{geometrical} shows that
the origin (A) is stable for $0< r < 1$ but unstable for $r > 1$,
while the latter steady states (B) and (C) are stable
for $1 < r< \sigma(\sigma+ b+3)/(\sigma- b-1) \equiv r_c$ but unstable for $r > r_c$.

In this
section,
we examine the non-linear stability around the origin (A) for $r\sim 1$.
In other words, we are interested in the case where $|r-1|$ is small,
say, less than $1$,
and hence
the amplitudes of the dynamical variables $(x,\, y,\,z)$ are small. Then
it is found convenient to rewrite  the control parameter $r$ 
with a small quantity $\epsilon$  as
\begin{eqnarray}
r = 1 + \chi \epsilon^2,
\end{eqnarray}
with $\chi=\pm 1$ depending on the sign of $r-1$. 
Furthermore, it turns out that  the amplitudes of the 
dynamical variables scale in the order of $\epsilon$. So we set them as
\begin{eqnarray}
(x,\,y,\,z) = \epsilon\,(X,\,Y,\,Z).
\end{eqnarray}
Then
the Lorenz model given by Eqs. (\ref{eq:lorentzeq_for_x})-(\ref{eq:lorentzeq_for_z}) 
is converted
into
\begin{eqnarray}
\label{eq:lorentzeq2}
\frac{\mathrm{d}}{\mathrm{d}t}\Vec{X}
= A \, \Vec{X}
+
\epsilon\,\left(
\begin{array}{c}
0 \\
-X\,Z \\
X\,Y
\end{array}
\right)
+
\epsilon^2\,\left(
\begin{array}{c}
0 \\
\chi\,X \\
0
\end{array}
\right),
\end{eqnarray}
with
$\Vec{X} \equiv {}^t(X,\,Y,\,Z)$
and
\begin{eqnarray}
\label{eq:evolutionop_L}
A &\equiv&
\left(
\begin{array}{ccc}
-\sigma & \sigma & 0\\
1 & -1 & 0 \\
0 & 0 & -b
\end{array}
\right).
\end{eqnarray}
The
eigenvalues
of $A$
are found to be
\begin{eqnarray}
\label{eq:eigenvalue_of_L}
\lambda_1 = 0,\,\,\,\lambda_2 = -1 - \sigma,\,\,\,\lambda_3 = -b,
\end{eqnarray}
whose respective (right-)eigenvectors are
\begin{eqnarray}
\label{eq:eigenvector_of_L}
\Vec{U}_1
=
\left(
\begin{array}{c}
1 \\
1 \\
0
\end{array}
\right),\,\,\,
\Vec{U}_2
=
\left(
\begin{array}{c}
\sigma \\
-1 \\
0
\end{array}
\right),\,\,\,
\Vec{U}_3
=
\left(
\begin{array}{c}
0 \\
0 \\
1
\end{array}
\right).
\end{eqnarray}

Since all the eigenvalues are real numbers,
it is easily verified that
the apparently asymmetric linear operator
$A$ is made symmetric with respect to a suitably defined inner product, 
which can be given with respect to the left-eigenvectors of $A$.
Thus, we see that 
the Lorenz model (\ref{eq:lorentzeq2}) can be analyzed by 
the doublet scheme in the RG method developed in the last section
where the symmetric property of the linear operator $A$
is assumed and utilized.

We pick up a point $\Vec{X}(t_0)$ on an exact solution yet to be determined with some unspecified initial condition,
and try to construct a perturbative solution $\tilde{\Vec{X}}(t;t_0)$  
around $t=t_0$ with $\Vec{X}(t_0)$ being set up with the initial value.
We assume that the
initial value
i.e., the exact solution
and the approximate solution
can be expanded with respect to $\epsilon$ as follows:
$\Vec{X}(t_0)
= \Vec{X}_0(t_0) + \epsilon\,\Vec{X}_1(t_0) + \epsilon^2\,\Vec{X}_2(t_0) + \cdots$
and
$\tilde{\Vec{X}}(t;t_0)
= \tilde{\Vec{X}}_0(t;t_0) + \epsilon\,\tilde{\Vec{X}}_1(t;t_0) + \epsilon^2\,\tilde{\Vec{X}}_2(t;t_0) + \cdots$,
which satisfy respective initial conditions
$\tilde{\Vec{X}}_l(t=t_0;t_0) = \Vec{X}_l(t_0)$ with $l=0,1,2,\cdots$.
Substituting these expansions into Eq. (\ref{eq:lorentzeq2}),
we have a series of perturbative equations.

The zeroth-order equation reads
\begin{eqnarray}
\label{eq:lorentzeq_order0}
\frac{\mathrm{d}}{\mathrm{d}t}\tilde{\Vec{X}}_0(t;t_0)
&=& A \tilde{\Vec{X}}_0(t;t_0).
\end{eqnarray}
Since we are interested in the asymptotic state as $t \rightarrow \infty$,
we take the neutrally stable solution
\begin{eqnarray}
\label{eq:solution_to_lorentzeq_order0}
\tilde{\Vec{X}}_0(t;t_0) = C(t_0)\Vec{U}_1,
\end{eqnarray}
where $C(t_0)$ is an integral constant
and we have made it explicit that the solution may depend on the initial time $t_0$.
We note that
$C(t_0)$ corresponds
to $C_\alpha(t_0)$ in the general case discussed in Sec. \ref{sec:ChapB-3}
with $M_0 = 1$.
The initial value corresponding to the solution (\ref{eq:solution_to_lorentzeq_order0}) is
\begin{eqnarray}
\label{eq:initialvalue_to_lorentzeq_order0}
\Vec{X}_0(t_0) = \tilde{\Vec{X}}_0(t=t_0;t_0) = C(t_0)\Vec{U}_1.
\end{eqnarray}
We note that
the P${}_0$ space is spanned by the zero mode $\Vec{U}_1$.

The first-order equation reads
\begin{eqnarray}
\label{eq:lorentzeq_order1}
\frac{\mathrm{d}}{\mathrm{d}t}\tilde{\Vec{X}}_1(t;t_0)
= A \tilde{\Vec{X}}_1(t;t_0) + C^2(t_0)\Vec{U}_3,
\end{eqnarray}
whose
general
solution
may be written as
\begin{eqnarray}
\label{eq:solution_to_lorentzeq_order1}
\tilde{\Vec{X}}_1(t;t_0)
&=& \mathrm{e}^{L(t-t_0)} \Vec{X}_1(t_0)
+ (\mathrm{e}^{A(t-t_0)}-1) A^{-1} C^2(t_0)\Vec{U}_3\nonumber\\
&=& \Vec{X}_1(t_0) + (t-t_0)(A\Vec{X}_1(t_0) + C^2(t_0)\Vec{U}_3) + O((t-t_0)^2).
\end{eqnarray}
As in the
general
case discussed in Sec. \ref{sec:ChapB-3},
we specify the initial value $\Vec{X}_1(t_0)$
so that
the dimension
of the tangent space given by the term proportional 
to $t-t_0$ of the solution (\ref{eq:solution_to_lorentzeq_order1})
is as small as possible.
This requirement is satisfied
if $A\Vec{X}_1(t_0)$ belongs to
a space spanned by $\Vec{U}_3$. 
Thus we set
\begin{eqnarray}
\label{eq:initialvalue_to_lorentzeq_order1}
\Vec{X}_1(t_0) = A^{-1} \Vec{U}_3 C^\prime(t_0) = - \frac{1}{b}\Vec{U}_3 C^\prime(t_0).
\end{eqnarray}
Here,
$C^\prime(t_0)$ is an integral constant
corresponding to $C^\prime_\mu(t_0)$ in Sec. \ref{sec:ChapB-3}
with $M_1 = 1$.
In accordance with the general scheme, the P${}_1$ space is spanned by the doublet modes
\begin{eqnarray}
\label{eq:doubletmodes_for_lorentzeq}
\Vec{U}_3,\,\,\,A^{-1} \Vec{U}_3.
\end{eqnarray}
One notice that
the doublet modes 
in the present simple case happen to
belong to a common space;
recall that
$\Vec{U}_3$ is
an
eigenvector of $A$.
Thus,
the left vector $\Vec{U}_2$ belongs to the Q${}_1$ space complement to the P${}_0$ and P${}_1$ spaces. 
The
structure of the vector space 
is summarized
as follows:
\begin{eqnarray}
\label{eq:structure_fo_P0_P1_Q1}
\mathrm{P}_0 \,\,\,&:&\,\,\, \Vec{U}_1,\\
\mathrm{P}_1 \,\,\,&:&\,\,\, \Vec{U}_3,\\
\mathrm{Q}_1 \,\,\,&:&\,\,\, \Vec{U}_2.
\end{eqnarray}

The second-order equation reads
\begin{eqnarray}
\label{eq:lorentzeq_order2}
\frac{\mathrm{d}}{\mathrm{d}t}\tilde{\Vec{X}}_2(t;t_0)
= A \tilde{\Vec{X}}_2(t;t_0) + K(t-t_0)\frac{1}{1+\sigma}(\sigma\Vec{U}_1 - \Vec{U}_2),
\end{eqnarray}
with the time-dependent inhomogeneous term
\begin{eqnarray}
\label{eq:K_for_lorentz}
K(t-t_0) \equiv \chi C(t_0) 
- C(t_0) \Bigg[ \frac{C^2(t_0)}{b}(1-\mathrm{e}^{-b(t-t_0)})
-\mathrm{e}^{-b(t-t_0)}\frac{1}{b}C^\prime(t_0) \Bigg].
\end{eqnarray}
A general
solution to Eq. (\ref{eq:lorentzeq_order2}) is given by
\begin{eqnarray}
\label{eq:comp_solution_to_lorentzeq_order2}
\tilde{\Vec{X}}_2(t;t_0)
&=&
\mathrm{e}^{A(t-t_0)}
\Big[ \Vec{X}_2(t_0) +
Q_1(A-\partial/\partial s)^{-1}
Q_0K(s)\frac{1}{1+\sigma}(\sigma\Vec{U}_1 - \Vec{U}_2)\Big|_{s=0}\Big]\nonumber\\
&&{}+
(1-\mathrm{e}^{(t-t_0)\partial/\partial s})(-\partial/\partial s)^{-1}
P_0K(s)\frac{1}{1+\sigma}(\sigma\Vec{U}_1 - \Vec{U}_2)\Big|_{s=0}\nonumber\\
&&{}+
(\mathrm{e}^{A(t-t_0)}-\mathrm{e}^{(t-t_0)\partial/\partial s})
P_1(A-\partial/\partial s)^{-1}
Q_0K(s)\frac{1}{1+\sigma}(\sigma\Vec{U}_1 - \Vec{U}_2)\Big|_{s=0}\nonumber\\
&&{}-
\mathrm{e}^{(t-t_0)\partial/\partial s}
Q_1(A-\partial/\partial s)^{-1}
Q_0K(s)\frac{1}{1+\sigma}(\sigma\Vec{U}_1 - \Vec{U}_2)\Big|_{s=0}.
\end{eqnarray}
We
can
utilize the initial value $\Vec{X}_2(t_0)$
to eliminate the
unwanted
fast motion caused by the Q${}_1$ space as 
\begin{eqnarray}
\label{eq:initialvalue_to_lorentzeq_order2}
\Vec{X}_2(t_0) =
\tilde{\Vec{X}}_2(t=t_0;t_0) &=&
-Q_1(A-\partial/\partial s)^{-1}
Q_0K(s)\frac{1}{1+\sigma}(\sigma\Vec{U}_1 - \Vec{U}_2)\Big|_{s=0}\nonumber\\
&=&
-((-1-\sigma)-\partial/\partial s)^{-1}
K(s)\Big|_{s=0}\frac{1}{1+\sigma}(- \Vec{U}_2),
\end{eqnarray}
which leads to
\begin{eqnarray}
\label{eq:solution_to_lorentzeq_order2}
\tilde{\Vec{X}}_2(t;t_0)
&=&
(1-\mathrm{e}^{(t-t_0)\partial/\partial s})(-\partial/\partial s)^{-1}
K(s)\Big|_{s=0}\frac{1}{1+\sigma}\sigma\Vec{U}_1\nonumber\\
&&{}-
\mathrm{e}^{(t-t_0)\partial/\partial s}
((-1-\sigma)-\partial/\partial s)^{-1}
K(s)\Big|_{s=0}\frac{1}{1+\sigma}(- \Vec{U}_2).
\end{eqnarray}
We stop the perturbative analysis in this order.

Summing up the above solutions,
the perturbative solutions and initial values
are given by
\begin{eqnarray}
\label{eq:solution_to_lorentzeq}
\tilde{\Vec{X}}(t;t_0)
&=&
C(t_0)\Vec{U}_1
+
\epsilon\Bigg[
\mathrm{e}^{A(t-t_0)} A^{-1} \Vec{U}_3 C^\prime(t_0)
+ (\mathrm{e}^{A(t-t_0)}-1) A^{-1} C^2(t_0)\Vec{U}_3
\Bigg]\nonumber\\
&&{}+
\epsilon^2
\Bigg[
(1-\mathrm{e}^{(t-t_0)\partial/\partial s})(-\partial/\partial s)^{-1}
K(s)\Big|_{s=0}\frac{1}{1+\sigma}\sigma\Vec{U}_1\nonumber\\
&&{}-
\mathrm{e}^{(t-t_0)\partial/\partial s}
((-1-\sigma)-\partial/\partial s)^{-1}
K(s)\Big|_{s=0}\frac{1}{1+\sigma}(- \Vec{U}_2)
\Bigg],\\
\label{eq:initialvalue_to_lorentzeq}
\Vec{X}(t_0)
&=&
C(t_0)\Vec{U}_1
+
\epsilon
A^{-1} \Vec{U}_3 C^\prime(t_0)
\nonumber\\
&&{}+
\epsilon^2
\Bigg[
-
((-1-\sigma)-\partial/\partial s)^{-1}
K(s)\Big|_{s=0}\frac{1}{1+\sigma}(- \Vec{U}_2)
\Bigg],
\end{eqnarray}
respectively.
Applying the RG equation
$\partial \tilde{\Vec{X}}(t;t_0)/\partial t_0|_{t_0=t}=0$
to Eq. (\ref{eq:solution_to_lorentzeq}),
we have
\begin{eqnarray}
\label{eq:RG_for_lorentzeq}
&&\dot{C}\Vec{U}_1
+
\epsilon\Bigg[
- \Vec{U}_3 C^\prime
+ A^{-1} \Vec{U}_3 \dot{C^\prime}
- C^2\Vec{U}_3
\Bigg]\nonumber\\
&&{}+
\epsilon^2
\Bigg[
-K(0)\frac{1}{1+\sigma}\sigma\Vec{U}_1
-
(-\partial/\partial s)
((-1-\sigma)-\partial/\partial s)^{-1}
K(s)\Big|_{s=0}\frac{1}{1+\sigma}(- \Vec{U}_2)
\Bigg] = 0.\nonumber\\
\end{eqnarray}
We can read off the P${}_0$ and P${}_1$ components from this expression as
\begin{eqnarray}
\label{eq:Weq}
\dot{C} &=& \epsilon^2 \frac{\sigma}{1+\sigma}(\chi C + C C^\prime/b),\\
\label{eq:phieq}
\dot{C^\prime} &=& -bC^\prime-bC^2.
\end{eqnarray}
With these $C(t)$ and $C^\prime(t)$,
the
globally improved solution
defined on the invariant/attractive manifold is given by
\begin{eqnarray}
\label{eq:manifold}
\Vec{X}^{\mathrm{global}}(t)
&\equiv&
\Vec{X}(t_0=t)\nonumber\\
&=&
C(t)\Vec{U}_1
+
\epsilon
A^{-1} \Vec{U}_3 C^\prime(t)
\nonumber\\
&&{}+
\epsilon^2
\Bigg[
-
((-1-\sigma)-\partial/\partial s)^{-1}
K(s)\Big|_{s=0}\frac{1}{1+\sigma}(- \Vec{U}_2)
\Bigg]\Bigg|_{t_0=t},
\end{eqnarray}
or in terms of components
\begin{eqnarray}
\label{eq:manifold_x}
x &=& \epsilon C
+ \epsilon^3 \frac{\sigma}{1+\sigma}
\Bigg[-\frac{1}{1+\sigma}(\chi C - C^3/b)
+
\frac{1}{b - (1+\sigma)}\frac{1}{b}C(C^2 + C^\prime)\Bigg],\\
\label{eq:manifold_y}
y &=& \epsilon C
- \epsilon^3 \frac{1}{1+\sigma}
\Bigg[-\frac{1}{1+\sigma}(\chi C - C^3/b)
+
\frac{1}{b - (1+\sigma)}\frac{1}{b}C(C^2 + C^\prime)\Bigg],\\
\label{eq:manifold_z}
z &=& - \epsilon^2 \frac{1}{b}C^\prime,
\end{eqnarray}
with ${}^t(x,\,y,\,z) = \epsilon\,\Vec{X}^{\mathrm{global}}$.
In the derivations of Eqs. (\ref{eq:manifold_x}) and (\ref{eq:manifold_y}),
we have used the identity
\begin{eqnarray}
\label{eq:identity}
-((-1-\sigma)-\partial/\partial s)^{-1}
K(s)
=
\frac{1}{1+\sigma}(\chi C - C^3/b)
-
\mathrm{e}^{-bs}\frac{1}{b-(1+\sigma)}(C^2+C^\prime)C/b.
\end{eqnarray}
To see what has been obtained, let us see a limiting case.
From Eq. (\ref{eq:phieq}),
we identify the relaxation time of $C^\prime$ with $1/b$.
In fact,
after the time evolution from $t = 0$ to $t > 1/b$,
$C^\prime$ approaches to
\begin{eqnarray}
\label{eq:constitutiveeq}
C^\prime = -C^2.
\end{eqnarray}
Substituting $C^\prime = -C^2$
into Eqs. (\ref{eq:Weq}), (\ref{eq:manifold_x})-(\ref{eq:manifold_z}),
we obtain a closed equation with respect to $C$
and the invariant manifold parametrized by only $C$.
We note that
these equations written by $C$ are the same as
the reduced equations derived by employing the zero mode from the outset \cite{env002}.
We stress that
the set of Eqs. (\ref{eq:Weq}) and (\ref{eq:phieq}) governing the dynamics of $C$ and $C^\prime$
describes the \textit{mesoscopic dynamics} of the Lorenz model,
and the corresponding two-dimensional invariant/attractive manifold is given
by Eqs. (\ref{eq:manifold_x})-(\ref{eq:manifold_z}).
It might be interesting
to compare the present result with the previous ones
obtained by various reduction theories,
e.g., the center manifold theory \cite{geometrical,Carr}.
This is, however, beyond the scope of this section,
whose aim is to show the validity of the doublet scheme in the RG method.
Thus, the comparison with the previous works
and a further analysis of the Lorenz model by the doublet scheme
will be reported elsewhere.


\begin{figure}
 \begin{minipage}{1.0\hsize}
 \begin{center}
 \includegraphics[width=8cm]{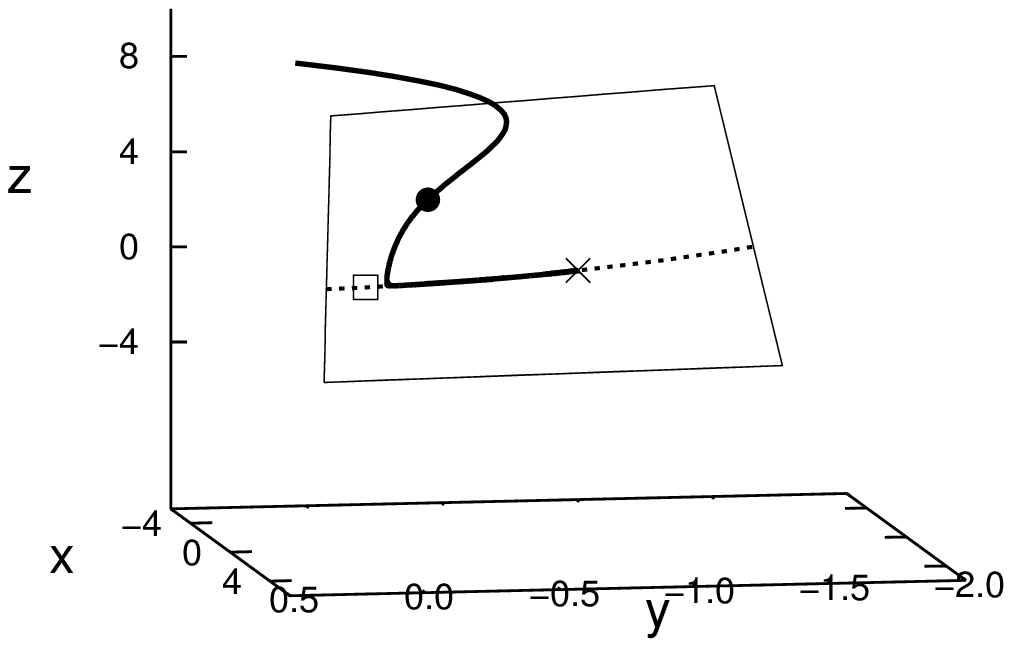}
 \end{center}
 \end{minipage}
 \caption{
 \label{fig:001}
 A comparison of the numerical solution of the Lorenz model (the solid line)
 with the invariant/attractive manifold given by the doublet scheme, i.e.,
 Eqs. (\ref{eq:manifold_x})-(\ref{eq:manifold_z}) for the parameter set
 $b = 8/3$, $\sigma = 10$, $\epsilon = 0.5$, and $\chi = +1$ (the surface).
 The square and cross denote the steady states (fixed points) (A) and (C), respectively,
 while the big dot with the coordinate $(x,\,y,\,z) = (-0.297,\,-0.212,\,3.437)$
 indicates the point
 from which the solution 
 gets to be confined on the invariant/attractive manifold.
 The dashed line shows the one-dimensional manifold
 given by imposing the constraint
 $C^\prime=-C^2$. 
 }
\end{figure}

\begin{figure}
 \begin{minipage}{\hsize}
 \begin{center}
 \includegraphics[width=16cm]{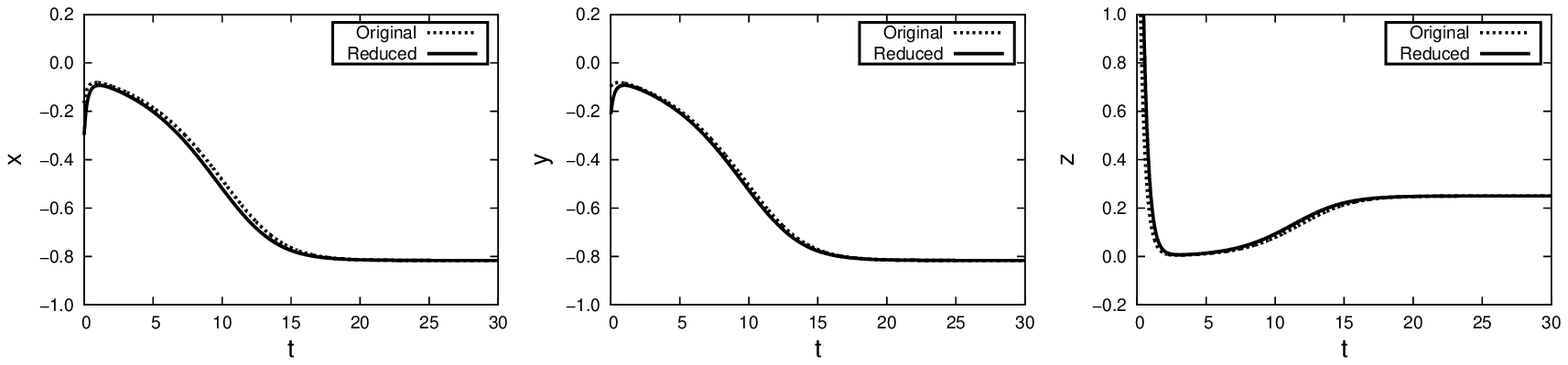}
 \end{center}
 \end{minipage}
 \caption{
 \label{fig:002}
 The time dependence of $x$, $y$, and $z$
 with the initial value set at the big dot in Fig. \ref{fig:001}
 for the same parameter set as that in Fig. \ref{fig:001}.
 The dashed lines denote the numerical solution to 
 the original equation (\ref{eq:lorentzeq2}),
 while the solid lines the solution
 to the reduced equations (\ref{eq:Weq}), (\ref{eq:phieq}), and (\ref{eq:manifold_x})-(\ref{eq:manifold_z}).
 The two solutions are hardly distinguishable.
 }
\end{figure}

Let us compare a solution to the Lorenz model (\ref{eq:lorentzeq2})
with
the
solution to
the reduced equations (\ref{eq:Weq}), (\ref{eq:phieq}), and (\ref{eq:manifold_x})-(\ref{eq:manifold_z})
under the same initial condition.
For this purpose,
we set the parameters of the Lorenz model as
$b = 8/3$, $\sigma = 10$, and $\epsilon = 0.5$ and $\chi = +1$,
which
gives
$r = 1.25$.
In this setting,
the origin (A) is unstable,
while the steady states (B) and (C) are stable
because $1 < r < r_c \sim 24.7$.
In Fig. \ref{fig:001},
we present a numerical solution to Eq. (\ref{eq:lorentzeq2})
whose initial values are $(x,\,y,\,z) = (1,\,5,\,15)$,
together with the two-dimensional manifold
described by Eqs. (\ref{eq:manifold_x})-(\ref{eq:manifold_z})
and the one-dimensional manifold
that we obtain by
imposing the constraint
 $C^\prime=-C^2$
on
Eqs. (\ref{eq:manifold_x})-(\ref{eq:manifold_z}).
It turns out that
the solution is attracted to the two-dimensional manifold 
at a rather early time, after which the solution
remains
on it.
Then it gets relaxed onto
the one-dimensional manifold,
and
finally
comes to the steady state (C) asymptotically.
In Fig. \ref{fig:002},
we demonstrate
the time dependence of $x$, $y$, and $z$ derived by the original equation (\ref{eq:lorentzeq2})
and that obtained by the reduced equations (\ref{eq:Weq}), (\ref{eq:phieq}), and (\ref{eq:manifold_x})-(\ref{eq:manifold_z}).
We find that
$x(t)$, $y(t)$, and $z(t)$ by Eqs. (\ref{eq:Weq}), (\ref{eq:phieq}), and (\ref{eq:manifold_x})-(\ref{eq:manifold_z})
are almost the same as those by Eq. (\ref{eq:lorentzeq2}).
Thus, we conclude that
this good agreement ensures the validity of the doublet scheme in the RG method.



\section{
Basics on Boltzmann equation}
\label{sec:ChapB-2}

In this section,
we give the basic notions and a brief summary of the previous
works as preliminaries
so that the significance of the present work will become clear.
First, 
a brief account is given on the basic properties of the Boltzmann equation
with a focus on those of the collision operator.
Then, 
we introduce Grad's moment method and its thirteen-moment approximation 
for the functional forms of the distribution function and the moments,
and present the explicit form of the Grad equation \cite{intro001}.

\subsection{
  Basic properties of Boltzmann equation
}
\label{sec:ChapB-2-1}
The Boltzmann equation that we consider in the present work reads
\begin{eqnarray}
  \label{eq:ChapB-2-1-001}
  \frac{\partial}{\partial t}f_{\Vec{v}}(t,\,\Vec{x})
  + \Vec{v}\cdot\Vec{\nabla}f_{\Vec{v}}(t,\,\Vec{x}) = C[f]_{\Vec{v}}(t,\,\Vec{x}).
\end{eqnarray}
Here,
$f_{\Vec{v}}(t,\,\Vec{x})$ denotes the distribution function
defined in
phase space 
$(\Vec{x},\,\Vec{v})$
with $t$ and $\Vec{x} = (x^1,\,x^2,\,x^3)$ being the space-time coordinate
and $\Vec{v} = (v^1,\,v^2,\,v^3)$ 
the velocity
of the one-shell particle
whose mass, momentum, and energy are given as $m$, $m\,\Vec{v}$, and $m\,|\Vec{v}|^2/2$, respectively.
The right-hand side of Eq. (\ref{eq:ChapB-2-1-001}) is the collision integral,
\begin{eqnarray}
  \label{eq:ChapB-2-1-002}
  C[f]_{\Vec{v}}(t,\,\Vec{x})
  &\equiv& \frac{1}{2!} \,
  \int_{\Vec{v_1}}
  \int_{\Vec{v_2}}
  \int_{\Vec{v_3}}
  \,
  \omega(\Vec{v},\,\Vec{v_1}|\Vec{v_2},\,\Vec{v_3})\nonumber\\
  &&{}\times \,
  \Big( f_{\Vec{v_2}}(t,\,\Vec{x}) \, f_{\Vec{v_3}}(t,\,\Vec{x})
  - f_{\Vec{v}}(t,\,\Vec{x}) \, f_{\Vec{v_1}}(t,\,\Vec{x}) \Big),
\end{eqnarray}
with $\int_{\Vec{v}} \equiv \int\!\!\mathrm{d}^3\Vec{v}$.
Here,
$\omega(\Vec{v},\,\Vec{v_1}|\Vec{v_2},\,\Vec{v_3})$ denotes
the transition probability due to the microscopic two particle interaction.
We note that
$\omega(\Vec{v},\,\Vec{v_1}|\Vec{v_2},\,\Vec{v_3})$ contains
the delta function representing the energy-momentum conservation,
\begin{eqnarray}
  \label{eq:ChapB-2-1-003}
  \omega(\Vec{v},\,\Vec{v_1}|\Vec{v_2},\,\Vec{v_3}) &\propto&
  \delta^3(m\,\Vec{v}+m\,\Vec{v_1} - m\,\Vec{v_2} -m\,\Vec{v_3})\nonumber\\
  &&{}\times \, \delta(m\,|\Vec{v}|^2/2 + m\,|\Vec{v_1}|^2/2
  - m\,|\Vec{v_2}|^2/2 - m\,|\Vec{v_3}|^2/2),
\end{eqnarray}
and also has the symmetric properties due to the indistinguishability of the particles
and the time reversal invariance of the microscopic transition probability,
\begin{eqnarray}
  \label{eq:ChapB-2-1-004}
  \omega(\Vec{v},\,\Vec{v_1}|\Vec{v_2},\,\Vec{v_3}) =
  \omega(\Vec{v_2},\,\Vec{v_3}|\Vec{v},\,\Vec{v_1}) =
  \omega(\Vec{v_1},\,\Vec{v}|\Vec{v_3},\,\Vec{v_2}) =
  \omega(\Vec{v_3},\,\Vec{v_2}|\Vec{v_1},\,\Vec{v}).
\end{eqnarray}
It should be stressed here that
we have confined ourselves to the case
in which the particle number 
is conserved in the collision process.
In the following,
we suppress the arguments
$(t,\,\Vec{x})$ when no misunderstanding is expected.


The property of the transition probability shown in Eq. (\ref{eq:ChapB-2-1-004})
leads to the following identity satisfied for an arbitrary vector $\varphi_{\Vec{v}}$,
\begin{eqnarray}
  \label{eq:ChapB-2-1-005}
  \int_{\Vec{v}}
  \,\varphi_{\Vec{v}}\,C[f]_{\Vec{v}}
  &=& \frac{1}{2\,!}\,\frac{1}{4}\,
  \int_{\Vec{v}}
  \int_{\Vec{v}_1}
  \int_{\Vec{v}_2}
  \int_{\Vec{v}_3}
  \,\omega(\Vec{v},\,\Vec{v_1}|\Vec{v_2},\,\Vec{v_3})\nonumber\\
  &&{}\times\,
  ( \varphi_{\Vec{v}} + \varphi_{\Vec{v}_1}
  - \varphi_{\Vec{v}_2} - \varphi_{\Vec{v}_3} )\,
  ( f_{\Vec{v_2}} \, f_{\Vec{v_3}}
  - f_{\Vec{v}} \, f_{\Vec{v_1}} ).
\end{eqnarray}

A function $\varphi_{\Vec{v}}$ is called a collision invariant
when it satisfies
\begin{eqnarray}
  \label{eq:ChapB-2-1-006}
  \int_{\Vec{v}}
  \,\varphi_{\Vec{v}}\,C[f]_{\Vec{v}} = 0.
\end{eqnarray}
As is easily confirmed using the identity (\ref{eq:ChapB-2-1-005}) and the property (\ref{eq:ChapB-2-1-003}),
$\varphi_{\Vec{v}} = 1$, $m\,\Vec{v}$, and $m\,|\Vec{v}|^2/2$
are collision invariants;
\begin{eqnarray}
  \label{eq:ChapB-2-1-007}
  \int_{\Vec{v}}
  \,(1,\,m\,\Vec{v},\,m\,|\Vec{v}|^2/2) \, C[f]_{\Vec{v}} = 0,
\end{eqnarray}
which represent the conservation of the particle number, momentum, and energy
by the collision process, respectively.
We see that
the linear combination of these collision invariants
given by
$\varphi_{\Vec{v}} = a + \Vec{b}\cdot m\,\Vec{v}
+ c \, m\,|\Vec{v}|^2/2$
is also a collision invariant with $a$, $\Vec{b}$, and $c$
being arbitrary functions of $t$ and $\Vec{x}$.

Using the Boltzmann equation (\ref{eq:ChapB-2-1-001})
together with the collision invariants $(1,\,m\,\Vec{v},\break\,m\,|\Vec{v}|^2/2)$,
we have the following balance equations,
\begin{eqnarray}
  \label{eq:ChapB-2-1-009}
  \int_{\Vec{v}}
  \, (1,\,m\,\Vec{v},\,m\,|\Vec{v}|^2/2) \, \Bigg[ \frac{\partial}{\partial t} + \Vec{v}\cdot\Vec{\nabla} \Bigg]
  f_{\Vec{v}}
  = 0,
\end{eqnarray}
which are reduced to the following forms expressed
with
the macroscopic variables
\begin{eqnarray}
  \label{eq:ChapB-2-1-010}
  \frac{\partial}{\partial t}\rho &=&  - \Vec{\nabla}\cdot ( \rho\,\Vec{V} ),\\
  \label{eq:ChapB-2-1-011}
  m\,\rho\,\frac{\partial}{\partial t}V^i
  &=&
  - m\,\rho\, \Vec{V} \cdot \Vec{\nabla} V^i - \nabla^j P^{ji},\\
  \label{eq:ChapB-2-1-012}
  \rho\,\frac{\partial}{\partial t}e
  &=&
  - \rho\, \Vec{V} \cdot \Vec{\nabla} e
  - P^{ij} \, \nabla^i V^j - \Vec{\nabla}\cdot\Vec{Q},
\end{eqnarray}
respectively.
Here, we have used the Einstein summation convention for the dummy indices
for the spatial components.
The particle density $\rho$, the fluid velocity $V^i$,
the internal energy density $e$, the pressure tensor $P^{ij}$,
and the heat current $Q^i$ have the following microscopic expressions, respectively;
\begin{eqnarray}
  \label{eq:ChapB-2-1-013}
  \rho &\equiv&
  \int_{\Vec{v}}
  \, f_{\Vec{v}},
  \,\,\,
  V^i \equiv
  \frac{1}{\rho} \,
  \int_{\Vec{v}}
  \, v^i
  \, f_{\Vec{v}},
  \,\,\,
  e \equiv
  \frac{1}{\rho} \,
  \int_{\Vec{v}}
  \, \frac{m}{2} \, |\Vec{v} - \Vec{V}|^2
  \, f_{\Vec{v}},\\
  \label{eq:ChapB-2-1-014}
  P^{ij} &\equiv&
  \int_{\Vec{v}}
  \, m \, (v^i - V^i) \, (v^j - V^j) \, f_{\Vec{v}},
  \,\,\,
  Q^i \equiv
  \int_{\Vec{v}}
  \, \frac{m}{2} \, |\Vec{v} - \Vec{V}|^2 \, (v^i - V^i) \, f_{\Vec{v}}.
\end{eqnarray}
It is noted that
while these equations have the same forms as the hydrodynamic equation,
nothing about the dynamical properties is contained in these equations
before the evolution of the distribution function $f_{\Vec{v}}$
is obtained by solving Eq. (\ref{eq:ChapB-2-1-001}).

In this kinetic theory,
the entropy density $s$ and current $\Vec{J}_s$ may be defined by
\begin{eqnarray}
  (s\,,\,\Vec{J}_s)
  \equiv -
  \int_{\Vec{v}}
  \, (1\,,\,\Vec{v}) \, f_{\Vec{v}} \, ( \ln f_{\Vec{v}} - 1 ).
\end{eqnarray}
Using Eq. (\ref{eq:ChapB-2-1-001}), we have
\begin{eqnarray}
  \label{eq:ChapB-2-1-021}
  \frac{\partial}{\partial t}s
  + \Vec{\nabla}\cdot\Vec{J}_s
  = -
  \int_{\Vec{v}}
  \,(\ln f_{\Vec{v}})\,
  C[f]_{\Vec{v}}.
\end{eqnarray}
The above equation tells us that
the entropy
$S(t) \equiv \int\!\!\mathrm{d}^3\Vec{x}\,s(t,\,\Vec{x})$
is conserved only if
$\ln f_{\Vec{v}}$ is a collision invariant,
or a linear combination of the basic collision invariants $(1,\,\Vec{v},\,m\,|\Vec{v}|^2/2)$.
In other words,
the entropy-conserving distribution function is parametrized as
\begin{eqnarray}
  \label{eq:ChapB-2-1-024}
  f_{\Vec{v}}
  = n \, \Bigg[ \frac{m}{2\,\pi\,T} \Bigg]^{\frac{3}{2}}
  \, \exp\Bigg[ - \frac{m \, |\Vec{v} - \Vec{u}|^2}{2\,T} \Bigg]
  \equiv f^{\mathrm{eq}}_{\Vec{v}},
\end{eqnarray}
which is identified with the Maxwellian, i.e., the local equilibrium distribution function.
The quantities 
$T=T(t,\,\Vec{x})$, $n=n(t,\,\Vec{x})$, and $\Vec{u}=\Vec{u}(t,\,\Vec{x})$
in Eq. (\ref{eq:ChapB-2-1-024}) 
are the temperature, density, and flow velocity with space- and time-dependence, respectively.

We note that
for $f_{\Vec{v}}=f^{\mathrm{eq}}_{\Vec{v}}$
the collision integral identically vanishes,
\begin{eqnarray}
  \label{eq:ChapB-2-1-025}
  C[f^{\mathrm{eq}}]_{\Vec{v}} = 0, 
\end{eqnarray}
because of the identity derived from Eq. (\ref{eq:ChapB-2-1-003}):
\begin{eqnarray}
  \omega(\Vec{v},\,\Vec{v_1}|\Vec{v_2},\,\Vec{v_3}) \,
  (f^{\mathrm{eq}}_{\Vec{v}}\,f^{\mathrm{eq}}_{\Vec{v_1}} - f^{\mathrm{eq}}_{\Vec{v_2}}\,f^{\mathrm{eq}}_{\Vec{v_3}}) = 0.
\end{eqnarray}

Substituting
$f_{\Vec{v}} = f^{\mathrm{eq}}_{\Vec{v}}$
into the balance equations (\ref{eq:ChapB-2-1-010})-(\ref{eq:ChapB-2-1-012}),
we have 
\begin{eqnarray}
  \label{eq:ChapB-2-1-026}
  \frac{\partial}{\partial t}n &=& - \Vec{\nabla}\cdot ( n\,\Vec{u} ),\\
  \label{eq:ChapB-2-1-027}
  m\,n\,\frac{\partial}{\partial t}u^i
  &=& - m\,n\, \Vec{u} \cdot \Vec{\nabla} u^i
  - \nabla^i (n\,T),\\
  \label{eq:ChapB-2-1-028}
  n\,\frac{\partial}{\partial t}(3\,T/2)
  &=& - n\, \Vec{u} \cdot \Vec{\nabla}(3\,T/2)
  - n\,T \, \Vec{\nabla}\cdot\Vec{u},
\end{eqnarray}
where we have used the fact that
Eqs. (\ref{eq:ChapB-2-1-013}) and (\ref{eq:ChapB-2-1-014}) are reduced to
$\rho =  n$,
$V^i = u^i$,
$e = 3\,T/2$,
$P^{ij} = \delta^{ij} \, n \, T$,
and $Q^i = 0$,
respectively.
We remark that
Eqs. (\ref{eq:ChapB-2-1-026})-(\ref{eq:ChapB-2-1-028}) are identical with the Euler equation,
which describes the fluid dynamics with no dissipative effects,
and $e$ and $P^{ij}$ are the equations of state of an isotropic dilute gas.
Since
the entropy-conserving distribution function $f^{\mathrm{eq}}_{\Vec{v}}$
reproduces the Euler equation,
we see that the dissipative effect is attributable to a deviation of $f_{\Vec{v}}$
from $f^{\mathrm{eq}}_{\Vec{v}}$.

\subsection{
  Grad's thirteen-moment approximation and Grad equation
}
\label{sec:ChapB-2-2}
In Grad's theory \cite{intro001}, 
the dissipative distribution function $f_{\Vec{v}}$ is first expanded around 
the local equilibrium one $f^{\mathrm{eq}}_{\Vec{v}}$ as
\begin{eqnarray}
  \label{eq:ChapB-2-2-001}
  f_{\Vec{v}} = f^{\mathrm{eq}}_{\Vec{v}}\,(1 +
  \Phi_{\Vec{v}}).
\end{eqnarray}
Substituting Eq. (\ref{eq:ChapB-2-2-001}) into the Boltzmann equation (\ref{eq:ChapB-2-1-001}),
we have
\begin{eqnarray}
  \label{eq:linboltzmanneq}
  (f^{\mathrm{eq}}_{\Vec{v}})^{-1} \, 
  \Bigg[ \frac{\partial}{\partial t} + \Vec{v}\cdot\Vec{\nabla} \Bigg]
  f^{\mathrm{eq}}_{\Vec{v}} \, (1 + \Phi_{\Vec{v}})
  =
  \int_{\Vec{k}}
  \, L_{\Vec{v}\Vec{k}} \, \Phi_{\Vec{k}}
  +
  (f^{\mathrm{eq}}_{\Vec{v}})^{-1}
  \frac{1}{2}\,
  \int_{\Vec{k}}\int_{\Vec{l}}
  \, B_{\Vec{v}\Vec{k}\Vec{l}} \, f^{\mathrm{eq}}_{\Vec{k}}\,\Phi_{\Vec{k}} \, f^{\mathrm{eq}}_{\Vec{l}}\,\Phi_{\Vec{l}},\nonumber\\
\end{eqnarray}
where 
$L_{\Vec{v}\Vec{k}}$ is the linearized collision operator
\begin{eqnarray}
  \label{eq:ChapB-2-2-012}
  L_{\Vec{v}\Vec{k}}
  &=& (f^{\mathrm{eq}}_{\Vec{v}})^{-1} \, \frac{\delta}{\delta
  f_{\Vec{k}}}C[f]_{\Vec{v}}\Bigg|_{f=f^{\mathrm{eq}}}
  \, f^{\mathrm{eq}}_{\Vec{k}}\nonumber\\
  &=&
  \frac{1}{2!}\,
  \int_{\Vec{v_1}}
  \int_{\Vec{v_2}}
  \int_{\Vec{v_3}}
  \,
  \omega(\Vec{v},\,\Vec{v_1}|\Vec{v_2},\,\Vec{v_3})\,f^{\mathrm{eq}}_{\Vec{v_1}}\,
  (\delta_{\Vec{v_2}\Vec{k}} + \delta_{\Vec{v_3}\Vec{k}} - \delta_{\Vec{v}\Vec{k}} - \delta_{\Vec{v_1}\Vec{k}}),
\end{eqnarray}
and $B_{\Vec{v}\Vec{k}\Vec{l}}$ is the second derivative of the collision integral
\begin{eqnarray}
  \label{eq:Btensor}
  B_{\Vec{v}\Vec{k}\Vec{l}}
  &=& \frac{\delta^2}{\delta f_{\Vec{k}} \delta f_{\Vec{l}}}C[f]_{\Vec{v}}\Bigg|_{f=f^{\mathrm{eq}}}\nonumber\\
  &=&
  \frac{1}{2!}\,
  \int_{\Vec{v_1}}
  \int_{\Vec{v_2}}
  \int_{\Vec{v_3}}
  \,
  \omega(\Vec{v},\,\Vec{v_1}|\Vec{v_2},\,\Vec{v_3})\,
  (\delta_{\Vec{v_2}\Vec{k}}\,\delta_{\Vec{v_3}\Vec{l}} + \delta_{\Vec{v_2}\Vec{l}}\,\delta_{\Vec{v_3}\Vec{k}}
  - \delta_{\Vec{v}\Vec{k}}\,\delta_{\Vec{v_1}\Vec{l}} - \delta_{\Vec{v}\Vec{l}}\delta_{\Vec{v_1}\Vec{k}}),\nonumber\\
\end{eqnarray}
with
$\delta_{\Vec{v}\Vec{k}} \equiv \delta^3(\Vec{v}-\Vec{k})$.

To express $\Phi_{\Vec{v}}$ in terms of hydrodynamic variables,
let us introduce $\Vec{v}$-dependent quantities
$\hat{\pi}^{ij}_{\Vec{v}}$ and $\hat{J}^i_{\Vec{v}}$ 
defined by
\begin{eqnarray}
  \label{eq:ChapB-2-2-003}
  \hat{\pi}^{ij}_{\Vec{v}} \equiv
  m \, \Delta^{ijkl} \, \delta v^k \, \delta v^l,
\end{eqnarray}
and
\begin{eqnarray}
  \label{eq:ChapB-2-2-004}
  \hat{J}^i_{\Vec{v}} \equiv
  \Big(\frac{m}{2}\,
  |\Vec{\delta v}|^2 - \frac{5}{2}\,T\Big)\,\delta v^i,
\end{eqnarray}
with the peculiar velocity
$\Vec{\delta v} \equiv \Vec{v} - \Vec{u}$
and
the projection matrix
\begin{eqnarray}
  \label{eq:ChapB-2-2-025}
  \Delta^{ijkl} \equiv
  \frac{1}{2} \, \Big(\delta^{ik}\,\delta^{jl} + \delta^{il}\,\delta^{jk} - \frac{2}{3}\,\delta^{ij}\,\delta^{kl}\Big).
\end{eqnarray}
Here,
$\hat{\pi}^{ij}_{\Vec{v}}$ and $\hat{J}^i_{\Vec{v}}$ are identified as
the microscopic representations of the viscous pressure and heat flux, respectively.
Thanks to the symmetry property of $\Delta^{ijkl}$,
$\hat{\pi}_{\Vec{v}}^{ij}$ is symmetric and traceless:
\begin{eqnarray}
  \label{eq:ChapB-2-2-005}
  \hat{\pi}_{\Vec{v}}^{ij} = \hat{\pi}_{\Vec{v}}^{ji},\,\,\,
  \delta^{ij} \, \hat{\pi}_{\Vec{v}}^{ij} = 0.
\end{eqnarray}
It is to be noted that
$\hat{\pi}^{ij}_{\Vec{v}}$ and $\hat{J}^i_{\Vec{v}}$ are orthogonal to the collision invariants as
\begin{eqnarray}
  {\langle\, \varphi\,,\, \hat{\pi}^{ij}\,\rangle}_{\mathrm{eq}} = {\langle\, \varphi\,,\, \hat{J}^{i}\,\rangle}_{\mathrm{eq}} = 0,\,\,\,\,\,\,
  \varphi_{\Vec{v}} = 1,\,m\,\Vec{v},\,m\,|\Vec{v}|^2/2,
\end{eqnarray}
where 
the inner product for two arbitrary functions $\psi_{\Vec{v}}$ and $\chi_{\Vec{v}}$ is defined by
\begin{eqnarray}
  \label{eq:ChapB-2-2-030}
  {\langle\, \psi\,,\, \chi\,\rangle}_{\mathrm{eq}} \equiv
  \int_{\Vec{v}}
  \, f^{\mathrm{eq}}_{\Vec{v}}
  \, \psi_{\Vec{v}} \, \chi_{\Vec{v}}.
\end{eqnarray}

With the use of the vector fields $\hat{\pi}^{ij}_{\Vec{v}}$ and $\hat{J}^i_{\Vec{v}}$,
the conventional ansatz for the form of $\Phi_{\Vec{v}}$ is expressed as
\begin{eqnarray}
  \label{eq:ChapB-2-2-002}
  \Phi_{\Vec{v}}
  = - \frac{\hat{\pi}^{ij}_{\Vec{v}} \, \pi^{ij}}{\frac{1}{5}\,{\langle\, \hat{\pi}^{kl}\,,\, \hat{\pi}^{kl}\,\rangle}_{\mathrm{eq}}}
  - \frac{\hat{J}^{i}_{\Vec{v}} \, J^{i}}{\frac{1}{3}\,{\langle\, \hat{J}^{k}\,,\, \hat{J}^{k}\,\rangle}_{\mathrm{eq}}}
  \equiv \Phi^{\mathrm{G}}_{\Vec{v}}.
\end{eqnarray}
Here,
$\pi^{ij}$ and $J^{i}$ are expansion coefficients
which have a dependence on  $(t,\,\Vec{x})$ as well as $T$, $n$, and $u^i$;
$\pi^{ij}=\pi^{ij}(t,\,\Vec{x})$ and $J^{i}=J^{i}(t,\,\Vec{x})$.
The coefficients $\pi^{ij}$ and $J^{i}$ should be interpreted as the viscous pressure and heat flux, respectively.
It is noted that
the total number of independent components of $T$, $n$, $u^i$, $\pi^{ij}$, and $J^{i}$ is thirteen
because
without loss of generality
we can suppose that
\begin{eqnarray}
  \pi^{ij} = \pi^{ji},\,\,\,
  \delta^{ij} \, \pi^{ij} = 0,
\end{eqnarray}
owning to the symmetric and traceless properties of $\hat{\pi}^{ij}_{\Vec{v}}$ 
in Eq. (\ref{eq:ChapB-2-2-005}).
The prefactors
$5/{\langle\, \hat{\pi}^{kl}\,,\, \hat{\pi}^{kl}\,\rangle}_{\mathrm{eq}}$
and
$3/{\langle\, \hat{J}^{k}\,,\, \hat{J}^{k}\,\rangle}_{\mathrm{eq}}$
in $\Phi^{\mathrm{G}}_{\Vec{v}}$
have been introduced so that 
the followings are satisfied:
\begin{eqnarray}
  \label{eq:pi_and_J}
  \pi^{ij} = -{\langle\, \hat{\pi}^{ij}\,,\, \Phi^{\mathrm{G}}\,\rangle}_{\mathrm{eq}},\,\,\,
  J^i = -{\langle\, \hat{J}^i\,,\, \Phi^{\mathrm{G}}\,\rangle}_{\mathrm{eq}},
\end{eqnarray}
where we have used the relations
\begin{eqnarray}
  {\langle\, \hat{\pi}^{ij}\,,\, \hat{\pi}^{kl}\,\rangle}_{\mathrm{eq}}
  &=& \frac{1}{5}\,\Delta^{ijkl}\,{\langle\, \hat{\pi}^{ab}\,,\, \hat{\pi}^{ab}\,\rangle}_{\mathrm{eq}},\,\,\,
  {\langle\, \hat{J}^i\,,\, \hat{J}^j\,\rangle}_{\mathrm{eq}}
  = \frac{1}{3}\,\delta^{ij}\,{\langle\, \hat{J}^a\,,\, \hat{J}^a\,\rangle}_{\mathrm{eq}},\\
  {\langle\, \hat{\pi}^{ij}\,,\, \hat{J}^{k}\,\rangle}_{\mathrm{eq}}
  &=&
  {\langle\, \hat{J}^i\,,\, \hat{\pi}^{kl}\,\rangle}_{\mathrm{eq}}
  = 0.
\end{eqnarray}

To determine the $(t,\,\Vec{x})$-dependence of the thirteen coefficients
$T$, $n$, $u^i$, $\pi^{ij}$, and $J^i$,
one may utilize the Boltzmann equation (\ref{eq:linboltzmanneq}),
the inner product of which with
\textit{any independent} thirteen variables dependent on $\Vec{v}$
would give a closed system of equations for the thirteen coefficients.
Let us denote such a set of the thirteen variables by $\vec{\phi}_{13\Vec{v}}$.
In Grad's thirteen-moment approximation,
the five collision invariants $(1,\,m\,\Vec{v},\,m\,|\Vec{v}|^2/2)$ and
the eight quantities $\hat{\pi}^{ij}_{\Vec{v}}$ and $\hat{J}^{i}_{\Vec{v}}$ are adopted as $\vec{\phi}_{13\Vec{v}}$:
\begin{eqnarray}
  \vec{\phi}_{13\Vec{v}} = \Big\{ 1,\,m\,\Vec{v},\,\frac{m}{2}\,|\Vec{v}|^2,\,\hat{\pi}^{ij}_{\Vec{v}},\,\hat{J}^{i}_{\Vec{v}} \Big\}
  \equiv \vec{\phi}^{\mathrm{G}}_{13\Vec{v}}.
\end{eqnarray}
Here,
it should be emphasized that this is merely a possible choice without any foundation.
The resultant closed equations consist of the five balance equations
\begin{eqnarray}
  \label{eq:ChapB-2-2-009}
  \int_{\Vec{v}}
  \, (1,\,m\,\Vec{v},\,m\,|\Vec{v}|^2/2) \, \Bigg[ \frac{\partial}{\partial t} + \Vec{v}\cdot\Vec{\nabla} \Bigg]
  f^{\mathrm{eq}}_{\Vec{v}} \, (1 + \Phi^{\mathrm{G}}_{\Vec{v}})
  = 0,
\end{eqnarray}
and the eight relaxation equations
\begin{eqnarray}
  \label{eq:ChapB-2-2-010}
  \int_{\Vec{v}}
  \, (\hat{\pi}^{ij}_{\Vec{v}},\,\hat{J}^{i}_{\Vec{v}})
  \, \Bigg[ \frac{\partial}{\partial t} + \Vec{v}\cdot\Vec{\nabla} \Bigg]
  f^{\mathrm{eq}}_{\Vec{v}} \, (1 + \Phi^{\mathrm{G}}_{\Vec{v}})
  &=& {\langle \, (\hat{\pi}^{ij},\,\hat{J}^{i}) \,,\, L \, \Phi^{\mathrm{G}} \, \rangle}_{\mathrm{eq}}\nonumber\\
  &+& \frac{1}{2}\,\int_{\Vec{v}}\int_{\Vec{k}}\int_{\Vec{l}}
  \,(\hat{\pi}^{ij}_{\Vec{v}},\,\hat{J}^{i}_{\Vec{v}})
  \,B_{\Vec{v}\Vec{k}\Vec{l}} \, f^{\mathrm{eq}}_{\Vec{k}} \, \Phi^{\mathrm{G}}_{\Vec{k}}
  \, f^{\mathrm{eq}}_{\Vec{l}} \, \Phi^{\mathrm{G}}_{\Vec{l}},\nonumber\\
\end{eqnarray}
where the following representation is used:
$\big[ L \, \Phi^{\mathrm{G}} \big]_{\Vec{v}}
=
\int_{\Vec{k}}
\,L_{\Vec{v}\Vec{k}} \, \Phi^{\mathrm{G}}_{\Vec{k}}$.

By carrying out the integration with respect to
the velocities,
Eqs. (\ref{eq:ChapB-2-2-009}) and (\ref{eq:ChapB-2-2-010}) are reduced to
the following equations
which govern the dynamics of $T$, $n$, $u^i$, $\pi^{ij}$, and $J^i$:
\begin{eqnarray}
  \label{eq:ChapB-2-2-018}
  \frac{\partial}{\partial t}n
  &=& -\Vec{\nabla}\cdot(n\,\Vec{u}),\\
  \label{eq:ChapB-2-2-019}
  m\,n\,\frac{\partial}{\partial t}u^i
  &=& -m\,n\,\Vec{u}\cdot\Vec{\nabla}u^i
  - \nabla^j(n\,T\,\delta^{ji} - \pi^{ji}),\\
%
  \label{eq:ChapB-2-2-020}
  n\,\frac{\partial}{\partial t}(3\,T/2)
  &=&
  -n\,\Vec{u}\cdot\Vec{\nabla}(3\,T/2)
  -
  (n\,T\,\delta^{ij} - \pi^{ij})\,\nabla^i u^j
  + \nabla^i J^i,\\
%
  \label{eq:ChapB-2-2-021}
%
  \pi^{ij}
  &=&
  2\,\eta^{\mathrm{G}} \, \sigma^{ij}
  - \tau^{\mathrm{G}}_\pi \, \Big(\frac{\partial}{\partial t}
  + \Vec{u}\cdot\Vec{\nabla}\Big)\pi^{ij}
  - \ell_{\pi J}^{\mathrm{G}} \, \nabla^{\langle i} J^{j \rangle}\nonumber\\
  &+&
  \kappa^{(1)\mathrm{G}}_{\pi\pi}\,\pi^{ij}\,\theta
  + \kappa^{(2)\mathrm{G}}_{\pi\pi} \, \pi^{k\langle i} \, \sigma^{j\rangle k}
  + \kappa^{(3)\mathrm{G}}_{\pi\pi}\,\pi^{k\langle i} \, \omega^{j\rangle k}
  + \kappa^{(1)\mathrm{G}}_{\pi J} \, J^{\langle i} \, \nabla^{j \rangle} T
  + \kappa^{(2)\mathrm{G}}_{\pi J} \, J^{\langle i} \, \nabla^{j \rangle} n\nonumber\\
  &+&
  b_{\pi\pi\pi}^{\mathrm{G}} \, \pi^{k\langle i} \, \pi^{j \rangle k}
  + b_{\pi JJ}^{\mathrm{G}} \, J^{\langle i} \, J^{j \rangle},\\
  \label{eq:ChapB-2-2-022}
%
  J^{i}
  &=&
  \lambda^{\mathrm{G}} \, \nabla^i T
  - \tau^{\mathrm{G}}_J \, \Big(\frac{\partial}{\partial t}+\Vec{u}\cdot\Vec{\nabla}\Big)J^{i}
  - \ell_{J\pi}^{\mathrm{G}} \, \nabla^k \pi^{ki}\nonumber\\
  &+& 
  \kappa^{(1)\mathrm{G}}_{J\pi}\,\pi^{ik}\,\nabla^k T
  + \kappa^{(2)\mathrm{G}}_{J\pi}\,\pi^{ik}\,\nabla^k n
  + \kappa^{(1)\mathrm{G}}_{JJ}\,J^{i}\,\theta
  + \kappa^{(2)\mathrm{G}}_{JJ}\,J^{k}\,\sigma^{ik} + \kappa^{(3)\mathrm{G}}_{JJ}\,J^{k}\,\omega^{ik}\nonumber\\
  &+&
  b_{J \pi J}^{\mathrm{G}} \, \pi^{ik} \, J^{k},
\end{eqnarray}
where $A^{\langle ij \rangle} \equiv \Delta^{ijkl}\,A^{kl}$ is traceless and symmetric tensor.
The scalar expansion $\theta \equiv \Vec{\nabla}\cdot\Vec{u}$,
the shear tensor $\sigma^{ij} \equiv \Delta^{ijkl}\,\nabla^k u^l$,
and the vorticity term $\omega^{ij} \equiv (\nabla^i u^j - \nabla^j u^i)/2$
have been introduced.
The characteristic properties of the moment method as a microscopic theory 
of the hydrodynamics is dictated in the microscopic expressions of 
the   transport coefficients and
relaxation times/lengths
in
Eqs. (\ref{eq:ChapB-2-2-021}) and (\ref{eq:ChapB-2-2-022}).
For the sake of later comparison with our results, let us just pick up
$\eta^{\mathrm{G}}$ and $\tau^{\mathrm{G}}_\pi$
($\lambda^{\mathrm{G}}$ and $\tau^{\mathrm{G}}_J$),
which
denote
the transport coefficient and relaxation time
associated with the viscous pressure $\pi^{ij}$ (the heat flux $J^{i}$), respectively;
their
microscopic representations
in the moment method
are given by
\begin{eqnarray}
  \label{eq:ChapB-2-2-026}
  \eta^{\mathrm{G}} &\equiv& - \frac{1}{10\,T}\,\frac{\big[{\langle\, \hat{\pi}^{ij}\,,\,\hat{\pi}^{ij} \,\rangle}_{\mathrm{eq}}\big]^2
  }{
  {\langle\, \hat{\pi}^{kl}\,,\,L \, \hat{\pi}^{kl} \,\rangle}_{\mathrm{eq}}},\,\,\,
  \lambda^{\mathrm{G}} \equiv - \frac{1}{3\,T^2}\,\frac{\big[{\langle\, \hat{J}^{i}\,,\,\hat{J}^{i} \,\rangle}_{\mathrm{eq}}\big]^2
  }{
  {\langle\, \hat{J}^{k}\,,\,L\,\hat{J}^k \,\rangle}_{\mathrm{eq}}},\\
  \label{eq:ChapB-2-2-028}
  \tau^{\mathrm{G}}_{\pi} &\equiv& -\frac{{\langle\, \hat{\pi}^{ij}\,,\,\hat{\pi}^{ij} \,\rangle}_{\mathrm{eq}}}
  {{\langle\, \hat{\pi}^{kl}\,,\,L\,\hat{\pi}^{kl} \,\rangle}_{\mathrm{eq}}},\,\,\,
  \tau^{\mathrm{G}}_{J} \equiv -\frac{{\langle\, \hat{J}^{i}\,,\,\hat{J}^{i} \,\rangle}_{\mathrm{eq}}}
  {{\langle\, \hat{J}^{k}\,,\,L\,\hat{J}^{k} \,\rangle}_{\mathrm{eq}}}.
\end{eqnarray}
The transport coefficients $\eta^{\mathrm{G}}$ and $\lambda^{\mathrm{G}}$ are to be identified with
the shear viscosity and thermal conductivity, respectively.
It is well known, however, that
the shear viscosity and thermal conductivity in the Grad equation
are not in accord with those by the Chapman-Enskog method \cite{chapman-enskog}.
In fact,
the microscopic representations by Chapman and Enskog read
\begin{eqnarray}
  \label{eq:ChapB-2-2-033}
  \eta^{\mathrm{CE}} \equiv - \frac{1}{10\,T}\,
  {\langle\, \hat{\pi}^{ij}\,,\,L^{-1} \, \hat{\pi}^{ij} \,\rangle}_{\mathrm{eq}},\,\,\,
  \lambda^{\mathrm{CE}} \equiv - \frac{1}{3\,T^2}\,{\langle\, \hat{J}^{i}\,,\,L^{-1} \, \hat{J}^{i} \,\rangle}_{\mathrm{eq}},
\end{eqnarray}
and one easily sees that
$\eta^{\mathrm{G}} \ne \eta^{\mathrm{CE}}$
and 
$\lambda^{\mathrm{G}} \ne \lambda^{\mathrm{CE}}$
where $L^{-1}_{\Vec{v}\Vec{k}}$ denotes the inverse matrix of $L_{\Vec{v}\Vec{k}}$.

\section{
  Reduction of Boltzmann equation to mesoscopic dynamics with doublet scheme in RG method
}  
\label{sec:ChapB-4}
In this section,
we 
show 
that
the doublet scheme
developed in Sec. \ref{sec:ChapB-3}
is naturally applied to
 the Boltzmann equation 
to derive
the causal hydrodynamic equation
in the mesoscopic scale,
without recourse to any ansatz, although
 straightforward but somewhat
tedious manipulations are inherently involved for
the explicit calculations of the excited modes of the linearized collision operator and so on;
the detailed computations are
presented in 
\ref{sec:ChapB-9}.
Then we clarify the desirable properties possessed by the resultant hydrodynamic equation
 and the microscopic expressions of the transport coefficients and relaxation times.

\subsection{
  Boltzmann equation with a form to which the doublet scheme can be applied
}
Since we are interested in the mesoscopic solution
whose space-time scales are coarse-grained from those in the kinetic regime,
we solve the Boltzmann equation (\ref{eq:ChapB-2-1-001}) in the mesoscopic regime
where the space-time variation of $f_{\Vec{v}}(t,\,\Vec{x})$ is small.
To make a coarse graining in a systematic manner,
we convert Eq. (\ref{eq:ChapB-2-1-001}) into
\begin{eqnarray}
  \label{eq:ChapB-4-001}
  \frac{\partial}{\partial t}f_{\Vec{v}}
  = C[f]_{\Vec{v}} - \epsilon \, \Vec{v}\cdot\Vec{\nabla}f_{\Vec{v}},
\end{eqnarray}
where a parameter $\epsilon$ has been introduced to express that
the space derivatives are small for the system that we are interested in.
Here, $\epsilon$ is identified with
the ratio of the average particle distance over the mean free path,
i.e., the Knudsen number.

In the present analysis,
the perturbative expansion of the distribution function 
is made
with respect to $\epsilon$
in the asymptotic regime where the system 
is supposed to show only
a
slow and long-wavelength motion.
Thus,  the zeroth-order solution will be given as a
local equilibrium distribution function
in this asymptotic regime, and then the first-order perturbation caused by the spatial inhomogeneity 
will give rise to
 a deformation of the distribution function,
which dictates the dissipative effects.
Thus, the above 
setting of the small quantity in the perturbative expansion in the asymptotic regime 
just
implements
the
physical assumption
that only the spatial inhomogeneity is the origin of the dissipation.
It
should be emphasized here
that
this asymptotic analysis combined with the perturbative expansion
with respect to $\epsilon$
successfully reduces the Boltzmann equation
to the Navier-Stokes equation
in
the
hydrodynamic regime \cite{kuramoto,env007,env008}.


It is easily recognized that
the Boltzmann equation (\ref{eq:ChapB-4-001})
has the same structure as
that of the generic equation
(\ref{eq:ChapB-3-1-002})
to which the doublet scheme
was developed.
Indeed we can make the following identifications
\begin{eqnarray}
  \label{eq:ChapB-4-1-001}
  \Vec{X} = \Big\{ f_{\Vec{v}} \Big\}_{\Vec{v}},\,\,\,
  \Vec{G}(\Vec{X}) = \Big\{ C[f]_{\Vec{v}} \Big\}_{\Vec{v}},\,\,\,
  \Vec{F}(\Vec{X}) = \Big\{ - \Vec{v}\cdot\Vec{\nabla}f_{\Vec{v}} \Big\}_{\Vec{v}},
\end{eqnarray}
by which
Eq. (\ref{eq:ChapB-4-001})
is converted
into Eq. (\ref{eq:ChapB-3-1-002}).
It is noted that
the velocity $\Vec{v}$ is interpreted as an index of the vector,
while we treat the space coordinate $\Vec{x}$
as a parameter,
in accordance with the
previous
works \cite{env007,env008}.
%
%
From now on, let us omit $\{\cdot\}_{\Vec{v}}$.

\subsection{
  Causal hydrodynamics with doublet scheme in RG method
  and microscopic representations of transport coefficients and relaxation times
}
\label{sec:ChapB-4-2}
%
%

Utilizing the identification
(\ref{eq:ChapB-4-1-001}),
we
can calculate straightforwardly the basic quantities appearing in the doublet scheme, i.e.,
$\Vec{X}^{\mathrm{eq}}$, $A$, $B$, $\Vec{F}_0$, $F_1$,
$f_{\mu}$, $C_{\alpha}$, $C^\prime_{\mu}$, $\Vec{\varphi}^{\alpha}_0$, 
and $\Vec{\varphi}^{\mu}_1$
for
the Boltzmann equation.
Then,
by substituting these into Eqs. (\ref{eq:ChapB-3-6-004}), (\ref{eq:ChapB-3-7-005}), 
and (\ref{eq:ChapB-3-7-006}),
we 
arrive at
the hydrodynamic equation as
 the mesoscopic dynamics of the Boltzmann equation. 

The resultant invariant/attractive manifold
is parameterized by the thirteen variables $T$, $n$, $u^i$, $\pi^{ij}$, and $J^i$ as
\begin{eqnarray}
  \label{eq:invariantmanifold}
  f^{\mathrm{global}}_{\Vec{v}}
  &=&f^{\mathrm{eq}}_{\Vec{v}}\,\Bigg[ 1
  + \epsilon\,\frac{\big[ L^{-1} \,\hat{\pi}^{ij} \big]_{\Vec{v}} \, \pi^{ij}}{2\,T\,\eta^{\mathrm{RG}}}
  + \epsilon\,\frac{\big[ L^{-1} \, \hat{J}^{i} \big]_{\Vec{v}} \, J^{i}}{T^2\,\lambda^{\mathrm{RG}}} \Bigg]
  + O(\epsilon^2),
\end{eqnarray}
and the equations that govern the dynamics of these variables are given by
\begin{eqnarray}
  \label{eq:causalhydrodynamics1}
  \frac{\partial}{\partial t}n
  &=& -\epsilon\,\nabla^i(n\,u^i) + O(\epsilon^3),\\
  \label{eq:causalhydrodynamics2}
  m\,n\,\frac{\partial}{\partial t}u^i
  &=& -\epsilon\,\Big[ m\,n\,u^j\,\nabla^j u^i
  + \nabla^i (n\,T) \Big]
  + \epsilon^2 \, \nabla^j\pi^{ji}
  + O(\epsilon^3),\\
  \label{eq:causalhydrodynamics3}
  n\,\frac{\partial}{\partial t} (3\,T/2)
  &=&
  -\epsilon\,\Big[ n\,u^i \, \nabla^i (3\,T/2)
  + n\,T\,\nabla^i u^i \Big]+ \epsilon^2 \, \Big[ \pi^{ij}\,\nabla^i u^j + \nabla^i J^i \Big]
  + O(\epsilon^3),\\
  \label{eq:causalhydrodynamics4}
  \epsilon\,\pi^{ij}
  &=&
  \epsilon\,2\,\eta^{\mathrm{RG}} \, \sigma^{ij}
  - \epsilon\,\tau^{\mathrm{RG}}_\pi \, \Big(\frac{\partial}{\partial t}
  + \epsilon\,\Vec{u}\cdot\Vec{\nabla}\Big)\pi^{ij}
  - \epsilon^2\,\ell_{\pi J}^{\mathrm{RG}} \, \nabla^{\langle i} J^{j \rangle}\nonumber\\
  &+&
  \epsilon^2\,\Big[
  \kappa^{(1)\mathrm{RG}}_{\pi\pi}\,\pi^{ij}\,\theta
  + \kappa^{(2)\mathrm{RG}}_{\pi\pi} \, \pi^{k\langle i} \, \sigma^{j\rangle k}
  + \kappa^{(3)\mathrm{RG}}_{\pi\pi}\,\pi^{k\langle i} \, \omega^{j\rangle k}\nonumber\\
  &+&
  \kappa^{(1)\mathrm{RG}}_{\pi J} \, J^{\langle i} \, \nabla^{j \rangle} T
  + \kappa^{(2)\mathrm{RG}}_{\pi J} \, J^{\langle i} \, \nabla^{j \rangle} n\nonumber\\
  &+&
  b_{\pi\pi\pi}^{\mathrm{RG}} \, \pi^{k\langle i} \, \pi^{j \rangle k}
  + b_{\pi JJ}^{\mathrm{RG}} \, J^{\langle i} \, J^{j \rangle}\Big] + O(\epsilon^3),\\
  \label{eq:causalhydrodynamics5}
  \epsilon\,J^{i}
  &=&
  \epsilon\,\lambda^{\mathrm{RG}} \, \nabla^i T
  - \epsilon\,\tau^{\mathrm{RG}}_J \, \Big(\frac{\partial}{\partial t}+\epsilon\,\Vec{u}\cdot\Vec{\nabla}\Big)J^{i}
  - \epsilon^2\,\ell_{J\pi}^{\mathrm{RG}} \, \nabla^k \pi^{ki}\nonumber\\
  &+& 
  \epsilon^2\,\Big[
  \kappa^{(1)\mathrm{RG}}_{J\pi}\,\pi^{ik}\,\nabla^k T
  + \kappa^{(2)\mathrm{RG}}_{J\pi}\,\pi^{ik}\,\nabla^k n\nonumber\\
  &+&
  \kappa^{(1)\mathrm{RG}}_{JJ}\,J^{i}\,\theta
  + \kappa^{(2)\mathrm{RG}}_{JJ}\,J^{k}\,\sigma^{ik} + \kappa^{(3)\mathrm{RG}}_{JJ}\,J^{k}\,\omega^{ik}
  + b_{J \pi J}^{\mathrm{RG}} \, \pi^{ik} \, J^{k}\Big] + O(\epsilon^3).
\end{eqnarray}
Here,
$\eta^{\mathrm{RG}}$, $\lambda^{\mathrm{RG}}$, $\tau^{\mathrm{RG}}_\pi$, and $\tau^{\mathrm{RG}}_J$
are the transport coefficients and relaxation times,
and
$\ell_{\pi J}^{\mathrm{RG}}$,
$\ell_{J\pi}^{\mathrm{RG}}$,
$\kappa^{(1,2,3)\mathrm{RG}}_{\pi\pi}$,
$\kappa^{(1,2)\mathrm{RG}}_{\pi J}$,
$\kappa^{(1,2)\mathrm{RG}}_{J\pi}$,
$\kappa^{(1,2,3)\mathrm{RG}}_{JJ}$,
$b_{\pi\pi\pi}^{\mathrm{RG}}$,
$b_{\pi JJ}^{\mathrm{RG}}$,
and
$b_{J \pi J}^{\mathrm{RG}}$
are coefficients of the $O(\epsilon^2)$ terms.
In Eq. (\ref{eq:invariantmanifold}),
we have presented the invariant/attractive manifold valid up to $O(\epsilon)$
for the sake of simplicity.
A full expression valid up to $O(\epsilon^2)$
is given in
\ref{sec:derivation_of_hydro}.

Setting $\epsilon=1$,
we find that
the form of Eqs. (\ref{eq:causalhydrodynamics1})-(\ref{eq:causalhydrodynamics5})
is the same as that of the Grad equation given by Eqs. (\ref{eq:ChapB-2-2-018})-(\ref{eq:ChapB-2-2-022}),
and hence our equation has the hyperbolic character 
with the causality being respected,
as the Grad equation does.
We stress that
Eqs. (\ref{eq:causalhydrodynamics1})-(\ref{eq:causalhydrodynamics5}) are
nothing but the causal hydrodynamic equation
consistent with the Boltzmann equation in the mesoscopic regime,
which has been long sought for.

The microscopic representations of
the transport coefficients and relaxation times in our causal hydrodynamic equations
read
\begin{eqnarray}
  \label{eq:ChapB-4-3-006}
  \eta^{\mathrm{RG}} &\equiv& - \frac{1}{10\,T} \, {\langle\, \hat{\pi}^{ij} \,,\, L^{-1} \,
  \hat{\pi}^{ij} \,\rangle}_{\mathrm{eq}},\,\,\,
  \lambda^{\mathrm{RG}} \equiv -\frac{1}{3\,T^2} \, {\langle\, \hat{J}^i \,,\, L^{-1} \,
  \hat{J}^i \,\rangle}_{\mathrm{eq}},\\
  \label{eq:ChapB-4-3-012}
  \tau_\pi^{\mathrm{RG}} &\equiv& - \frac{{\langle\,\hat{\pi}^{ij}\,,\,
  L^{-2} \, \hat{\pi}^{ij}\,\rangle}_{\mathrm{eq}}}{{\langle\,\hat{\pi}^{kl}\,,\,
  L^{-1} \, \hat{\pi}^{kl}\,\rangle}_{\mathrm{eq}}}\,,\,\,\,
  \tau_J^{\mathrm{RG}} \equiv - \frac{{\langle\,\hat{J}^{i}\,,\,
  L^{-2} \, \hat{J}^{i}\,\rangle}_{\mathrm{eq}}}{{\langle\,\hat{J}^{k}\,,\,
  L^{-1} \, \hat{J}^{k}\,\rangle}_{\mathrm{eq}}},
\end{eqnarray}
while those of the other coefficients
are presented in
\ref{sec:derivation_of_hydro}.
The transport coefficients in Eq. (\ref{eq:ChapB-4-3-006})
perfectly agree with those derived
in the Chapman-Enskog method
\begin{eqnarray}
  \eta^{\mathrm{RG}} = \eta^{\mathrm{CE}},\,\,\,
  \lambda^{\mathrm{RG}} = \lambda^{\mathrm{CE}}.
\end{eqnarray}
Thus, it is manifest that
our microscopic representation of the transport coefficients are
different from those
by
Grad given
in
Eq. (\ref{eq:ChapB-2-2-026}).
We note that
the relaxation times in Eq. (\ref{eq:ChapB-4-3-012}) differ from
those in the previous work
and have novel representations.

To make the physical meaning of
the microscopic representations of $\tau_\pi^{\mathrm{RG}}$ and $\tau_J^{\mathrm{RG}}$ clearer,
we convert the definitions in Eq. (\ref{eq:ChapB-4-3-012})
into the forms as the Green-Kubo formula in the linear response theory.
To this end,
we utilize the following identity:
\begin{eqnarray}
  \label{eq:ChapB-4-3-014}
  \big[ L^{-n} \, (\hat{\pi}^{ij},\,\hat{J}^i) \big]_{\Vec{v}}
  =
  \frac{(-1)^n}{(n-1)!}\,\int_0^\infty\!\!\mathrm{d}s\,s^{n-1}
  \, (\hat{\pi}^{ij}_{\Vec{v}}(s),\,\hat{J}^i_{\Vec{v}}(s)),
\end{eqnarray}
where we have defined
$(\hat{\pi}^{ij}_{\Vec{v}}(s),\,\hat{J}^i_{\Vec{v}}(s)) \equiv
\big[\mathrm{e}^{s\,L} \, (\hat{\pi}^{ij},\,\hat{J}^i)\big]_{\Vec{v}}$.
It is noted that
$\hat{\pi}^{ij}_{\Vec{v}}(s)$ ($\hat{J}^i_{\Vec{v}}(s)$) could be interpreted as a ``time-evolved'' vector
of $\hat{\pi}^{ij}_{\Vec{v}}$ ($\hat{J}^i_{\Vec{v}}$) by $L_{\Vec{v}\Vec{k}}$.
Using Eq. (\ref{eq:ChapB-4-3-014}) with $n=1$ or $2$,
we can obtain the compact forms for the transport coefficients and relaxation times as
\begin{eqnarray}
  \label{eq:ChapB-4-3-017}
  \eta^{\mathrm{RG}} &=& \int_0^\infty\!\!\mathrm{d}s \,R_\pi(s),\,\,\,
  \lambda^{\mathrm{RG}} = \int_0^\infty\!\!\mathrm{d}s \,R_J(s),\\
  \label{eq:ChapB-4-3-019}
  \tau_\pi^{\mathrm{RG}} &=& \frac{\int_0^\infty\!\!\mathrm{d}s \,s \, R_\pi(s)}
	{\int_0^\infty\!\!\mathrm{d}s \,R_\pi(s)},\,\,\,
  \tau_J^{\mathrm{RG}} = \frac{\int_0^\infty\!\!\mathrm{d}s \,s \, R_J(s)}
	{\int_0^\infty\!\!\mathrm{d}s \,R_J(s)},
\end{eqnarray}
where $R_\pi(s)$ and $R_J(s)$ are defined by
\begin{eqnarray}
  \label{eq:ChapB-4-3-021}
  R_\pi(s) \equiv
  \frac{1}{10\,T}\,
  {\langle\,\hat{\pi}^{ij}(0)\,,\,\hat{\pi}^{ij}(s)\,\rangle}_{\mathrm{eq}},\,\,\,
  R_J(s) \equiv
  \frac{1}{3\,T^2}\,
  {\langle\,\hat{J}^{i}(0)\,,\,\hat{J}^{i}(s)\,\rangle}_{\mathrm{eq}}.
\end{eqnarray}
It is noted that
$R_\pi(s)$ and $R_J(s)$ denote the relaxation functions introduced in the linear response theory.
We remark that
the formulae of $\tau^{\mathrm{RG}}_\pi$ and $\tau^{\mathrm{RG}}_J$
allow the natural interpretation of them  as the correlation times of $R_\pi(s)$ and $R_J(s)$, respectively.

Here,
we discuss the reason why
the novel and natural microscopic expressions of the relaxation times $\tau^{\mathrm{RG}}_\pi$ and $\tau^{\mathrm{RG}}_J$
have been obtained, together with 
those of the transport coefficients in agreement with the Chapman-Enskog formulae.
First of all, it should be noted that our method is based on a faithful solution
of the Boltzmann equation in the perturbation theory as the Chapman-Enskog theory is,
with the secular terms being resummed by the RG method or multiple-scale method,
although the latter method fails in deriving the
causal hydrodynamic
equation.  
As for the relaxation times appearing in the
causal hydrodynamic
equation, 
our microscopic expressions 
come from the forms of the excited modes,
i.e.,
$[ L^{-1} \, \hat{\pi}^{ij} ]_{\Vec{v}}$ and $[ L^{-1} \, \hat{J}^{i} ]_{\Vec{v}}$,
which are clearly different from $\hat{\pi}^{ij}_{\Vec{v}}$ and $\hat{J}^{i}_{\Vec{v}}$
adopted just as ansatz in the conventional approaches. 
The present forms of the excited modes are \textit{derived}
by solving the Boltzmann equation in a faithful manner based on the perturbation theory 
with the secular terms resummed by the RG method.
Indeed,
the analytical forms of our excited modes are 
constructed so as to 
solve the Boltzmann equation and
represent the relaxation process
to the local equilibrium distribution function:
The doublet scheme in the RG method developed in
Sec.
\ref{sec:ChapB-3}
provides us with the powerful scheme for
describing the relaxation dynamics in the mesoscopic scale,
and hence deriving the natural microscopic representations of 
the relaxation times $\tau^{\mathrm{RG}}_\pi$ and $\tau^{\mathrm{RG}}_J$.
Thus, we are confident that we have arrived at the correct formulae of the relaxation times 
for the first time.

\subsection{
  Functional forms of distribution function and moments
  to reproduce causal hydrodynamics by RG method
}
\label{sec:ChapB-4-3}
We can read off the form of the derivation $\Phi_{\Vec{v}}$
and the thirteen quantities $\vec{\phi}_{13\Vec{v}}$
that reproduce 
the causal hydrodynamic equations (\ref{eq:causalhydrodynamics1})-(\ref{eq:causalhydrodynamics5})
as
\begin{eqnarray}
  \label{eq:ChapB-4-2-004}
  \Phi_{\Vec{v}} &=&
  - \frac{\big[ L^{-1} \,\hat{\pi}^{ij} \big]_{\Vec{v}} \, \pi^{ij}}{\frac{1}{5}\,{\langle\, \hat{\pi}^{kl}\,,\, L^{-1} \, \hat{\pi}^{kl}\,\rangle}_{\mathrm{eq}}}
  - \frac{\big[ L^{-1} \, \hat{J}^{i} \big]_{\Vec{v}} \, J^{i}}{\frac{1}{3}\,{\langle\, \hat{J}^{k}\,,\, L^{-1} \, \hat{J}^{k}\,\rangle}_{\mathrm{eq}}}
  \equiv \Phi^{\mathrm{RG}}_{\Vec{v}},\\
  \vec{\phi}_{13\Vec{v}} &=& \Big\{ 1,\,m\,\Vec{v},\,\frac{m}{2}\,|\Vec{v}|^2,\,\big[ L^{-1} \, \hat{\pi}^{ij}\big]_{\Vec{v}},\,\big[ L^{-1} \, \hat{J}^{i}\big]_{\Vec{v}} \Big\}
  \equiv \vec{\phi}^{\mathrm{RG}}_{13\Vec{v}}.
\end{eqnarray}
It is obvious that
$\Phi^{\mathrm{RG}}_{\Vec{v}}$ and 
$\vec{\phi}^{\mathrm{RG}}_{13\Vec{v}}$
are different from $\Phi^{\mathrm{G}}_{\Vec{v}}$
and $\vec{\phi}^{\mathrm{G}}_{13\Vec{v}}$, respectively.
We stress that
the set of $\Phi^{\mathrm{RG}}_{\Vec{v}}$ and $\vec{\phi}^{\mathrm{RG}}_{13\Vec{v}}$ provides
the correct functional forms of the distribution function and the moments to be used in the method of moments,
which thereby should lead to the causal hydrodynamic equation
compatible with the Boltzmann equation in the mesoscopic regime.

As a summary,
we compare in Table \ref{tab:002}
the basic variables and the functional forms of the deformation of the distribution function
to describe the mesoscopic dynamics
in the moment method formulated by Grad
and that by the present doublet scheme,
together with the respective microscopic representations of the transport coefficients and relaxation times.

\begin{table}
\begin{center}
\caption{
  \label{tab:002}
In the first raw: 
A comparison of the functional forms of
deformation of
the distribution function $f_{\Vec{v}}=f^{\mathrm{eq}}_{\Vec{v}} (1 + \Phi_{\Vec{v}})$
  from the local one and the moments $\vec{\phi}_{13\Vec{v}}$ as the basic variables to describe
the mesoscopic dynamics in the present  and Grad's works.
In the other raws:
  The resultant microscopic representations of
  the shear viscosity $\eta$,
  the thermal conductivity $\lambda$,
  and the relaxation times $\tau_\pi$ and $\tau_J$
  obtained
in the two works.
}
\begin{tabular}{ccc}
\hline
\hline
&
This work
&
Grad's work
\\
\hline
$\Phi_{\Vec{v}}$
&
$\displaystyle
- \frac{\big[ L^{-1} \,\hat{\pi}^{ij} \big]_{\Vec{v}} \, \pi^{ij}}{\frac{1}{5}\,{\langle\, \hat{\pi}^{kl}\,,\, L^{-1} \, \, \hat{\pi}^{kl}\,\rangle}_{\mathrm{eq}}}
- \frac{\big[ L^{-1} \, \hat{J}^{i} \big]_{\Vec{v}} \, J^{i}}{\frac{1}{3}\,{\langle\, \hat{J}^{k}\,,\, L^{-1} \, \hat{J}^{k}\,\rangle}_{\mathrm{eq}}}$
&
$\displaystyle
- \frac{\hat{\pi}^{ij}_{\Vec{v}} \, \pi^{ij}}{\frac{1}{5}\,{\langle\, \hat{\pi}^{kl}\,,\, \hat{\pi}^{kl}\,\rangle}_{\mathrm{eq}}}
- \frac{\hat{J}^{i}_{\Vec{v}} \, J^{i}}{\frac{1}{3}\,{\langle\, \hat{J}^{k}\,,\, \hat{J}^{k}\,\rangle}_{\mathrm{eq}}}$
\\
$\vec{\phi}_{13\Vec{v}}$
&
$\Big\{ 1,\,m\,\Vec{v},\,\frac{m}{2}\,|\Vec{v}|^2,\,\big[ L^{-1} \, \hat{\pi}^{ij}\big]_{\Vec{v}},\,\big[ L^{-1} \, \hat{J}^{i}\big]_{\Vec{v}} \Big\}$
&
$\Big\{ 1,\,m\,\Vec{v},\,\frac{m}{2}\,|\Vec{v}|^2,\,\hat{\pi}^{ij}_{\Vec{v}},\,\hat{J}^{i}_{\Vec{v}} \Big\}$
\\
\hline
$\eta$
&
$\displaystyle
-\frac{1}{10\,T} \, {\langle\, \hat{\pi}^{ij} \,,\, L^{-1} \,
\hat{\pi}^{ij} \,\rangle}_{\mathrm{eq}}$
&
$\displaystyle
- \frac{1}{10\,T}\,\frac{\big[{\langle\, \hat{\pi}^{ij}\,,\,\hat{\pi}^{ij} \,\rangle}_{\mathrm{eq}}\big]^2
}{
{\langle\, \hat{\pi}^{kl}\,,\,L \, \hat{\pi}^{kl} \,\rangle}_{\mathrm{eq}}}$
\\
$\lambda$
&
$\displaystyle
-\frac{1}{3\,T^2} \, {\langle\, \hat{J}^{i} \,,\, L^{-1} \,
\hat{J}^{i} \,\rangle}_{\mathrm{eq}}$
&
$\displaystyle
- \frac{1}{3\,T^2}\,\frac{\big[{\langle\, \hat{J}^{i}\,,\,\hat{J}^{i} \,\rangle}_{\mathrm{eq}}\big]^2
}{
{\langle\, \hat{J}^{k}\,,\,L\,\hat{J}^k \,\rangle}_{\mathrm{eq}}}$
\\
$\tau_\pi$
&
$\displaystyle
- \frac{{\langle\,\hat{\pi}^{ij}\,,\,
L^{-2} \, \hat{\pi}^{ij}\,\rangle}_{\mathrm{eq}}}{{\langle\,\hat{\pi}^{kl}\,,\,
L^{-1} \, \hat{\pi}^{kl}\,\rangle}_{\mathrm{eq}}}$
&
$\displaystyle
-\frac{{\langle\, \hat{\pi}^{ij}\,,\,\hat{\pi}^{ij} \,\rangle}_{\mathrm{eq}}}
{{\langle\, \hat{\pi}^{kl}\,,\,L\,\hat{\pi}^{kl} \,\rangle}_{\mathrm{eq}}}$
\\
$\tau_J$
&
$\displaystyle
- \frac{{\langle\,\hat{J}^{i}\,,\,
L^{-2} \, \hat{J}^{i}\,\rangle}_{\mathrm{eq}}}{{\langle\,\hat{J}^{k}\,,\,
L^{-1} \, \hat{J}^{k}\,\rangle}_{\mathrm{eq}}}$
&
$\displaystyle
-\frac{{\langle\, \hat{J}^{i}\,,\,\hat{J}^{i} \,\rangle}_{\mathrm{eq}}}
{{\langle\, \hat{J}^{k}\,,\,L\,\hat{J}^{k} \,\rangle}_{\mathrm{eq}}}$
\\
\hline
\hline
\end{tabular}
\end{center}
\end{table}

Finally,
we point out that 
since the linearized collision operator $L_{\Vec{v}\Vec{k}}$
is specified by the microscopic transition probablity $\omega(\Vec{v},\,\Vec{v_1}|\Vec{v_2},\,\Vec{v_3})$,
our $\Phi^{\mathrm{RG}}_{\Vec{v}}$ and $\vec{\phi}^{\mathrm{RG}}_{13\Vec{v}}$ 
may happen to coincide with
$\Phi^{\mathrm{G}}_{\Vec{v}}$ and $\vec{\phi}^{\mathrm{G}}_{13\Vec{v}}$
when the transition probability has some peculiar properties.
We find that 
such a 
coincidence
is realized
only when both $\hat{\pi}^{ij}_{\Vec{v}}$ and $\hat{J}^{i}_{\Vec{v}}$
are eigenvectors of $L_{\Vec{v}\Vec{k}}$.
It is well known that
the linearized collision operator 
for the Maxwell molecules has such a property \cite{intro013-A,intro013-B}.
Thus,
we conjecture that
the method of moments 
with the use of our $\Phi^{\mathrm{RG}}_{\Vec{v}}$ and $\vec{\phi}^{\mathrm{RG}}_{13\Vec{v}}$
can provide the causal hydrodynamic equation for generic systems
with the particle interaction not restricted to that of the Maxwell molecules type,
wheras that of Grad
may be at most compatible with
the Boltzmann equation only for the Maxwell molecules.
It is left as a future work
to show that the conjecture is true,
which may imply that
the present theory makes
the correct and general method
for constructing mesoscopic dynamics for a given microscopic dynamics.


\section{
  Summary and concluding remarks
}
\label{sec:ChapB-5}
In this paper,
we have derived the mesoscopic dynamics from the Boltzmann equation
on the basis of the renormalization group (RG) method
in a systematic manner with no ad-hoc assumption:
The mesoscopic dynamics as a reduced dynamics consists of two ingredients, i.e.,
(a)~the invariant/attractive manifold of which the reduced number of the variables 
constitute a natural coordinate system and (b)~a set of differential equations
that governs the time evolution of the variables.
A basic observation presented in the extraction of the mesoscopic dynamics
from the Boltzmann equation 
is to include some excited (fast) modes
of the linearized collision operator
as additional components for the invariant manifold
spanned by the zero modes, where the hydrodynamics is defined.
We have newly developed a general theory for extracting the mesoscopic dynamics on the basis
of the RG method, which is based on a simple but
basic principle in the reduction theory of the dynamics:
The resultant dynamics should be as simple as possible 
because we are interested to reduce the dynamics to a simpler one.
The newly developed theory is called the doublet scheme.
We have shown that
the number and form of the excited modes that should be included in the invariant/attractive manifold
is uniquely determined by the doublet scheme.
We have used the Lorenz model
to demonstrate
the validity of the doublet scheme
for constructing the invariant/attractive manifold 
and the reduced differential equation for the mesoscopic dynamics:
The validness of the scheme is verified also numerically.

We have also demonstrated that
the mesoscopic dynamics of the Boltzmann equation obtained by the doublet scheme in the RG method
respects the hyperbolic character, i.e., the causality,
where the number of the dynamical variables is thirteen.
We have shown that
the form of the resultant equation
is the same as that of the Grad equation \cite{intro001},
but the microscopic formulae of the coefficients,
e.g., the transport coefficients and relaxation times, are different.
It has turned out that
our theory leads to the same expressions for the transport coefficients
as given by the Chapman-Enskog method \cite{chapman-enskog}.
We have found that
our microscopic representations of the relaxation times
differ from those of the previous work,
and can be converted into formulae
written in terms of relaxation functions,
which allow a natural physical interpretation of the relaxation times.
We have shown that
the distribution function
and the moments
which are explicitly
constructed in our theory
provides a new ansatz for the functional forms of the distribution function and the moments 
in the method of moments proposed by Grad.
Furthermore,
we have conjectured that
the functional forms of the distribution function and 
the moments in the previous
work of Grad
are valid only for a specific interacting systems such as the Maxwell molecules,
while our functional forms can be applied to
generic interacting systems 
not restricted to the Maxwell molecules.

It is interesting and important to present numerical simulations
to elucidate that
a solution of the causal hydrodynamic equation obtained in this work
is actually consistent
with that of the Boltzmann equation in the mesoscopic regime,
even if the systems of interest are generic interacting systems instead of the Maxwell molecules.

It is also interesting to apply the doublet scheme in the RG method
to extract the mesoscopic dynamics of the relativistic Boltzmann equation.
This is because
the fourteen-moment approximation for the distribution function
of the relativistic Boltzmann equation proposed by Israel and Stewart \cite{intro006,intro007},
i.e., the most famous relativistically covariant extension of Grad's thirteen-moment approximation,
has encountered the same difficulty as in the non-relativistic case:
Numerical simulations by several groups \cite{intro008,intro009}
show that
the dynamics described by Israel-Stewart' s fourteen-moment equation is not consistent with
that of the relativistic Boltzmann equation in the mesoscopic regime.
Indeed,
we can show \cite{TK2012PTP,TK2012EPJA,RBE_next} that
the equation obtained by the RG method,
which ensures the consistency with the mesoscopic dynamics of the relativistic Boltzmann equation,
respects the causality,
and the number of the dynamical variables are fourteen.
We can also show that
the form of our fourteen-moment causal equation
is the same as the Israel-Stewart' s fourteen-moment equation,
but the formulae of the coefficients, i.e., the transport coefficients and relaxation times, include in the equation
are different
and
the microscopic representations of the coefficients are
given as natural forms in terms of the relaxation functions
as in the non-relativistic case shown in this work.

Finally, we note that
the doublet scheme in the RG method itself
has a universal nature
and can be applied to derive a mesoscopic dynamics
from kinetic equations other than the simple Boltzmann equation,
e.g., Kadanoff-Baym equation \cite{kadanoff}.

\section*{Acknowledgments}
Y.K. is supported by the Grants-in-Aid for JSPS fellows (No.15J01626). 
T.K. was partially supported by a
Grant-in-Aid for Scientific Research from the Ministry of Education,
Culture, Sports, Science and Technology (MEXT) of Japan
(Nos. 20540265 and 23340067),
by the Yukawa International Program for Quark-Hadron Sciences.

\appendix

\section{
  Solutions to linear differential equation
  with time dependent inhomogeneous term
}
\label{sec:ChapB-6}
In this Appendix,
we present
solutions of the linear differential equations
with a time-dependent inhomogeneous term.

Let us consider the solution of the equation given by
\begin{eqnarray}
  \label{eq:ChapB-6-001}
  \frac{\partial}{\partial t}\Vec{Y}(t) &=& A \, \Vec{Y}(t) + \Vec{R}(t-t_0).
\end{eqnarray}
The solution reads
\begin{eqnarray}
  \label{eq:ChapB-6-002}
  \Vec{Y}(t)
  = \mathrm{e}^{A(t-t_0)}\,\Vec{Y}(t_0) +
  \int_{t_0}^t\!\!\mathrm{d}t^\prime \, P_0\,\Vec{R}(t^\prime-t_0)
  +
  \int_{t_0}^t\!\!\mathrm{d}t^\prime \,
  \mathrm{e}^{A(t-t^\prime)}\,Q_0\,\Vec{R}(t^\prime-t_0),
\end{eqnarray}
where we have inserted $1 = P_0 + Q_0$
in front of $\Vec{R}(t^\prime-t_0)$.
Substituting the following Taylor expansion,
$\Vec{R}(t^\prime - t_0) =
\mathrm{e}^{(t^\prime - t_0)\partial/\partial s} \, \Vec{R}(s)\Big|_{s=0}$,
into Eq. (\ref{eq:ChapB-6-002})
and carrying out integration with respect to $t^\prime$,
we have
\begin{eqnarray}
  \label{eq:ChapB-6-004}
  \Vec{Y}(t) &=& \mathrm{e}^{A(t-t_0)}\,
  \Vec{Y}(t_0)
  +
  (1 - \mathrm{e}^{(t-t_0)\partial/\partial s})
  \,
  (- \partial/\partial s)^{-1}
  \, P_0 \, \Vec{R}(s)\Big|_{s=0}\nonumber\\
  &&{}+
  (\mathrm{e}^{A(t-t_0)} -
  \mathrm{e}^{(t-t_0)\partial/\partial s})\,
  (A - \partial/\partial s)^{-1} \,
  Q_0 \, \Vec{R}(s)\Big|_{s=0}\nonumber\\
  &=& \mathrm{e}^{A(t-t_0)}\,
  \Big[ 
  \Vec{Y}(t_0)
  +
  Q_1 \, (A - \partial/\partial s)^{-1} \,
  Q_0 \, \Vec{R}(s)\Big|_{s=0}
  \Big]\nonumber\\
  &&{}+
  (1 - \mathrm{e}^{(t-t_0)\partial/\partial s})
  \,
  (- \partial/\partial s)^{-1}
  \, P_0 \, \Vec{R}(s)\Big|_{s=0}\nonumber\\
  &&{}+
  (\mathrm{e}^{A(t-t_0)} -
  \mathrm{e}^{(t-t_0)\partial/\partial s})\,
  P_1 \, (A - \partial/\partial s)^{-1} \,
  Q_0 \, \Vec{R}(s)\Big|_{s=0}\nonumber\\
  &&{}-
  \mathrm{e}^{(t-t_0)\partial/\partial s}\,
  Q_1 \, (A - \partial/\partial s)^{-1} \,
  Q_0 \, \Vec{R}(s)\Big|_{s=0},
\end{eqnarray}
where $1 = P_0 + P_1 + Q_1$ has been inserted in front of $(A - \partial/\partial s)^{-1} \, Q_0 \, \Vec{R}(s)$
in the second line of Eq. (\ref{eq:ChapB-6-004}).
We note that
the contributions from the inhomogeneous term $\Vec{R}(t-t_0)$
are decomposed into two parts,
whose time dependencies are given by
$\mathrm{e}^{A(t-t_0)}$
and
$\mathrm{e}^{(t-t_0)\partial/\partial s}$,
respectively.
The former gives a fast motion characterized by the eigenvalues of $A$
acting on Q${}_1$ space,
while the time dependence of the latter is
independent of the dynamics due to the absence of $A$.
Since we are interested in the motion coming from the P${}_0$ and P${}_1$ spaces,
we eliminate the former associated with the Q${}_1$ space
with 
a choice
of the initial value $\Vec{Y}(t_0)$ that has not yet
been
specified
as follows:
\begin{eqnarray}
  \label{eq:ChapB-6-005}
  \Vec{Y}(t_0) =
  - Q_1 \, (A - \partial/\partial s)^{-1} \,
  Q_0 \, \Vec{R}(s)\Big|_{s=0},
\end{eqnarray}
which reduces Eq. (\ref{eq:ChapB-6-004}) to
\begin{eqnarray}
  \label{eq:ChapB-6-006}
  \Vec{Y}(t) &=&
  (1 - \mathrm{e}^{(t-t_0)\partial/\partial s})
  \,
  (- \partial/\partial s)^{-1}
  \, P_0 \, \Vec{R}(s)\Big|_{s=0}\nonumber\\
  &&{}+
  (\mathrm{e}^{A(t-t_0)} -
  \mathrm{e}^{(t-t_0)\partial/\partial s})\,
  P_1 \, (A - \partial/\partial s)^{-1} \,
  Q_0 \, \Vec{R}(s)\Big|_{s=0}\nonumber\\
  &&{}-
  \mathrm{e}^{(t-t_0)\partial/\partial s}\,
  Q_1 \, (A - \partial/\partial s)^{-1} \,
  Q_0 \, \Vec{R}(s)\Big|_{s=0}.
\end{eqnarray}
Equations (\ref{eq:ChapB-6-005}) and (\ref{eq:ChapB-6-006})
are nothing but the formulae we wanted.

\section{
 Naturalness of mesoscopic dynamics:
 Consistency with slow dynamics as described with only zero modes in asymptotic regime
}
\label{sec:appA}
In this Appendix,
first we derive the slow dynamics
described  only by
the zero modes
from the generic evolution equation (\ref{eq:ChapB-3-1-002}) with the RG method for completeness,
although a detailed derivation can be seen in Ref. \cite{env006}. 
Then,
we demonstrate that
the mesoscopic dynamics given by Eqs. (\ref{eq:ChapB-3-7-005}) and (\ref{eq:ChapB-3-7-006})
approaches asymptotically to
the slow dynamics.

\subsection{
  Slow dynamics as described with would-be zero modes
}
\label{sec:zeromode}
As mentioned in
Sec.
\ref{sec:ChapB-3},
we first try to obtain the perturbative solution $\Vec{\tilde{X}}$ to Eq. (\ref{eq:ChapB-3-1-002})
around an arbitrary initial time $t=t_0$ with the initial value $\Vec{X}(t_0)$;
$\Vec{\tilde{X}}(t=t\,;\,t_0) = \Vec{X}(t_0)$.
We expand the initial value as well as the solution with respect to $\epsilon$
as shown in Eqs. (\ref{eq:ChapB-3-2-002}) and (\ref{eq:ChapB-3-2-003}),
and obtain the series of the perturbative equations with respect to $\epsilon$.

The zeroth-order equation is the same as Eq. (\ref{eq:ChapB-3-2-005}).
Since we are interested in the slow motion
realized asymptotically for $t\rightarrow\infty$,
we adopt the
static
solution $\Vec{X}^{\mathrm{eq}}$ as the zeroth-order solution:
$\Vec{\tilde{X}}_0(t\,;\,t_0) = \Vec{X}^{\mathrm{eq}}$,
which means that
the zeroth-order initial value reads
$\Vec{X}_0(t_0) = \Vec{\tilde{X}}_0(t_0\,;\,t_0) = \Vec{X}^{\mathrm{eq}}$.

The first-order equation is
$\frac{\partial}{\partial t}\Vec{\tilde{X}}_1(t\,;\,t_0) =
A \, \Vec{\tilde{X}}_1(t\,;\,t_0) + \Vec{F}_0$,
where $A$ and $\Vec{F}_0$ have been defined
in Eqs. (\ref{eq:ChapB-3-1-007}) and (\ref{eq:inhomo}), respectively.
A solution to
the first-order equation
reads
\begin{eqnarray}
  \label{eq:zeromodeRG-004}
  \Vec{\tilde{X}}_1(t\,;\,t_0)
  =
  \mathrm{e}^{A(t-t_0)}\,\Big[ \Vec{\tilde{X}}_1(t_0\,;\,t_0) + A^{-1}\,Q_0\,\Vec{F}_0 \Big]
  + (t-t_0)\,P_0\,\Vec{F}_0
  - A^{-1}\,Q_0\,\Vec{F}_0.
\end{eqnarray}
Here, $P_0$ denotes the projection operator onto the P${}_0$ space spanned
by the zero modes $\Vec{\varphi}^{\alpha}_0$, i.e., the kernel space of $A$,
and $Q_0$ the projection operator onto the Q${}_0$ space as the complement to the P${}_0$ space.

Since we are interested in the slow motion caused by the P${}_0$ space,
we eliminate the fast motion coming from the Q${}_0$ space
with the use of the initial value $\Vec{\tilde{X}}_1(t_0\,;\,t_0)$
that has not yet
been
specified as follows:
$\Vec{X}_1(t_0) = \Vec{\tilde{X}}_1(t_0\,;\,t_0) = -  A^{-1}\,Q_0\,\Vec{F}_0$,
which reduces Eq. (\ref{eq:zeromodeRG-004}) to
$\Vec{\tilde{X}}_1(t\,;\,t_0) = (t-t_0)\,P_0\,\Vec{F}_0 - A^{-1}\,Q_0\,\Vec{F}_0$.

The second-order equation is
$\frac{\partial}{\partial t}\Vec{\tilde{X}}_2(t\,;\,t_0) =
A \, \Vec{\tilde{X}}_2(t\,;\,t_0) + \Vec{U}(t-t_0)$,
with
\begin{eqnarray}
  \label{eq:zeromodeRG-008}
  \Vec{U}(s) \equiv \frac{1}{2}\,B\,\big[ s\,P_0\,\Vec{F}_0 - A^{-1}\,Q_0\,\Vec{F}_0\,,\,s\,P_0\,\Vec{F}_0 - A^{-1}\,Q_0\,\Vec{F}_0 \big]
  + F_1 \, (s\,P_0\,\Vec{F}_0 - A^{-1}\,Q_0\,\Vec{F}_0),\nonumber\\
\end{eqnarray}
where
$B$ and $F_1$ have been defined
in Eq. (\ref{eq:ChapB-3-5-002}).
With the use of the method developed in
\ref{sec:ChapB-6},
we have a solution to
the second-order equation
as
\begin{eqnarray}
  \label{eq:zeromodeRG-009}
  \Vec{\tilde{X}}_2(t\,;\,t_0)
  &=& \mathrm{e}^{A(t-t_0)}\,
  \Big[ 
  \Vec{\tilde{X}}_2(t_0\,;\,t_0)
  +
  (A - \partial/\partial s)^{-1} \,
  Q_0 \, \Vec{U}(s)\Big|_{s=0}
  \Big]\nonumber\\
  &&{}+
  (1 - \mathrm{e}^{(t-t_0)\partial/\partial s})
  \,
  (- \partial/\partial s)^{-1}
  \, P_0 \, \Vec{U}(s)\Big|_{s=0}\nonumber\\
  &&{}-
  \mathrm{e}^{(t-t_0)\partial/\partial s}\,
  (A - \partial/\partial s)^{-1} \,
  Q_0 \, \Vec{U}(s)\Big|_{s=0}.
\end{eqnarray}
As in the case of the first order,
we eliminate the fast motion caused by the Q${}_0$ space
using
the initial value $\Vec{\tilde{X}}_2(t_0\,;\,t_0)$ as
follows:
$\Vec{X}_2(t_0) = \Vec{\tilde{X}}_2(t_0\,;\,t_0) =
-(A - \partial/\partial s)^{-1} \, Q_0 \, \Vec{U}(s)\Big|_{s=0}$,
which leads to
$\Vec{\tilde{X}}_2(t\,;\,t_0) = (1 - \mathrm{e}^{(t-t_0)\partial/\partial s})
\,
(- \partial/\partial s)^{-1}
\, P_0 \,\Vec{U}(s)\Big|_{s=0}
- \mathrm{e}^{(t-t_0)\partial/\partial s}\,
(A - \partial/\partial s)^{-1} \,
\break 
Q_0 \, \Vec{U}(s)\Big|_{s=0}$.

Summing up the solutions and initial values
constructed in the perturbative analysis up to $O(\epsilon^2)$,
we have
\begin{eqnarray}
  \label{eq:zeromodeRG-012}
  \Vec{X}(t_0) &=& \Vec{X}^{\mathrm{eq}} -  \epsilon \, A^{-1}\,Q_0\,\Vec{F}_0
  - \epsilon^2 \, (A - \partial/\partial s)^{-1} \, Q_0 \, \Vec{U}(s)\Big|_{s=0}
  + O(\epsilon^3),\\
  \label{eq:zeromodeRG-013}
  \Vec{\tilde{X}}(t\,;\,t_0) &=& \Vec{X}^{\mathrm{eq}}
  + \epsilon \, \Bigg[ (t-t_0)\,P_0\,\Vec{F}_0 - A^{-1}\,Q_0\,\Vec{F}_0 \Bigg]\nonumber\\
  &&{}+ \epsilon^2 \, \Bigg[ (1 - \mathrm{e}^{(t-t_0)\partial/\partial s})
  \,
  (- \partial/\partial s)^{-1}
  \, P_0 \, \Vec{U}(s)\Big|_{s=0}\nonumber\\
  &&{}-
  \mathrm{e}^{(t-t_0)\partial/\partial s}\,
  (A - \partial/\partial s)^{-1} \,
  Q_0 \, \Vec{U}(s)\Big|_{s=0} \Bigg]
  + O(\epsilon^3).
\end{eqnarray}
We note the appearance of the secular term proportional to $t-t_0$,
which invalidates the perturbative solution when $|t-t_0|$ becomes large.
For obtaining the
globally improved solution
from this local perturbative solution,
we apply the RG equation
$\partial\Vec{\tilde{X}}_1(t\,;\,t_0)/\partial t_0|_{t_0=t}=0$
to Eq. (\ref{eq:zeromodeRG-013}):
The RG equation reads
\begin{eqnarray}
  \label{eq:zeromodeRG-014}
  &&\frac{\partial}{\partial t}(\Vec{X}^{\mathrm{eq}} - \epsilon \, A^{-1}\,Q_0\,\Vec{F}_0)
  - \epsilon \, P_0\,\Vec{F}_0\nonumber\\
  &&{}+ \epsilon^2 \, \Bigg[ - P_0 \, \Vec{U}(0)
  - (- \partial/\partial s)\,(A - \partial/\partial s)^{-1} \,
  Q_0 \, \Vec{U}(s)\Big|_{s=0}
  \Bigg]
  + O(\epsilon^3) = 0,
\end{eqnarray}
which is the equation governing the slow motion of $C_{\alpha}$ in $\Vec{X}^{\mathrm{eq}}$.

By taking the inner product with the zero modes $\Vec{\varphi}^{\alpha}_0$,
we can convert Eq. (\ref{eq:zeromodeRG-014}) into
\begin{eqnarray}
  \label{eq:ChapB-add2}
  &&\langle\,\Vec{\varphi}^{\alpha}_0\,,\,\frac{\partial}{\partial t}(\Vec{X}^{\mathrm{eq}}-\epsilon\,A^{-1}\,Q_0\,\Vec{F}_0)
  \,\rangle
  - \epsilon \,
  \langle\,\Vec{\varphi}^{\alpha}_0\,,\,
  \Vec{F}_0 - \epsilon\,F_1\,A^{-1}\,Q_0\,\Vec{F}_0\,\rangle\nonumber\\
  &=& \epsilon^2\,\frac{1}{2}\, \langle\,\Vec{\varphi}^{\alpha}_0\,,\,
  B\, \big[ A^{-1} \, Q_0 \, \Vec{F}_0 \,,\, A^{-1} \, Q_0 \, \Vec{F}_0 \big] \,\rangle + O(\epsilon^3).
\end{eqnarray}
Here, we have used
$\Vec{U}(0) = \frac{1}{2}\,B\,\big[ A^{-1}\,Q_0\,\Vec{F}_0 \,,\, A^{-1}\,Q_0\,\Vec{F}_0 \big]
- F_1 \, A^{-1}\,Q_0\,\Vec{F}_0$,
which can be derived from Eq. (\ref{eq:zeromodeRG-008}).

\subsection{
Proof for consistency of mesoscopic dynamics with slow dynamics
}
\label{sec:ChapB-3-8-0}
We should notice the \textit{time-scale separation}
between the fast motion of $C^\prime_{\mu}$ caused by the P${}_1$ space
and the slow motion of $C_{\alpha}$ by the P${}_0$ space.
Thanks to this separation which becomes significant in the asymptotic regime,
we can solve Eq. (\ref{eq:ChapB-3-7-006}) with respect to $C^\prime_{\mu}$
with $C_{\alpha}$ being a constant,
and obtain the closed equations with respect to $C_{\alpha}$
by eliminating $C^\prime_{\mu}$
from Eq. (\ref{eq:ChapB-3-7-005}).

Here, let us construct
the solution $C^\prime_{\mu}$ valid up to $O(1)$,
which is sufficient for the derivation of the closed equation valid up to $O(\epsilon^2)$,
because
$C^\prime_{\mu}$ enters Eq. (\ref{eq:ChapB-3-7-005}) as the $O(\epsilon^2)$ terms.
Such a $C^\prime_{\mu}$ is governed by
\begin{eqnarray}
  \sum_{\nu=1}^{M_1}\,
  \langle\,A^{-1}\,\Vec{\varphi}^{\mu}_1\,,\,A^{-1}\,\Vec{\varphi}_1^{\nu}\,\rangle\,\frac{\partial}{\partial t}C^\prime_{\nu}
  =
  \sum_{\nu=1}^{M_1}\,
  \langle\,\Vec{\varphi}^{\mu}_1\,,\,A^{-1}\,\Vec{\varphi}_1^{\nu}\,\rangle\,
  (C^\prime_{\nu} + f_{\nu}) + O(\epsilon),
\end{eqnarray}
where
$C_{\alpha}$ is treated as a constant.
We note that
$\langle\,A^{-1}\,\Vec{\varphi}^{\mu}_1\,,\,A^{-1}\,\Vec{\varphi}_1^{\nu}\,\rangle$
is a positive definite matrix,
while
$\langle\,\Vec{\varphi}^{\mu}_1\,,\,A^{-1}\,\Vec{\varphi}_1^{\nu}\,\rangle$
is a negative definite matrix
because the eigenvalues of $A$ except for the zero are supposed to be real negative
as mentioned in Sec. \ref{sec:ChapB-3-1}. 
Thus,
we find that
$C^\prime_{\mu}$ approaches asymptotically to $-f_{\mu}$:
\begin{eqnarray}
  \label{eq:ChapB-add1}
  C^\prime_{\mu} = - f_{\mu} + O(\epsilon),
\end{eqnarray}
which is equivalent to 
\begin{eqnarray}
  \label{eq:ChapB-add3}
  \Vec{\phi} = - A^{-1} \, Q_0 \, \Vec{F}_0 + O(\epsilon).
\end{eqnarray}
Substituting Eq. (\ref{eq:ChapB-add3}) into Eq. (\ref{eq:ChapB-3-7-006}),
we have the closed equation for $C_{\alpha}$,
which is the same as Eq. (\ref{eq:ChapB-add2}).
We stress that
the mesoscopic dynamics derived by the doublet scheme
has a natural property that 
it is consistent with the slow dynamics
as described with only the zero modes
in the asymptotic regime.


\section{
  Detailed derivation of explicit form of mesoscopic dynamics of Boltzmann equation
}
\label{sec:ChapB-9}
In this
Appendix,
we present
a detailed derivation
of the mesoscopic dynamics of the Boltzmann equation (\ref{eq:ChapB-4-001})
based on the doublet scheme in the RG method developed in
Sec.
\ref{sec:ChapB-3}.

\subsection{
   Set up suited for doublet scheme
}
\label{sec:ChapB-4-1}
We build $\Vec{X}^{\mathrm{eq}}$, $A$, $B$, $\Vec{F}_0$, $F_1$,
$f_{\mu}$, $C_{\alpha}$, $C^\prime_{\mu}$, $\Vec{\varphi}^{\alpha}_0$, and $\Vec{\varphi}^{\mu}_1$
in the case of the Boltzmann equation (\ref{eq:ChapB-4-001}) which are required for the doublet scheme.

With the use of Eq. (\ref{eq:ChapB-3-1-004}),
we find that
the static solution reads
\begin{eqnarray}
  \label{eq:ChapB-4-1-004}
  \Vec{X}^{\mathrm{eq}}(t_0) =
  f^{\mathrm{eq}}_{\Vec{v}}(\Vec{x}\,;\,t_0)
  = n(\Vec{x}\,;\,t_0) \,
  \Bigg[\frac{m}{2\,\pi\,T(\Vec{x}\,;\,t_0)}\Bigg]^{\frac{3}{2}}
  \,
  \exp\Bigg[- \frac{m\,|\Vec{v} -
  \Vec{u}(\Vec{x}\,;\,t_0)|^2}{2\,T(\Vec{x}\,;\,t_0)}\Bigg],
\end{eqnarray}
which is nothing but the Maxwellian (\ref{eq:ChapB-2-1-024}) and satisfies
$C[f^{\mathrm{eq}}]_{\Vec{v}} = 0$ as discussed in Eq. (\ref{eq:ChapB-2-1-025}).
We note that
the five would-be integral constants
$n(\Vec{x}\,;\,t_0)$, $T(\Vec{x}\,;\,t_0)$, and $\Vec{u}(\Vec{x}\,;\,t_0)$
corresponding to $C_{\alpha}(t_0)$ in
Sec.
\ref{sec:ChapB-3}
are lifted to the dynamical variables by applying the RG equation.
In the following,
we suppress $(\Vec{x}\,;\,t_0)$ when no misunderstanding is expected.

Using Eq. (\ref{eq:ChapB-3-1-007}),
we have the linearized evolution operator $A$ as
\begin{eqnarray}
  \label{eq:ChapB-4-1-005}
  A &=&
  \frac{\delta}{\delta f_{\Vec{k}}}
  C[f]_{\Vec{v}}\Bigg|_{f=f^{\mathrm{eq}}}
  = f^{\mathrm{eq}}_{\Vec{v}} \, L_{\Vec{v}\Vec{k}} \, (f^{\mathrm{eq}}_{\Vec{k}})^{-1}\nonumber\\
  &=& \frac{1}{2!} \,
  \int_{\Vec{v_1}}\int_{\Vec{v_2}}\int_{\Vec{v_3}}
  \,
  \omega(\Vec{v},\,\Vec{v_1}|\Vec{v_2},\,\Vec{v_3}) \, (
  \delta_{\Vec{v_2}\Vec{k}} \, f^{\mathrm{eq}}_{\Vec{v_3}}
  + f^{\mathrm{eq}}_{\Vec{v_2}} \,\delta_{\Vec{v_3}\Vec{k}}
  - \delta_{\Vec{v}\Vec{k}} \, f^{\mathrm{eq}}_{\Vec{v_1}}
  - f^{\mathrm{eq}}_{\Vec{v}} \,\delta_{\Vec{v_1}\Vec{k}} ).\nonumber\\
\end{eqnarray}
Here, let us examine the properties of $A$.
We define the inner product by
\begin{eqnarray}
  \label{eq:ChapB-4-1-006}
  \langle\,\psi\,,\,\chi\,\rangle \equiv
  \int_{\Vec{v}}
  \,(f^{\mathrm{eq}}_{\Vec{v}})^{-1} \, \psi_{\Vec{v}} \, \chi_{\Vec{v}}
  = {\langle\,(f^{\mathrm{eq}})^{-1}\,\psi\,,\,(f^{\mathrm{eq}})^{-1}\,\chi\,\rangle}_{\mathrm{eq}},
\end{eqnarray}
with $\psi_{\Vec{v}}$ and $\chi_{\Vec{v}}$ being arbitrary vectors
and
the diagonal matrix $f^{\mathrm{eq}}_{\Vec{v}\Vec{k}}
\equiv \delta_{\Vec{v}\Vec{k}} \, f^{\mathrm{eq}}_{\Vec{v}}$.
We note that the norm through this inner product is positive definite,
\begin{eqnarray}
  \label{eq:ChapB-4-1-007}
  \langle\,\psi\,,\,\psi\,\rangle
  =
  \int_{\Vec{v}}
  \,(f^{\mathrm{eq}}_{\Vec{v}})^{-1} \, (\psi_{\Vec{v}})^2
  > 0,\,\,\,\,\,\,\,\,\,\psi\neq 0.
\end{eqnarray}
It is noteworthy that
$A$ is self-adjoint with respect this inner product as
\begin{eqnarray}
  \label{eq:ChapB-4-1-008}
  \langle\,\psi\,,\,A\,\chi\,\rangle
  &=&
  - \frac{1}{2!} \, \frac{1}{4} \,
  \int_{\Vec{v}}\int_{\Vec{v_1}}\int_{\Vec{v_2}}\int_{\Vec{v_3}}
  \,
  \omega(\Vec{v},\,\Vec{v_1}|\Vec{v_2},\,\Vec{v_3})\,
  f^{\mathrm{eq}}_{\Vec{v}}\,f^{\mathrm{eq}}_{\Vec{v_1}}\nonumber\\
  &&{}\times\,\Big((f^{\mathrm{eq}}_{\Vec{v}})^{-1}\,\psi_{\Vec{v}}
  + (f^{\mathrm{eq}}_{\Vec{v_1}})^{-1}\,\psi_{\Vec{v_1}}
  - (f^{\mathrm{eq}}_{\Vec{v_2}})^{-1}\,\psi_{\Vec{v_2}}
  - (f^{\mathrm{eq}}_{\Vec{v_3}})^{-1}\,\psi_{\Vec{v_3}}\Big)\nonumber\\
  &&{}\times\,\Big((f^{\mathrm{eq}}_{\Vec{v}})^{-1}\,\chi_{\Vec{v}}
  + (f^{\mathrm{eq}}_{\Vec{v_1}})^{-1}\,\chi_{\Vec{v_1}}
  - (f^{\mathrm{eq}}_{\Vec{v_2}})^{-1}\,\chi_{\Vec{v_2}}
  - (f^{\mathrm{eq}}_{\Vec{v_3}})^{-1}\,\chi_{\Vec{v_3}}\Big)\nonumber\\
  &=& \langle\,A\,\psi\,,\,\chi\,\rangle,
\end{eqnarray}
and real semi-negative definite;
\begin{eqnarray}
  \label{eq:ChapB-4-1-009}
  \langle\,\psi\,,\,A\,\psi\,\rangle
  &=&
  - \frac{1}{2!} \, \frac{1}{4} \,
  \int_{\Vec{v}}\int_{\Vec{v_1}}\int_{\Vec{v_2}}\int_{\Vec{v_3}}
  \,
  \omega(\Vec{v},\,\Vec{v_1}|\Vec{v_2},\,\Vec{v_3})\,
  f^{\mathrm{eq}}_{\Vec{v}}\,f^{\mathrm{eq}}_{\Vec{v_1}}\nonumber\\
  &&{}\times\,\Big((f^{\mathrm{eq}}_{\Vec{v}})^{-1}\,\psi_{\Vec{v}}
  + (f^{\mathrm{eq}}_{\Vec{v_1}})^{-1}\,\psi_{\Vec{v_1}}
  - (f^{\mathrm{eq}}_{\Vec{v_2}})^{-1}\,\psi_{\Vec{v_2}}
  - (f^{\mathrm{eq}}_{\Vec{v_3}})^{-1}\,\psi_{\Vec{v_3}}\Big)^2\nonumber\\
  &\le& 0.
\end{eqnarray}
Thanks to these properties of $A$,
we can apply the
doublet scheme
presented in
Sec.
\ref{sec:ChapB-3}
to extract the mesoscopic dynamics from Eq. (\ref{eq:ChapB-4-001}).

By differentiating $f^{\mathrm{eq}}_{\Vec{v}}$ with respect to $n$, $u^i$, and $T$,
we have the zero modes of $A$,
\begin{eqnarray}
  \label{eq:ChapB-4-1-010}
  \Vec{\varphi}^{\alpha}_{0} = f^{\mathrm{eq}}_{\Vec{v}} \,
  \varphi^\alpha_{0\Vec{v}}
  = \big[ f^{\mathrm{eq}} \, \varphi^\alpha_{0} \big]_{\Vec{v}},
  \,\,\,\,\,\,\,\,\,\alpha=0,\,1,\,2,\,3,\,4,
\end{eqnarray}
where
\begin{eqnarray}
  \label{eq:ChapB-4-1-011}
  \varphi^{\alpha}_{0\Vec{v}} \equiv
  \left\{
  \begin{array}{ll}
  \displaystyle{
    1,
  }
  &
  \displaystyle{
    \alpha=0,
  }
  \\[2mm]
  \displaystyle{
    m\,\delta v^{i},
  }
  &
  \displaystyle{
    \alpha=i,
  }
  \\[2mm]
  \displaystyle{
    \frac{m}{2}\,|\Vec{\delta v}|^2 - \frac{3}{2}\,T,
  }
  &
  \displaystyle{
    \alpha=4,
  }
  \end{array}
  \right.
\end{eqnarray}
with the peculiar velocity $\Vec{\delta v} = \Vec{v} - \Vec{u}$.
It is noted that
$\varphi^\alpha_{0\Vec{v}}$ with $\alpha = 0,\,\cdots,\,4$
coincide with the collision invariants
shown in Eq. (\ref{eq:ChapB-2-1-007}),
and the dimension of the kernel space of $A$ is five, i.e., $M_0 = 5$.

With the use of $\varphi^\alpha_{0\Vec{v}}$ in Eq. (\ref{eq:ChapB-4-1-010}) 
and the inner product in Eq. (\ref{eq:ChapB-4-1-006}),
we have the P${}_0$-space metric matrix as follows:
\begin{eqnarray}
  \label{eq:ChapB-4-1-013}
  \eta^{\alpha\beta}_0 =
  \langle\,f^{\mathrm{eq}}\,\varphi^\alpha_0\,,\,
  f^{\mathrm{eq}}\,\varphi^\beta_0\,\rangle
  =
  \int_{\Vec{v}}
  \,f^{\mathrm{eq}}_{\Vec{v}} \, \varphi^\alpha_{0\Vec{v}} \, \varphi^\beta_{0\Vec{v}} =
  c^\alpha \, \delta^{\alpha\beta},
\end{eqnarray}
with
\begin{eqnarray}
  c^0 \equiv n,\,\,\,
  c^{i=1,2,3} \equiv m\,n\,T,\,\,\,
  c^4 \equiv \frac{3}{2}\,n\,T^2.
\end{eqnarray}
We note that
$\eta^{\alpha\beta}_0$ is a diagonal matrix,
and hence $\eta_{0\alpha\beta}^{-1} = \delta^{\alpha\beta}/c^\alpha$.
Thus, we have
the projection operators $P_0$ and $Q_0$ given as
\begin{eqnarray}
  \label{eq:ChapB-4-1-014}
  \big[ P_0 \, \psi \big]_{\Vec{v}} &=& \sum_{\alpha=0}^4 \, f^{\mathrm{eq}}_{\Vec{v}} \,
  \varphi^\alpha_{0\Vec{v}} \,
  \frac{1}{c^\alpha}\,
  \langle\,f^{\mathrm{eq}}\,\varphi^\alpha_0\,,\,\psi\,\rangle,
\end{eqnarray}
and
$Q_0 = 1 - P_0$.

The perturbative term defined in Eq. (\ref{eq:inhomo}) now takes the form
\begin{eqnarray}
  \label{eq:ChapB-4-1-015}
  \Vec{F}_0 = - v^i \, f^{\mathrm{eq}}_{\Vec{v}} \, \Bigg[ \frac{1}{n}\,\nabla^i n
  +\Big(\frac{m}{2\,T}\,|\Vec{\delta v}|^2 - \frac{3}{2}\Big) \,\frac{1}{T}\,\nabla^i T
  + \frac{m}{T}\,\delta v^j \, \nabla^i u^j \Bigg].
\end{eqnarray}
Through the straightforward calculation shown in \ref{sec:ChapB-8},
we have
\begin{eqnarray}
  \label{eq:A-1Q0F0}
  A^{-1} \, Q_0 \, \Vec{F}_0
  =
  - \big[ A^{-1}\,f^{\mathrm{eq}}\,\hat{\pi}^{ij} \big]_{\Vec{v}} \,\frac{\sigma^{ij}}{T}
  - \big[ A^{-1}\,f^{\mathrm{eq}}\,\hat{J}^i \big]_{\Vec{v}}\,\frac{\nabla^i T}{T^2},
\end{eqnarray}
with $\sigma^{ij} = \Delta^{ijkl}\,\nabla^k u^l$.
We remark that
$\hat{J}^i_{\Vec{v}}$ and $\hat{\pi}^{ij}_{\Vec{v}}$ are identical to
the vector fields introduced in the method of moments
in Eqs. (\ref{eq:ChapB-2-2-003}) and (\ref{eq:ChapB-2-2-004}), respectively.
It is obvious that
$-\sigma^{ij}/T$ and $-(\nabla^i T) /T^2$
are linear independent functions of the hydrodynamic variables 
$n$, $T$, and $u^i$.
Thus,
we can read off $\Vec{\varphi}^{\mu}_1$ and $f_{\mu}$ as
\begin{eqnarray}
  \label{eq:ChapB-4-1-016}
  \Vec{\varphi}^{\mu}_1 &=&
  (
  f^{\mathrm{eq}}_{\Vec{v}}\,\hat{\pi}^{ij}_{\Vec{v}},\,
  f^{\mathrm{eq}}_{\Vec{v}}\,\hat{J}^{i}_{\Vec{v}}
  ),\\
  f_{\mu} &=&
  (-\sigma^{ij}/T,\,
  -(\nabla^i T)/T^2),
\end{eqnarray}
respectively.
It is noted that
the number of independent vector components of
$A^{-1}\,Q_0\,\Vec{F}_0$ is eight, i.e., $M_1 = 8$,
because of $\sigma^{ij}=\sigma^{ji}$ and $\delta^{ij}\,\sigma^{ij}=0$.

Correspondingly,
we introduce eight integral constants
\begin{eqnarray}
  \pi^{ij}(\Vec{x}\,;\,t_0),\,\,\,J^i(\Vec{x}\,;\,t_0),
\end{eqnarray}
with the constraints
\begin{eqnarray}
  \label{eq:ChapB-4-1-034}
  \pi^{ij} = \pi^{ji},\,\,\,
  \delta^{ij} \, \pi^{ij} = 0.
\end{eqnarray}
We note that
$\pi^{ij}$ and $J^i$
should be interpreted as the viscous pressure and heat flux,
respectively.
Using $\pi^{ij}$ and $J^i$,
we can set $C^\prime_{\mu}$ equal to
\begin{eqnarray}
  \label{eq:ChapB-4-1-027}
  C^\prime_{\mu} = (-5\,\pi^{ij}/\langle\,f^{\mathrm{eq}}\,\hat{\pi}^{ab}\,,\,A^{-1} \, f^{\mathrm{eq}}\,\hat{\pi}^{ab}\,\rangle
  ,\,-3\,J^i/\langle\,f^{\mathrm{eq}}\,\hat{J}^{a}\,,\,A^{-1} \, f^{\mathrm{eq}}\,\hat{J}^{a}\,\rangle),
\end{eqnarray}
which makes the deviation $\Vec{\phi}$ to be
\begin{eqnarray}
    \label{eq:ChapB-4-1-029}
    \Vec{\phi}
    =
    -
   \Bigg[
      \frac{
        \big[A^{-1}\,f^{\mathrm{eq}}\,\hat{\pi}^{ij}\big]_{\Vec{v}}
      }{
        \frac{1}{5}\,\langle\,f^{\mathrm{eq}}\,\hat{\pi}^{ab}\,,\,A^{-1} \, f^{\mathrm{eq}}\,\hat{\pi}^{ab}\,\rangle
      }
    \Bigg]
    \,\pi^{ij}
    -
    \Bigg[
      \frac{
        \big[A^{-1}\,f^{\mathrm{eq}}\,\hat{J}^{i}\big]_{\Vec{v}}
      }{
        \frac{1}{3}\,\langle\,f^{\mathrm{eq}}\,\hat{J}^{a}\,,\,A^{-1} \, f^{\mathrm{eq}}\,\hat{J}^{a}\,\rangle
      }
    \Bigg]
    \, J^{i}.
\end{eqnarray}
We note that
the coefficients are needed for the normalizations
$\langle\,f^{\mathrm{eq}}\,\hat{\pi}^{ij}\,,\,\Vec{\phi}\,\rangle = - \pi^{ij}$
and 
$\langle\,f^{\mathrm{eq}}\,\hat{J}^{i}\,,\,\Vec{\phi}\,\rangle = - J^{i}$
being satisfied.


A remark is in order here.
A total number of the would-be integral constants 
$T$, $n$, $u^i$, $\pi^{ij}$ and $J^{i}$ are thirteen.
Although this number is the same as that of the dynamical variables
introduced in the thirteen-moment approximation proposed by Grad,
we mention that
this number and form of the dynamical variables have been
automatically determined from the Boltzmann equation
by the doublet scheme in the RG method developed in Sec. \ref{sec:ChapB-3},
which does not demand any ansatz at all
in contrast to the traditional approaches.
This agreement strongly suggests the reliability
of the doublet scheme in the RG method.

We find that
the P${}_1$ space is spanned by the doublet modes
$(f^{\mathrm{eq}}_{\Vec{v}}\,\hat{\pi}^{ij}_{\Vec{v}},\,\big[ A^{-1}\,f^{\mathrm{eq}}\,\hat{\pi}^{ij} \big]_{\Vec{v}})$
and
$(f^{\mathrm{eq}}_{\Vec{v}}\,\hat{J}^{i}_{\Vec{v}},\,\big[ A^{-1}\,f^{\mathrm{eq}}\,\hat{J}^{i} \big]_{\Vec{v}})$.
Using the P${}_1$-space vectors,
we have the projection operators $P_1$ given as
\begin{eqnarray}
  \label{eq:ChapB-4-1-026}
  \big[ P_1 \, \psi \big]_{\Vec{v}}
  &=&
  \sum_{m,n=0,1}\,
  \big[A^{-m}\,f^{\mathrm{eq}}\,\hat{\pi}^{ij}\big]_{\Vec{v}}\,
  \eta^{-1}_{\pi mn}\,
  \langle\, A^{-n}\,f^{\mathrm{eq}}\,\hat{\pi}^{ij}\,,\,\psi \,\rangle\nonumber\\
  &&{}+
  \sum_{m,n=0,1}\,
  \big[A^{-m}\,f^{\mathrm{eq}}\,\hat{J}^{i}\big]_{\Vec{v}}\,
  \eta^{-1}_{J mn}\,
  \langle\, A^{-n}\,f^{\mathrm{eq}}\,\hat{J}^{i}\,,\,\psi \,\rangle,
\end{eqnarray}
and $Q_1 = Q_0 - P_1$.
Here,
$\eta^{-1}_{\pi mn}$ and $\eta^{-1}_{J mn}$ denote inverse matrices
of $\eta_\pi^{mn}$ and $\eta_J^{mn}$
which are the P${}_1$-space metric matrices given by
\begin{eqnarray}
  \eta_\pi^{mn}
  &\equiv&
  \frac{1}{5}\,\langle\, A^{-m}\,f^{\mathrm{eq}}\,\hat{\pi}^{ij}\,,\,A^{-n}\,f^{\mathrm{eq}}\,\hat{\pi}^{ij} \,\rangle,\\
  \eta_J^{mn}
  &\equiv&
  \frac{1}{3}\,\langle\, A^{-m}\,f^{\mathrm{eq}}\,\hat{J}^{i}\,,\,A^{-n}\,f^{\mathrm{eq}}\,\hat{J}^{i} \,\rangle,
\end{eqnarray}
respectively.

The definitions presented in Eq. (\ref{eq:ChapB-3-5-002}) lead to
\begin{eqnarray}
  \label{eq:ChapB-4-1-039}
  B &=&
  \frac{\delta^2}{\delta f_{\Vec{k}}\delta f_{\Vec{l}}}
  C[f]_{\Vec{v}}\Bigg|_{f=f^{\mathrm{eq}}}
  = B_{\Vec{v}\Vec{k}\Vec{l}},\\
  \label{eq:ChapB-4-1-040}
  F_1 &=& - \Vec{v}\cdot\Vec{\nabla} \delta_{\Vec{v}\Vec{k}}.
\end{eqnarray}
An explicit form of $B_{\Vec{v}\Vec{k}\Vec{l}}$ is given in Eq. (\ref{eq:Btensor}).

\subsection{
  Mesoscopic dynamics of Boltzmann equation with doublet scheme
}
\label{sec:derivation_of_hydro}
Substituting
$\Vec{X}^{\mathrm{eq}}$, 
$A$, $B$, $\Vec{F}_0$, $F_1$,
$f_{\mu}$, $C_{\alpha}$, $C^\prime_{\mu}$, $\Vec{\varphi}^{\alpha}_0$, and $\Vec{\varphi}^{\mu}_1$ constructed in
\ref{sec:ChapB-4-1}
into Eqs. (\ref{eq:ChapB-3-6-004}), (\ref{eq:ChapB-3-7-005}), and (\ref{eq:ChapB-3-7-006}),
we obtain the mesoscopic dynamics of the Boltzmann equation.
In this section,
we use new vectors
\begin{eqnarray}
  \tilde{\pi}^{ij}_{\Vec{v}}
  &\equiv&
  -\frac{\big[A^{-1}f^{\mathrm{eq}}\hat{\pi}^{ij}\big]_{\Vec{v}}
  }{
  \frac{1}{5}\langle\, f^{\mathrm{eq}}\hat{\pi}^{kl}\,,\,A^{-1}f^{\mathrm{eq}}\hat{\pi}^{kl} \,\rangle
  },\\
  \tilde{J}^{i}_{\Vec{v}}
  &\equiv&
  -\frac{\big[A^{-1}f^{\mathrm{eq}}\hat{J}^{i}\big]_{\Vec{v}}
  }{
  \frac{1}{3}\langle\, f^{\mathrm{eq}}\hat{J}^{k}\,,\,A^{-1}f^{\mathrm{eq}}\hat{J}^{k} \,\rangle
  },
\end{eqnarray}
which reduce $\Vec{\phi}$ to the following simple form:
\begin{eqnarray}
  \Vec{\phi}
  = \tilde{\pi}^{ij}_{\Vec{v}} \pi^{ij} + \tilde{J}^{i}_{\Vec{v}} J^{i}.
\end{eqnarray}

First, we start with Eqs. (\ref{eq:ChapB-3-7-005}) and (\ref{eq:ChapB-3-7-006}),
which can be reduced to
\begin{eqnarray}
  \label{eq:ChapB-9-001}
  \langle \, f^\mathrm{eq} \, \varphi^\alpha_0\,,\,\frac{\partial}{\partial t}f^\mathrm{eq}\, \rangle
  -\epsilon \,
  \langle \, f^\mathrm{eq} \, \varphi^\alpha_0\,,\,
  \Vec{F}_0 \, \rangle
  =
  -\epsilon \,
  \langle\,f^\mathrm{eq}\,\varphi^\alpha_0\,,\,K^\mu \, \partial_\mu
  (
  \tilde{\pi}^{kl}\,\pi^{kl}
  +
  \tilde{J}^k\,J^k
  )\,\rangle + O(\epsilon^3),
\end{eqnarray}
and
\begin{eqnarray}
  \label{eq:ChapB-9-002}
  &&\epsilon \,
  \langle\,A^{-1}\,f^\mathrm{eq}\,(\hat{\pi}^{ij},\,\hat{J}^i)\,,\,
  K^\mu \, \partial_\mu
  (
  \tilde{\pi}^{kl}\,\pi^{kl}
  +
  \tilde{J}^k\,J^k
  )\,\rangle\nonumber\\
  &=&
  \epsilon \, \langle \, A^{-1}\,f^\mathrm{eq} \, (\hat{\pi}^{ij},\,\hat{J}^i)\,,\,
  \Vec{F}_0
  \, \rangle
  +
  \epsilon \, \langle \, A^{-1}\,f^\mathrm{eq} \, (\hat{\pi}^{ij},\,\hat{J}^i)\,,\,
  A\,(
  \tilde{\pi}^{kl}\,\pi^{kl}
  +
  \tilde{J}^k\,J^k
  )
  \, \rangle\nonumber\\
  &&{}+
  \epsilon^2 \, \frac{1}{2}\,
  \langle \, A^{-1}\,f^\mathrm{eq} \, 
  (\hat{\pi}^{ij},\,\hat{J}^i)
  \,,\,
  B\,\big[
  \tilde{\pi}^{kl}\,\pi^{kl}
  +
  \tilde{J}^k\,J^k
  \,,\,
  \tilde{\pi}^{mn}\,\pi^{mn}
  +
  \tilde{J}^m\,J^m
  \big] \, \rangle
  + O(\epsilon^3),\nonumber\\
\end{eqnarray}
respectively,
where we have used the relation
\begin{eqnarray}
  \label{eq:ChapB-9-004}
  \Bigg[\frac{\partial}{\partial t} - \epsilon\,F_1\Bigg]_{\Vec{v}\Vec{k}}
  =
  \Bigg[\frac{\partial}{\partial t} + \epsilon\,\Vec{v}\cdot\Vec{\nabla}\Bigg]\,\delta_{\Vec{v}\Vec{k}}
  =
  K_{\Vec{v}\Vec{k}}^\mu \, \partial_\mu,
\end{eqnarray}
with the definitions
\begin{eqnarray}
  \label{eq:ChapB-9-005}
  (\partial_0,\,\partial_1,\,\partial_2,\,\partial_3)
  &\equiv& (\partial/\partial t,\,\epsilon\,\nabla^1,\,\epsilon\,\nabla^2,\,\epsilon\,\nabla^3),\\
  \label{eq:ChapB-9-006}
  (K_{\Vec{v}\Vec{k}}^0,\,K_{\Vec{v}\Vec{k}}^1,\,K_{\Vec{v}\Vec{k}}^2,\,K_{\Vec{v}\Vec{k}}^3)
  &\equiv& (1,\,v^1,\,v^2,\,v^3)\,\delta_{\Vec{v}\Vec{k}}.
\end{eqnarray}
We note the presence of $\epsilon$ in front of the spatial derivatives $\nabla^1$, $\nabla^2$, and $\nabla^3$.
In the derivation of
Eq. (\ref{eq:ChapB-9-001}),
we have used the identity
\begin{eqnarray}
  &&\frac{1}{2}\, \langle \, f^\mathrm{eq} \, \varphi^{\alpha}_0\,,\,
  B\,\big[
  \tilde{\pi}^{kl}\,\pi^{kl}
  +
  \tilde{J}^k\,J^k
  \,,\,
  \tilde{\pi}^{mn}\,\pi^{mn}
  +
  \tilde{J}^m\,J^m
  \big] \, \rangle\nonumber\\
  &=& \int_{\Vec{v}} \, \varphi^{\alpha}_{0\Vec{v}} \, C[\tilde{\pi}^{kl}\,\pi^{kl}
  +
  \tilde{J}^k\,J^k]_{\Vec{v}}
  = 0,
\end{eqnarray}
where we have used the fact that
$\varphi^{\alpha}_{0\Vec{v}}$ are collision invariants shown in Eq. (\ref{eq:ChapB-2-1-007}).

Now we show the explicit form of each terms
in Eqs. (\ref{eq:ChapB-9-001}) and (\ref{eq:ChapB-9-002}) one by one:
The first and second terms in the left-hand side of Eq. (\ref{eq:ChapB-9-001}) read
\begin{eqnarray}
  \label{eq:ChapB-9-007}
  \langle \, f^\mathrm{eq} \, \varphi^\alpha_0\,,\,\frac{\partial}{\partial t}f^\mathrm{eq} \, \rangle
  &=& \left\{
  \begin{array}{ll}
    \displaystyle{
      \frac{\partial}{\partial t}n,
    }
    &
    \displaystyle{
      \alpha=0,
    }
    \\[2mm]
    \displaystyle{
      m \, n \, \frac{\partial}{\partial t}u^i,
    }
    &
    \displaystyle{
      \alpha=i,
    }
    \\[2mm]
    \displaystyle{
      n \, \frac{\partial}{\partial t}(3\,T/2),
    }
    &
    \displaystyle{
      \alpha=4,
    }
  \end{array}
  \right.\\
  \label{eq:ChapB-9-008}
  \epsilon \,
  \langle \, f^\mathrm{eq} \, \varphi^\alpha_0\,,\,\Vec{F}_0 \, \rangle
  &=& \left\{
  \begin{array}{ll}
    \displaystyle{
      -\epsilon \, \Vec{\nabla}\cdot(n\,\Vec{u}),
    }
    &
    \displaystyle{
      \alpha=0,
    }
    \\[2mm]
    \displaystyle{
      -\epsilon \, m \, n \, \Vec{u}\cdot\Vec{\nabla}u^i - \epsilon \, \nabla^i (n\,T),
    }
    &
    \displaystyle{
      \alpha=i,
    }
    \\[2mm]
    \displaystyle{
      - \epsilon \, n \, \Vec{u}\cdot\Vec{\nabla}(3\,T/2)
      - \epsilon \, n\,T \, \Vec{\nabla}\cdot\Vec{u},
    }
    &
    \displaystyle{
      \alpha=4,
    }
  \end{array}
  \right.
\end{eqnarray}
respectively.
The first and  second terms in the right-hand side of Eq. (\ref{eq:ChapB-9-002}) read
\begin{eqnarray}
  \label{eq:ChapB-9-011}
  \epsilon \, \langle \, 
  A^{-1}\,f^\mathrm{eq} \, (\hat{\pi}^{ij},\,\hat{J}^i)\,,\,
  \Vec{F}_0
  \, \rangle
  &=&
  \epsilon \, \langle \, 
  f^\mathrm{eq} \, (\hat{\pi}^{ij},\,\hat{J}^i)\,,\,
  A^{-1}\,Q_0\,\Vec{F}_0
  \, \rangle\nonumber\\
  &=&
  (
  \epsilon \, 2\,\eta \, \sigma^{ij},\,
  \epsilon \, \lambda \, \nabla^i T
  ),\\
  \epsilon \, \langle \, 
  A^{-1}\,f^\mathrm{eq} \, (\hat{\pi}^{ij},\,\hat{J}^i)\,,\,
  A\,(
  \tilde{\pi}^{kl}\,\pi^{kl}
  +
  \tilde{J}^k\,J^k
  )
  \, \rangle
  &=&
  (
  - \epsilon \, \pi^{ij},\,
  - \epsilon \, J^i
  ),
\end{eqnarray}
where we have defined
\begin{eqnarray}
  \label{eq:ChapB-9-012}
  \eta \equiv - \frac{1}{10\,T} \, \langle\, f^\mathrm{eq} \, \hat{\pi}^{ij} \,,\, A^{-1} \,
  f^\mathrm{eq} \, \hat{\pi}^{ij} \,\rangle,\,\,\,
  \lambda \equiv -\frac{1}{3\,T^2} \, \langle\, f^\mathrm{eq} \, \hat{J}^i \,,\, A^{-1} \,
  f^\mathrm{eq} \, \hat{J}^i \,\rangle.
\end{eqnarray}
We note that
the transport coefficients
$\eta$ and $\lambda$ given by Eq. (\ref{eq:ChapB-9-012})
accord with
$\eta^{\mathrm{RG}}$ and $\lambda^{\mathrm{RG}}$
in Eq. (\ref{eq:ChapB-4-3-006})
on account of
the inner product defined in Eq. (\ref{eq:ChapB-2-2-030})
and $A^{-1} = f^{\mathrm{eq}} \, L^{-1} \, (f^{\mathrm{eq}})^{-1}$.

The term in the right-hand side of Eq. (\ref{eq:ChapB-9-001})
is more complicated than the other terms.
First, we expand this term as
\begin{eqnarray}
  \label{eq:ChapB-9-016}
  \epsilon \,
  \langle\,f^\mathrm{eq}\,\varphi^\alpha_0\,,\,K^\mu \, \partial_\mu
  (\tilde{\pi}^{kl}\,\pi^{kl}
  +
  \tilde{J}^k\,J^k)
  \,\rangle
  &=&
  \epsilon \, \Bigg[
  \partial_\mu \langle\,Q_0\,f^\mathrm{eq}\,K^\mu\,\varphi^\alpha_0\,,\,
  \tilde{\pi}^{kl}\,\pi^{kl}
  +
  \tilde{J}^k\,J^k
  \,\rangle\nonumber\\
  &&{}-
  \langle\,Q_0\,f^\mathrm{eq}\,K^\mu\,\partial_\mu \varphi^\alpha_0\,,\,
  \tilde{\pi}^{kl}\,\pi^{kl}
  +
  \tilde{J}^k\,J^k
  \,\rangle
  \Bigg],
\end{eqnarray}
where we have inserted $Q_0$
in the final stage of the expansion
because
$\tilde{\pi}^{ij}_{\Vec{v}}$ and $\tilde{J}^{i}_{\Vec{v}}$ belong to the Q${}_0$ space
and this insertion does not change the results.

Then,
with the direct manipulation based on the definitions
(\ref{eq:ChapB-4-1-011}),
(\ref{eq:ChapB-4-1-026}),
(\ref{eq:ChapB-9-005}),
and (\ref{eq:ChapB-9-006}),
we can show the following identities:
\begin{eqnarray}
  \label{eq:ChapB-9-017}
%
  \big[ Q_0\,f^\mathrm{eq}\,K^0\,\varphi^\alpha_0 \big]_{\Vec{v}}
  &=&
  0,\\
  \label{eq:Q0fKiphi0}
  \big[ Q_0\,f^\mathrm{eq}\,K^i\,\varphi^\alpha_0 \big]_{\Vec{v}}
  &=& \left\{
  \begin{array}{ll}
    \displaystyle{
      0,
    }
    &
    \displaystyle{
      \alpha=0,
    }
    \\[2mm]
    \displaystyle{
      f^{\mathrm{eq}}_{\Vec{v}}\,\hat{\pi}^{ij}_{\Vec{v}},
    }
    &
    \displaystyle{
      \alpha=j,
    }
    \\[2mm]
    \displaystyle{
      f^{\mathrm{eq}}_{\Vec{v}}\,\hat{J}^{i}_{\Vec{v}},
    }
    &
    \displaystyle{
      \alpha=4.
    }
  \end{array}
  \right.
  \\
  \label{eq:ChapB-9-018}
  \big[ Q_0 \, f^\mathrm{eq} \, K^\mu \,
  \partial_\mu \varphi^\alpha_0\big]_{\Vec{v}}
  &=& \left\{
  \begin{array}{ll}
    \displaystyle{
      0,
    }
    &
    \displaystyle{
      \alpha=0,
    }
    \\[2mm]
    \displaystyle{
      0,
    }
    &
    \displaystyle{
      \alpha=i,
    }
    \\[2mm]
    \displaystyle{
      - \epsilon\,f^\mathrm{eq}_{\Vec{v}}\,\hat{\pi}^{jk}_{\Vec{v}} \,\sigma^{jk},
    }
    &
    \displaystyle{
      \alpha=4.
    }
  \end{array}
  \right.
\end{eqnarray}
We note that
a detailed derivation of Eq. (\ref{eq:Q0fKiphi0})
is presented in \ref{sec:ChapB-8}.
Substituting these into Eq. (\ref{eq:ChapB-9-016}),
we have
\begin{eqnarray}
  \label{eq:ChapB-9-019}
  \epsilon \,
  \langle\,f^\mathrm{eq}\,\varphi^\alpha_0\,,\,K^\mu \, \partial_\mu
  (
  \tilde{\pi}^{kl}\,\pi^{kl}
  +
  \tilde{J}^k\,J^k
  )\,\rangle
  = \left\{
  \begin{array}{ll}
    \displaystyle{
      0,
    }
    &
    \displaystyle{
      \alpha=0,
    }
    \\[2mm]
    \displaystyle{
      -\epsilon^2\,\nabla^j \pi^{ji},
    }
    &
    \displaystyle{
      \alpha=i,
    }
    \\[2mm]
    \displaystyle{
      -\epsilon^2\,(\nabla^j J^j + \pi^{jk}\,\sigma^{jk}),
    }
    &
    \displaystyle{
      \alpha=4.
    }
  \end{array}
  \right.
\end{eqnarray}

Substituting Eqs. (\ref{eq:ChapB-9-007}), (\ref{eq:ChapB-9-008}), and (\ref{eq:ChapB-9-019})
into Eq. (\ref{eq:ChapB-9-001}),
we have the balance equations as
\begin{eqnarray}
  \label{eq:ChapB-9-020}
  \frac{\partial}{\partial t}n
  &=& - \epsilon\,\Vec{\nabla}\cdot(n\,\Vec{u}) + O(\epsilon^3),\\
  \label{eq:ChapB-9-021}
  m\,n\,\frac{\partial}{\partial t}u^i
  &=&
  -\epsilon\,\Big[ m\,n\,\Vec{u}\cdot\Vec{\nabla}u^i + \nabla^i(n\,T) \Big]
  +\epsilon^2\,\nabla^j\pi^{ji} + O(\epsilon^3),\\
  \label{eq:ChapB-9-022}
  n\,\frac{\partial}{\partial t}(3\,T/2)
  &=&
  -\epsilon\,\Big[ n\,\Vec{u}\cdot\Vec{\nabla}(3\,T/2) + n\,T\,\Vec{\nabla}\cdot \Vec{u} \Big]
  +\epsilon^2\,\Big[ \nabla^j J^j + \pi^{jk}\,\sigma^{jk} \Big] + O(\epsilon^3).\nonumber\\
\end{eqnarray}
We emphasize that
Eqs. (\ref{eq:ChapB-9-020})-(\ref{eq:ChapB-9-022})
describe the slow motion of $n$, $u^i$, and $T$,
because the time derivative of them is
$O(\epsilon)$, as is manifest.

The term in the left-hand side of Eq. (\ref{eq:ChapB-9-002}) reads
\begin{eqnarray}
  \label{eq:ChapB-9-023}
  &&\epsilon\,
  \langle\,A^{-1}\,f^\mathrm{eq}\,(\hat{\pi}^{ij},\,\hat{J}^{i})\,,\,
  K^\mu \, \partial_\mu
  (
  \tilde{\pi}^{kl}\,\pi^{kl}
  +
  \tilde{J}^k\,J^k
  )\,\rangle\nonumber\\
  &=& 
  \epsilon\,
  (2\,T\,\eta\,\langle\,\tilde{\pi}^{ij}\,,\,
  K^\mu \, \partial_\mu
  (\tilde{\pi}^{kl}\,\pi^{kl}+\tilde{J}^{k}\,J^{k})\,\rangle,\,
  T^2\,\lambda\,\langle\,\tilde{J}^{i}\,,\,
  K^\mu \, \partial_\mu
  (\tilde{\pi}^{kl}\,\pi^{kl}+\tilde{J}^{k}\,J^{k})\,\rangle).
\end{eqnarray}
Using the expansions
\begin{eqnarray}
  \label{eq:ChapB-9-026}
  \langle\,\tilde{\pi}^{ij}\,,\,
  K^\mu \, \partial_\mu
  (\tilde{\pi}^{kl}\,\pi^{kl} + \tilde{J}^{k}\,J^{k})\,\rangle
  &=&
  \langle\,\tilde{\pi}^{ij}\,,\,
  K^\mu \,
  \tilde{\pi}^{kl}\,\rangle \,\partial_\mu\pi^{kl}
  +
  \langle\,\tilde{\pi}^{ij}\,,\,
  K^\mu \, \partial_\mu
  \tilde{\pi}^{kl}\,\rangle \,\pi^{kl}\nonumber\\
  &&{}+
  \langle\,\tilde{\pi}^{ij}\,,\,
  K^\mu \,
  \tilde{J}^{k}\,\rangle \,\partial_\mu J^{k}
  +
  \langle\,\tilde{\pi}^{ij}\,,\,
  K^\mu \, \partial_\mu
  \tilde{J}^{k}\,\rangle \,J^{k},\\
  \label{eq:ChapB-9-027}
  \langle\,\tilde{J}^{i}\,,\,
  K^\mu \, \partial_\mu
  (\tilde{\pi}^{kl}\,\pi^{kl} + \tilde{J}^{k}\,J^{k})\,\rangle
  &=&
  \langle\,\tilde{J}^{i}\,,\,
  K^\mu \,
  \tilde{\pi}^{kl}\,\rangle \,\partial_\mu\pi^{kl}
  +
  \langle\,\tilde{J}^{i}\,,\,
  K^\mu \, \partial_\mu
  \tilde{\pi}^{kl}\,\rangle \,\pi^{kl}\nonumber\\
  &&{}+
  \langle\,\tilde{J}^{i}\,,\,
  K^\mu \,
  \tilde{J}^{k}\,\rangle \,\partial_\mu J^{k}
  +
  \langle\,\tilde{J}^{i}\,,\,
  K^\mu \, \partial_\mu
  \tilde{J}^{k}\,\rangle \,J^{k},
\end{eqnarray}
we proceed to the further analysis of Eq. (\ref{eq:ChapB-9-023}).
First, the first and third terms in the right-hand side of Eqs. (\ref{eq:ChapB-9-026}) and (\ref{eq:ChapB-9-027}) read
\begin{eqnarray}
  \label{eq:ChapB-9-028}
  \langle\,\tilde{\pi}^{ij}\,,\,
  K^\mu \,
  \tilde{\pi}^{kl}\,\rangle \,\partial_\mu\pi^{kl}
  &=&
  \frac{1}{2\,T\,\eta}\,
  \tau_\pi\, \Big(\frac{\partial}{\partial t}+\epsilon\,\Vec{u}\cdot\Vec{\nabla}\Big)\pi^{ij},\\
  \label{eq:ChapB-9-029}
  \langle\,\tilde{\pi}^{ij}\,,\,
  K^\mu \,
  \tilde{J}^{k}\,\rangle \,\partial_\mu J^{k}
  &=& \epsilon\,
  \frac{1}{2\,T\,\eta}\,
  \ell_{\pi J}\, 
  \Delta^{ijmk}\,
  \nabla^m J^{k},\\
  \label{eq:ChapB-9-030}
  \langle\,\tilde{J}^{i}\,,\,
  K^\mu \,
  \tilde{\pi}^{kl}\,\rangle \,\partial_\mu\pi^{kl}
  &=& \epsilon\,
  \frac{1}{T^2\,\lambda}\,
  \ell_{J\pi}\,
  \Delta^{imkl}\,
  \nabla^m \pi^{kl},\\
  \label{eq:ChapB-9-031}
  \langle\,\tilde{J}^{i}\,,\,
  K^\mu \,
  \tilde{J}^{k}\,\rangle \,\partial_\mu J^{k}
  &=&
  \frac{1}{T^2\,\lambda}\,
  \tau_{J}\,\Big(\frac{\partial}{\partial t}+\epsilon\,\Vec{u}\cdot\Vec{\nabla}\Big)J^{i},
\end{eqnarray}
respectively,
where
\begin{eqnarray}
  \label{eq:ChapB-9-032}
  \tau_\pi &\equiv& \frac{2\,T\,\eta}{5} \,
  \langle\,\tilde{\pi}^{ij}\,,\,\tilde{\pi}^{ij}\,\rangle
  = \frac{1}{10\,T\,\eta}\,\langle\,A^{-1} \, f^\mathrm{eq}\,\hat{\pi}^{ij}\,,\,
  A^{-1} \, f^\mathrm{eq} \,
  \hat{\pi}^{ij}\,\rangle,\\
  \label{eq:ChapB-9-033}
  \tau_J &\equiv&
  \frac{T^2\,\lambda}{3}\,\langle\,\tilde{J}^i\,,\,\tilde{J}^i\,\rangle
  = \frac{1}{3\,T^2\,\lambda}\,\langle\,A^{-1} \, f^\mathrm{eq}\,\hat{J}^{i}\,,\,
  A^{-1} \, f^\mathrm{eq} \,
  \hat{J}^{i}\,\rangle,\\
  \label{eq:ChapB-9-034}
  \ell_{\pi J} &\equiv& \frac{2\,T\,\eta}{5} \,
  \langle\,\tilde{\pi}^{ij}\,,\,\delta K^i \, \tilde{J}^{j}\,\rangle
  =
  \frac{1}{5\,T^2\,\lambda} \,
  \langle\,A^{-1}\,f^{\mathrm{eq}}\,\hat{\pi}^{ij}\,,\,\delta
  K^i\,A^{-1}\,f^{\mathrm{eq}}
  \, \hat{J}^{j}\,\rangle
  ,\\
  \label{eq:ChapB-9-035}
  \ell_{J \pi} &\equiv& \frac{T^2\,\lambda}{5} \,
  \langle\,\tilde{J}^{i}\,,\,\delta K^j \, \tilde{\pi}^{ij}\,\rangle
  =
  \frac{1}{10\,T\,\eta} \,
  \langle\,A^{-1}\,f^{\mathrm{eq}}\,\hat{J}^{i}\,,\,\delta
  K^j\,A^{-1}\,f^{\mathrm{eq}}
  \, \hat{\pi}^{ij}\,\rangle,
\end{eqnarray}
with $\delta K_{\Vec{v}\Vec{k}}^i \equiv \delta v^i \,\delta_{\Vec{v}\Vec{k}}$.
In Eqs. (\ref{eq:ChapB-9-032})-(\ref{eq:ChapB-9-035}),
we have used the following identities:
\begin{eqnarray}
  \label{eq:ChapB-9-036}
  \langle\,\tilde{\pi}^{ij}\,,\, \tilde{\pi}^{kl}\,\rangle
  &=&
  \frac{1}{5}\,\Delta^{ijkl}\,\langle\,\tilde{\pi}^{ab}\,,\,\tilde{\pi}^{ab}\,\rangle,\,\,\,
  \langle\,\tilde{J}^{i}\,,\,\tilde{J}^{k}\,\rangle
  =
  \frac{1}{3}\,\delta^{ik}\,
  \langle\,\tilde{J}^{a}\,,\,
  \tilde{J}^{a}\,\rangle,\\
  \label{eq:ChapB-9-041}
  \langle\,\tilde{\pi}^{ij}\,,\,\delta K^m\,\tilde{J}^{k}\,\rangle
  &=&
  \frac{1}{5}\,\Delta^{ijmk}\,
  \langle\,\tilde{\pi}^{ab}\,,\,\delta K^a\,
  \tilde{J}^{b}\,\rangle,\\
  \langle\,\tilde{J}^{i}\,,\,\delta K^m\,\tilde{\pi}^{kl}\,\rangle
  &=&
  \frac{1}{5}\,\Delta^{imkl}\,
  \langle\,\tilde{J}^{a}\,,\,\delta K^b\,
  \tilde{\pi}^{ab}\,\rangle,\\
  \label{eq:ChapB-9-037}
  \langle\,\tilde{\pi}^{ij}\,,\,\tilde{J}^{k}\,\rangle
  &=&
  \langle\,\tilde{J}^{i}\,,\,\tilde{\pi}^{kl}\,\rangle
  =
  \langle\,\tilde{\pi}^{ij}\,,\,\delta K^m\,\tilde{\pi}^{kl}\,\rangle
  =
  \langle\,\tilde{J}^{i}\,,\,\delta K^m\,\tilde{J}^{k}\,\rangle
  =
  0.
\end{eqnarray}
We
find
that
the relaxation times $\tau_\pi$ and $\tau_J$ given by Eqs. (\ref{eq:ChapB-9-032}) and (\ref{eq:ChapB-9-033})
agree with
$\tau^{\mathrm{RG}}_\pi$ and $\tau^{\mathrm{RG}}_J$
in Eq. (\ref{eq:ChapB-4-3-012}), respectively,
using the inner product (\ref{eq:ChapB-2-2-030})
and $A^{-1} = f^{\mathrm{eq}} \, L^{-1} \, (f^{\mathrm{eq}})^{-1}$.
The coefficients
$\ell_{\pi J}$ and $\ell_{J\pi}$ defined in Eqs. (\ref{eq:ChapB-9-034}) and (\ref{eq:ChapB-9-035})
denote the relaxation lengths.

Next,
we consider
the second and fourth terms
in the right-hand side of Eqs. (\ref{eq:ChapB-9-026}) and (\ref{eq:ChapB-9-027}).
These terms
contain the temporal and spatial first-order derivatives of $n$, $T$, and $u^i$.
The temporal derivatives can be converted into the spatial derivatives with the use of the balance equations
(\ref{eq:ChapB-9-020})-(\ref{eq:ChapB-9-022}),
and there exists $\epsilon$ in front of the spatial derivatives.
Thus,
we can represent the above terms as the quantities of $O(\epsilon)$,
\begin{eqnarray}
  \label{eq:ChapB-9-049}
  \langle\,\tilde{\pi}^{ij}\,,\,K^\mu \, \partial_\mu \tilde{\pi}^{kl}\,\rangle
  &\equiv& -\epsilon\,\frac{1}{2\,T\,\eta}\,\bar{X}^{ijkl}_{\pi\pi},\,\,\,
  \langle\,\tilde{\pi}^{ij}\,,\,K^\mu \, \partial_\mu \tilde{J}^{k}\,\rangle
  \equiv - \epsilon\,\frac{1}{2\,T\,\eta}\,\bar{X}^{ijk}_{\pi J},\\
  \label{eq:ChapB-9-051}
  \langle\,\tilde{J}^{i}\,,\,K^\mu \, \partial_\mu \tilde{\pi}^{kl}\,\rangle
  &\equiv& - \epsilon\,\frac{1}{T^2\,\lambda}\,\bar{X}^{ikl}_{J \pi},\,\,\,
  \langle\,\tilde{J}^{i}\,,\,K^\mu \, \partial_\mu \tilde{J}^{k}\,\rangle
  \equiv - \epsilon\,\frac{1}{T^2\,\lambda}\,\bar{X}^{ik}_{J J}.
\end{eqnarray}
Their explicit forms are given by
\begin{eqnarray}
  \bar{X}^{ijkl}_{\pi\pi} &=& \Delta^{ijkl}\,\kappa^{(1)}_{\pi\pi}\,\theta
  + \Delta^{ijac}\,\Delta^{cbkl}\,(\kappa^{(2)}_{\pi\pi}\,\sigma^{ab}
  + \kappa^{(3)}_{\pi\pi}\,\omega^{ab}),\\
  \bar{X}^{ijk}_{\pi J} &=& \Delta^{ijak} \, (\kappa^{(1)}_{\pi J}\,\nabla^a T + \kappa^{(2)}_{\pi J}\,\nabla^a n),\\
  \bar{X}^{ikl}_{J \pi} &=& \Delta^{iakl} \, (\kappa^{(1)}_{J \pi}\,\nabla^a T + \kappa^{(2)}_{J \pi}\,\nabla^a n),\\
%
  \bar{X}^{ik}_{JJ} &=& \delta^{ik}\,\kappa^{(1)}_{JJ}\,\theta
  + \kappa^{(2)}_{JJ}\,\sigma^{ik}
  + \kappa^{(3)}_{JJ}\,\omega^{ik},
\end{eqnarray}
with
$\theta = \Vec{\nabla}\cdot\Vec{u}$,
$\sigma^{ij} = \Delta^{ijkl} \, \nabla^k u^l$,
and
$\omega^{ij} = (\nabla^i u^j - \nabla^j u^i)/2$.
Here,
the coefficients
$\kappa^{(1)}_{\pi\pi}$,
$\kappa^{(2)}_{\pi\pi}$,
$\kappa^{(3)}_{\pi\pi}$,
$\kappa^{(1)}_{\pi J}$,
$\kappa^{(2)}_{\pi J}$,
$\kappa^{(1)}_{J \pi}$,
$\kappa^{(2)}_{J \pi}$,
$\kappa^{(1)}_{JJ}$,
$\kappa^{(2)}_{JJ}$,
and $\kappa^{(3)}_{JJ}$
are defined by
\begin{eqnarray}
  \kappa^{(1)}_{\pi\pi} &\equiv& \frac{\Delta^{ijkl}}{-\frac{5}{2\,T\,\eta}}\,
  \langle\, \tilde{\pi}^{ij} \,,\, ((\delta^{ab}/3)\,\delta K^a \, \partial/\partial u^b - (2\,T/3)\,\partial/\partial T - n\,\partial/\partial n) \, \tilde{\pi}^{kl} \,\rangle,\\
  \kappa^{(2)}_{\pi\pi} &\equiv& \frac{\Delta^{ijac}\,\Delta^{cbkl}+\Delta^{ijbc}\,\Delta^{cakl}-\frac{2}{3}\,\Delta^{ijkl}\,\delta^{ab}}{-\frac{35}{12\,T\,\eta}}\,
  \langle\, \tilde{\pi}^{ij} \,,\, \delta K^a \, \partial/\partial u^b \, \tilde{\pi}^{kl} \,\rangle,\\
  \kappa^{(3)}_{\pi\pi} &\equiv& \frac{\Delta^{ijac}\,\Delta^{cbkl}-\Delta^{ijbc}\,\Delta^{cakl}}{-\frac{15}{4\,T\,\eta}}\,
  \langle\, \tilde{\pi}^{ij} \,,\, \delta K^a \, \partial/\partial u^b \, \tilde{\pi}^{kl} \,\rangle,\\
  \kappa^{(1)}_{\pi J} &\equiv& \frac{\Delta^{ijak}}{-\frac{5}{T^2\,\lambda}}\,
  \langle\, \tilde{\pi}^{ij} \,,\, (\delta K^a \, \partial/\partial T - (1/m)\,\partial/\partial u^a) \, \tilde{J}^k \,\rangle,\\
  \kappa^{(2)}_{\pi J} &\equiv& \frac{\Delta^{ijak}}{-\frac{5}{T^2\,\lambda}}\,
  \langle\, \tilde{\pi}^{ij} \,,\, (\delta K^a \, \partial/\partial n - (T/m\,n)\,\partial/\partial u^a) \, \tilde{J}^k \,\rangle,\\
  \kappa^{(1)}_{J \pi} &\equiv& \frac{\Delta^{iakl}}{-\frac{5}{2\,T\,\eta}}\,
  \langle\, \tilde{J}^{i} \,,\, (\delta K^a \, \partial/\partial T - (1/m)\,\partial/\partial u^a) \, \tilde{\pi}^{kl} \,\rangle,\\
  \kappa^{(2)}_{J \pi} &\equiv& \frac{\Delta^{iakl}}{-\frac{5}{2\,T\,\eta}}\,
  \langle\, \tilde{J}^{i} \,,\, (\delta K^a \, \partial/\partial n - (T/m\,n)\,\partial/\partial u^a) \, \tilde{\pi}^{kl} \,\rangle,\\
  \kappa^{(1)}_{JJ} &\equiv& \frac{\delta^{ik}}{-\frac{3}{T^2\,\lambda}}\,
  \langle\, \tilde{J}^{i} \,,\, ((\delta^{ab}/3)\,\delta K^a \, \partial/\partial u^b - (2\,T/3)\,\partial/\partial T - n\,\partial/\partial n) \, \tilde{J}^k \,\rangle,\\
  \kappa^{(2)}_{JJ} &\equiv& \frac{\Delta^{ikab}}{-\frac{5}{T^2\,\lambda}}\,
  \langle\, \tilde{J}^{i} \,,\, \delta K^a \, \partial/\partial u^b \, \tilde{J}^k \,\rangle,\\
  \kappa^{(3)}_{JJ} &\equiv& \frac{\delta^{ia}\,\delta^{bk} - \delta^{ib}\,\delta^{ak}}{-\frac{6}{T^2\,\lambda}}\,
  \langle\, \tilde{J}^{i} \,,\, \delta K^a \, \partial/\partial u^b \, \tilde{J}^k \,\rangle.
\end{eqnarray}

The second term in the right-hand side of Eq. (\ref{eq:ChapB-9-002}) reads
\begin{eqnarray}
  \label{eq:ChapB-9-063}
  &&\epsilon^2 \, \frac{1}{2}\,\langle \, A^{-1} \, f^\mathrm{eq} \, (\hat{\pi}^{ij},\,\hat{J}^{i})\,,\,
  B\,\big[
  \tilde{\pi}^{kl}\,\pi^{kl} + \tilde{J}^k\,J^k
  \,,\,
  \tilde{\pi}^{mn}\,\pi^{mn} + \tilde{J}^m\,J^m
  \big] \, \rangle
  \nonumber\\
  &=&
(
\epsilon^2 \, b_{\pi\pi\pi} \, \Delta^{ijkl} \, \pi^{km} \, \pi^{ml} + \epsilon^2 \, b_{\pi JJ} \, \Delta^{ijkl} \, J^k \, J^l,\,
\epsilon^2 \, b_{J \pi J} \, \pi^{ik} \, J^{k}
).
\end{eqnarray}
We note that 
$b_{\pi\pi\pi}$, $b_{\pi JJ}$ and $b_{J \pi J}$
denote the coefficients in the non-linear terms of $\pi^{ij}$ and $J^i$,
whose definitions are given by
\begin{eqnarray}
  \label{eq:ChapB-9-064}
  b_{\pi\pi\pi} &\equiv& \frac{3}{70\,T^2\,\eta^2}\, \Delta^{ijkl} \, \langle \, A^{-1} \, f^\mathrm{eq} \, \hat{\pi}^{ij}\,,\,
  B\,\big[ A^{-1} \, f^\mathrm{eq} \, \hat{\pi}^{km} \,,\, A^{-1} \, f^\mathrm{eq} \, \hat{\pi}^{ml} \big] \, \rangle,\\
  b_{\pi JJ} &\equiv& \frac{1}{10\,T^4\,\lambda^2}\, \Delta^{ijkl} \, \langle \, A^{-1} \, f^\mathrm{eq} \, \hat{\pi}^{ij}\,,\,
  B\,\big[ A^{-1} \, f^\mathrm{eq} \, \hat{J}^{k} \,,\, A^{-1} \, f^\mathrm{eq} \, \hat{J}^{l} \big] \, \rangle,\\
  \label{eq:ChapB-9-065}
  b_{J \pi J} &\equiv& \frac{1}{10\,T^3\,\eta\,\lambda}\, \Delta^{ijkl} \, \langle \, A^{-1} \, f^\mathrm{eq} \, \hat{J}^{i}\,,\,
  B\,\big[ A^{-1} \, f^\mathrm{eq} \, \hat{\pi}^{kl} \,,\, A^{-1} \, f^\mathrm{eq} \, \hat{J}^{j} \big] \, \rangle.
\end{eqnarray}

By substituting Eqs. (\ref{eq:ChapB-9-011}) and (\ref{eq:ChapB-9-023})
together with Eqs. (\ref{eq:ChapB-9-026})-(\ref{eq:ChapB-9-031}),
(\ref{eq:ChapB-9-049}), (\ref{eq:ChapB-9-051}),
and (\ref{eq:ChapB-9-063}) into Eq. (\ref{eq:ChapB-9-002}),
we have the relaxation equations as
\begin{eqnarray}
  \label{eq:ChapB-9-053}
  \epsilon\,\pi^{ij}
  &=&
  \epsilon\,2\,\eta\,\sigma^{ij}
  - \epsilon\,\tau_\pi \, \Big(\frac{\partial}{\partial t}+\epsilon\,\Vec{u}\cdot\Vec{\nabla}\Big)\pi^{ij}
  - \epsilon^2\,\ell_{\pi J} \, \Delta^{ijmk} \, \nabla^m J^k
  + \epsilon^2\,\bar{X}^{ijkl}_{\pi\pi}\,\pi^{kl}
  + \epsilon^2\,\bar{X}^{ijk}_{\pi J}\,J^{k}\nonumber\\
  &&{}+ \epsilon^2 \, b_{\pi\pi\pi} \, \Delta^{ijkl} \, \pi^{km} \, \pi^{ml}
  + \epsilon^2 \, b_{\pi JJ} \, \Delta^{ijkl} \, J^k \, J^l
  + O(\epsilon^3),\\
  \label{eq:ChapB-9-054}
  \epsilon\,J^{i}
  &=&
  \epsilon\,\lambda\,\nabla^i T
  - \epsilon\,\tau_J \, \Big(\frac{\partial}{\partial t}+\epsilon\,\Vec{u}\cdot\Vec{\nabla}\Big)J^{i}
  - \epsilon^2\,\ell_{J\pi} \, \Delta^{imkl}\,\nabla^m \pi^{kl}
  + \epsilon^2\,\bar{X}^{ikl}_{J\pi}\,\pi^{kl}
  + \epsilon^2\,\bar{X}^{ik}_{JJ}\,J^{k}\nonumber\\
  &&{}+ \epsilon^2 \, b_{J \pi J} \, \pi^{ik} \, J^{k}
  + O(\epsilon^3).
\end{eqnarray}

Finally,
we present an explicit form of the invariant/attractive manifold.
Equation (\ref{eq:ChapB-3-6-004}) can be reduced to
\begin{eqnarray}
  \label{eq:ChapB-4-2-009}
  f^{\mathrm{global}}_{\Vec{v}}
  &=& f^{\mathrm{eq}}_{\Vec{v}}
  + \epsilon \,\Bigg[
  \frac{\big[ A^{-1} \, f^{\mathrm{eq}} \, \hat{\pi}^{ij} \big]_{\Vec{v}}\,\pi^{ij}}{2\,T\,\eta}
  + \frac{\big[ A^{-1} \, f^{\mathrm{eq}} \, \hat{J}^{i} \big]_{\Vec{v}}\,J^{i}}{T^2\,\lambda}
  \Bigg]\nonumber\\
  &&{}- \epsilon^2 \, \big[ Q_1 \, (A - \partial/\partial s)^{-1}\,
  Q_0 \, (\frac{1}{2}\,B \, \big[ \psi(s)\,,\,\psi(s) \big]
  + F_1 \, \psi(s))\Big|_{s=0} \big]_{\Vec{v}}
  + O(\epsilon^3),\nonumber\\
\end{eqnarray}
where
\begin{eqnarray}
  \psi_{\Vec{v}}(s)
  &\equiv&
  \frac{\big[ \mathrm{e}^{As}\,A^{-1} \, f^{\mathrm{eq}} \, \hat{\pi}^{ij} \big]_{\Vec{v}}\,\pi^{ij}}{2\,T\,\eta}
  + \frac{\big[ \mathrm{e}^{As}\,A^{-1} \, f^{\mathrm{eq}} \, \hat{J}^{i} \big]_{\Vec{v}}\,J^{i}}{T^2\,\lambda}
  + s \, \big[P_0 \, \Vec{F}_0\big]_{\Vec{v}}\nonumber\\
  &&{}
  -
  \frac{\big[
  (\mathrm{e}^{As} - 1) \, A^{-1} \, f^{\mathrm{eq}}
  \, \hat{\pi}^{ij}
  \big]_{\Vec{v}}\,\sigma^{ij}}{T}
  -
  \frac{\big[
  (\mathrm{e}^{As} - 1) \, A^{-1} \, f^{\mathrm{eq}}
  \, \hat{J}^{i}
  \big]_{\Vec{v}}\,\nabla^i T}{T^2}.
\end{eqnarray}

Substituting
$\eta = \eta^{\mathrm{RG}}$,
$\lambda = \lambda^{\mathrm{RG}}$,
$\tau_\pi = \tau_\pi^{\mathrm{RG}}$,
$\tau_J = \tau_J^{\mathrm{RG}}$,
$\ell_{\pi J} = \ell_{\pi J}^{\mathrm{RG}}$,
$\ell_{J\pi} = \ell_{J\pi}^{\mathrm{RG}}$,
$\kappa^{(1,2,3)}_{\pi\pi} = \kappa^{(1,2,3)\mathrm{RG}}_{\pi\pi}$,
$\kappa^{(1,2)}_{\pi J} = \kappa^{(1,2)\mathrm{RG}}_{\pi J}$,
$\kappa^{(1,2)}_{J\pi} = \kappa^{(1,2)\mathrm{RG}}_{J\pi}$,
$\kappa^{(1,2,3)}_{JJ} = \kappa^{(1,2,3)\mathrm{RG}}_{JJ}$,
$b_{\pi\pi\pi} = b_{\pi\pi\pi}^{\mathrm{RG}}$,
$b_{\pi JJ} = b_{\pi JJ}^{\mathrm{RG}}$,
$b_{J \pi J} = b_{J \pi J}^{\mathrm{RG}}$,
$A^{-1} = f^{\mathrm{eq}} \, L^{-1} \, (f^{\mathrm{eq}})^{-1}$,
and explicit forms of
$\bar{X}^{ijkl}_{\pi\pi}$,
$\bar{X}^{ijk}_{\pi J}$,
$\bar{X}^{ikl}_{J\pi}$,
and $\bar{X}^{ik}_{JJ}$
into the balance equations (\ref{eq:ChapB-9-020})-(\ref{eq:ChapB-9-022}),
relaxation equations (\ref{eq:ChapB-9-053}) and (\ref{eq:ChapB-9-054}),
and invariant/attractive manifold (\ref{eq:ChapB-4-2-009}),
we arrive at Eqs. (\ref{eq:causalhydrodynamics1})-(\ref{eq:causalhydrodynamics5}),
and (\ref{eq:invariantmanifold}).

\subsection{
  Detailed derivation of explicit form of excited modes
}
\label{sec:ChapB-8}
In this section,
we present the detailed derivation of $A^{-1}\,Q_0\,\Vec{F}_0$,
whose explicit form is given by Eq. (\ref{eq:A-1Q0F0}).
Because $Q_0\,\Vec{F}_0$ takes the form
\begin{eqnarray}
  \label{eq:Q0F0}
  Q_0\,\Vec{F}_0
  = - \big[ Q_0\,K^i \, f^{\mathrm{eq}} \, \varphi^{0}_0 \big]_{\Vec{v}} \, \frac{1}{n}\,\nabla^i n
  - \big[ Q_0\,K^i \, f^{\mathrm{eq}} \, \varphi^{4}_0 \big]_{\Vec{v}} \, \frac{1}{T^2}\,\nabla^i T
  - \big[ Q_0\,K^i \, f^{\mathrm{eq}} \, \varphi^{j}_0 \big]_{\Vec{v}} \, \frac{1}{T}\,\nabla^i u^j,\nonumber\\
\end{eqnarray}
with $K^i_{\Vec{v}\Vec{k}} = v^i\,\delta_{\Vec{v}\Vec{k}}$,
the calculation of $A^{-1}\,Q_0\,\Vec{F}_0$ can be reduced to that of
$\big[ Q_0 \, K^i \, f^{\mathrm{eq}} \, \varphi^{\alpha}_0 \big]_{\Vec{v}}$.

First,
we utilize $P_0 = 1 - Q_0$ in Eq. (\ref{eq:ChapB-4-1-014})
and $K^i_{\Vec{v}\Vec{k}} = u^i\,\delta_{\Vec{v}\Vec{k}} + \delta K^i_{\Vec{v}\Vec{k}}$
to expand $\big[ Q_0 \, K^i \, f^{\mathrm{eq}} \, \varphi^{\alpha}_0 \big]_{\Vec{v}}$ as
\begin{eqnarray}
  \label{eq:Q0fKiphi0_2}
  \big[ Q_0 \, K^i \, f^{\mathrm{eq}} \, \varphi^{\alpha}_0 \big]_{\Vec{v}}
  =
  \big[ Q_0 \, \delta K^i \, f^{\mathrm{eq}} \, \varphi^{\alpha}_0 \big]_{\Vec{v}}
  =
  \big[ \delta K^i \, f^{\mathrm{eq}} \, \varphi^{\alpha}_0 \big]_{\Vec{v}}
  -
  \sum_{\beta=0}^4 \, f^{\mathrm{eq}}_{\Vec{v}} \, \varphi^\beta_{0\Vec{v}} \,
  \frac{1}{c^\beta} \,
  M^{\beta i \alpha},
\end{eqnarray}
with
\begin{eqnarray}
  M^{\beta i \alpha} \equiv \langle\,  f^{\mathrm{eq}} \, \varphi^\beta_{0} \,,\,
  \delta K^i \, f^{\mathrm{eq}} \, \varphi^{\alpha}_0
  \,\rangle
  = \int_{\Vec{v}}\,
  f^{\mathrm{eq}}_{\Vec{v}}\,
  \varphi^\beta_{0\Vec{v}} \,
  \delta v^i\,
  \varphi^\alpha_{0\Vec{v}}.
\end{eqnarray}
We note that
$M^{\beta i \alpha}$ can be calculated as
\begin{eqnarray}
  \label{eq:ChapB-8-004}
  M^{0i0} &=& M^{0i4} = M^{jik} = M^{4i0} = M^{4i4} = 0,\\
  \label{eq:ChapB-8-005}
  M^{0ij} &=& M^{ji0} = n\,T\,\delta^{ij},\,\,\,
  M^{ji4} = M^{4ij} = n\,T^2 \, \delta^{ji}.
\end{eqnarray}

Then,
substituting the above $M^{\beta i \alpha}$ into Eq. (\ref{eq:Q0fKiphi0_2}),
we have
\begin{eqnarray}
  \label{eq:Q0fKiphi0_3}
  \big[ Q_0 \, K^i \, f^{\mathrm{eq}} \, \varphi^{\alpha}_0 \big]_{\Vec{v}}
  =
  \left\{
  \begin{array}{ll}
    \displaystyle{
      0,
    }
    &
    \displaystyle{
      \alpha=0,
    }
    \\[2mm]
    \displaystyle{
      f^{\mathrm{eq}}_{\Vec{v}}\,m\,\Delta^{ijkl}\,\delta v^k\,\delta v^l,
    }
    &
    \displaystyle{
      \alpha=j,
    }
    \\[2mm]
    \displaystyle{
      f^{\mathrm{eq}}_{\Vec{v}}\,\Big(\frac{m}{2}\,|\Vec{\delta v}|^2 - \frac{5}{2}\,T\Big)\,\delta v^i,
    }
    &
    \displaystyle{
      \alpha=4,
    }
  \end{array}
  \right.
\end{eqnarray}
which is identical to Eq. (\ref{eq:Q0fKiphi0}).
By combining Eq. (\ref{eq:Q0fKiphi0_3}) with Eq. (\ref{eq:Q0F0}),
we obtain $A^{-1}\,Q_0\,\Vec{F}_0$ in Eq. (\ref{eq:A-1Q0F0}).

\end{document}